\documentclass[11pt,a4paper]{article}
\usepackage{epsfig}
\usepackage{float}
\usepackage{jheppub}

\newcommand{\ba}{\begin{array}}
\newcommand{\ea}{\end{array}}
\newcommand{\bd}{\begin{displaymath}}
\newcommand{\ed}{\end{displaymath}}
\newcommand{\bi}{\begin{itemize}}
\newcommand{\ei}{\end{itemize}}
\newcommand{\benu}{\begin{enumerate}}
\newcommand{\eenu}{\end{enumerate}}
\newcommand{\be}{\begin{equation}}
\newcommand{\ee}{\end{equation}}
\newcommand{\bea}{\begin{eqnarray}}
\newcommand{\eea}{\end{eqnarray}}
\def\ltap{\ \raisebox{-.4ex}{\rlap{$\sim$}} \raisebox{.4ex}{$<$}\ }
\def\gtap{\ \raisebox{-.4ex}{\rlap{$\sim$}} \raisebox{.4ex}{$>$}\ }
\newcommand{\dmsol}{\mbox{$\Delta m^2_{\odot}$}}

\newcommand{\pmns}{\mbox{$U_{\rm PMNS}$}}
\newcommand{\bad}{\begin{array}{ccc}}
\newcommand{\eV}{\mbox{$ \  \mathrm{eV} \ $}}

\newcommand{\beq}{\begin{equation}}
\newcommand{\eeq}{\end{equation}}
\newcommand{\deltaatm}{\mbox{$\Delta m^2_{31}$}}
\newcommand{\deltasol}{\mbox{$ \Delta m^2_{21}$}}

\title{Neutrino Mass Hierarchy Determination Using Reactor Antineutrinos}

\author[a]{Pomita Ghoshal}
\author[a,b,1]{and S. T. Petcov%
\note{Also at: Institute of Nuclear Research and
Nuclear Energy, Bulgarian Academy of Sciences, 1784 Sofia, Bulgaria.}}

\affiliation[a]{SISSA and INFN-Sezione di Trieste,\\ 
34136 Trieste, Italy}
\affiliation[b]{IPMU, University of Tokyo,\\ 
Tokyo, Japan}

\emailAdd{pghoshal@sissa.it}
\emailAdd{petcov@sissa.it}

\abstract{Building on earlier studies,
we investigate the possibility to 
determine the type of neutrino mass spectrum
(i.e., ``the neutrino mass hierarchy'')
in a high statistics reactor $\bar{\nu}_e$ experiment 
with a relatively large KamLAND-like detector 
and an optimal baseline of 60 Km.
We analyze systematically the 
Fourier Sine and Cosine Transforms (FST and FCT) 
of simulated reactor antineutrino data with reference to their
specific mass hierarchy-dependent features 
discussed earlier in the literature. 
We perform also a binned 
$\chi^2$ analysis of the sensitivity 
of simulated reactor $\bar{\nu}_e$ 
event spectrum data to the 
neutrino mass hierarchy, and 
determine, in particular,
the characteristics of the detector 
and the experiment (energy resolution, visible 
energy threshold, exposure, systematic errors, binning of data, etc.),
which would allow us to get significant 
information on, or even determine, 
the type of the neutrino mass spectrum.
We find that if 
$\sin^2 2\theta_{13}$ is sufficiently large, 
$\sin^2 2\theta_{13} \gtap 0.02$,
the requirements on the set-up of interest 
are very challenging, but not impossible to realize.} 


\arxivnumber{1011.1646}

\begin{document}
\maketitle

\flushbottom

\section{Introduction}

\vspace{-0.6cm}
\hskip 1cm 

  The experiments with solar,
atmospheric, reactor and accelerator neutrinos
\cite{cleveland98,fukuda96,abd09,anselmann92,SKsolar02,ahmad01,
SKatm98,SKdip04,KL162,BOREX,ahn06,Michael06}
have provided compelling evidences for the
existence of flavour neutrino oscillations \cite{BPont57,MNS62} 
caused by nonzero neutrino masses and neutrino mixing.
The data imply the presence of neutrino mixing in the weak charged 
lepton current:
\begin{equation}
\nu_{l \mathrm{L}}(x)  = \sum_{j} U_{l j} \, \nu_{j \mathrm{L}}(x),~~
l  = e,\mu,\tau, 
\label{3numixCC}
\end{equation}

\noindent where 
$\nu_{lL}$ are the flavour neutrino fields, 
$\nu_{j \mathrm{L}}(x)$ is the 
left-handed (LH)
component of the field of 
the neutrino $\nu_j$ possessing a mass $m_j$ and $U$ 
is a unitary matrix - the
Pontecorvo-Maki-Nakagawa-Sakata (PMNS)
neutrino mixing matrix \cite{BPont57,MNS62,BPont67}. 

\vskip.3cm

   All compelling  neutrino oscillation data
can be described assuming 
3-flavour neutrino mixing in vacuum.
The data on the invisible decay width of the $Z^0$-boson 
is compatible with only 3 light flavour neutrinos 
 coupled to $Z^0$ (see, e.g. \cite{Znu}). 
The number of massive neutrinos $\nu_j$, $n$, 
can, in general, be greater than 3, $n>3$,
if, for instance, there exist right-handed (RH)
sterile neutrinos \cite{BPont67} 
and they mix with the LH flavour neutrinos. 
It follows from the existing data that 
at least 3 of the neutrinos $\nu_j$, say 
$\nu_1$, $\nu_2$, $\nu_3$, must be light,
$m_{1,2,3} \ltap 1$ eV, and must have different 
masses, $m_1\neq m_2 \neq m_3$. 
At present there are no compelling 
experimental evidences for the existence  
of more than 3 light neutrinos. 

\vskip.3cm

 Being electrically neutral, the 
massive neutrinos $\nu_j$ can be 
Dirac fermions (possessing distinctive 
antiparticles), or Majorana particles 
(which are identical with their respective 
antiparticles, see, e.g., \cite{BiPet87}).
On the basis of the existing neutrino 
data it is impossible to 
determine whether the massive neutrinos 
are Dirac or Majorana fermions.

\vskip.3cm

 In the case of 3 light neutrinos, the 
neutrino mixing matrix $U$ can be 
parametrized by 3 angles and, 
depending on whether 
the  massive neutrinos $\nu_j$ are Dirac 
or Majorana particles, 
by 1 or 3 CP violation (CPV) phases 
\cite{BHP80}: 
\bea \label{eq:Upara}
\pmns = \left( \bad 
c_{12} c_{13} & s_{12} c_{13} & s_{13}e^{-i \delta}  \\[0.2cm] 
 -s_{12} c_{23} - c_{12} s_{23} s_{13} e^{i \delta} 
 & c_{12} c_{23} - s_{12} s_{23} s_{13} e^{i \delta} 
& s_{23} c_{13}  \\[0.2cm] 
 s_{12} s_{23} - c_{12} c_{23} s_{13} e^{i \delta} & 
 - c_{12} s_{23} - s_{12} c_{23} s_{13} e^{i \delta} 
 & c_{23} c_{13} \\ 
                \ea   \right) 
~{\rm diag}(1, e^{i \frac{\alpha_{21}}{2}}, e^{i \frac{\alpha_{31}}{2}})
\eea

\noindent where 
$c_{ij} \equiv \cos\theta_{ij}$,
$s_{ij} \equiv \sin\theta_{ij}$, 
$\theta_{ij} = [0,\pi/2]$,
$\delta = [0,2\pi]$ is the 
Dirac CP-violation (CPV) phase and $\alpha_{21},\alpha_{31}$ are 
two Majorana  CPV phases 
\footnote{The two Majorana 
CP-violation phases~\cite{BHP80}
do not enter into the expressions for the oscillation 
probabilities of interest~\cite{BHP80,Lang87} and we are 
not going to discuss them further. They play important 
role in the phenomenology of neutrinoless double 
beta decay (see, e.g., \cite{BPP1}). 
The phases $\alpha_{21,31}$ 
can affect significantly the predictions for 
the rates of the (LFV) decays $\mu \rightarrow e + \gamma$,
$\tau \rightarrow \mu + \gamma$, etc.
in a large class of supersymmetric theories
incorporating the see-saw mechanism 
\cite{PPY03}. The Majorana phases 
can provide the CP violation, necessary 
for the generation of the baryon asymmetry 
of the Universe 
in the leptogenesis scenario 
of the asymmetry origins.
\cite{PPRio106}. 
}.
If one identifies $\Delta m^2_{21} > 0$ and $\Delta m^2_{31}$
(or  $\Delta m^2_{32}$)  
with the neutrino mass squared differences
which drive the solar and atmospheric 
neutrino oscillations, 
$\theta_{12}$ and $\theta_{23}$
represent the solar and atmospheric 
neutrino mixing angles, 
while $\theta_{13}$ is the 
CHOOZ angle 
\cite{CHOOZ}.
The existing  
oscillation data allow us to determine 
$\Delta m^2_{21} \equiv \dmsol$, $\theta_{12}$, and
$|\Delta m^2_{31}| \equiv |\Delta m^2_{\rm atm}|$,
$\theta_{23}$, with a relatively good precision 
\cite{BCGPRKL2,Fogli08,TSchw08}, 
and to obtain rather stringent limits on the angle 
$\theta_{13}$ \cite{CHOOZ}.
The best fit values and the 99.73\% C.L. allowed ranges of 
$\Delta m^2_{21}$, $\sin^2\theta_{12}$, 
$|\Delta m^2_{31(32)}|$ and $\sin^2\theta_{23}$,
read \cite{TSchw08}:
\vspace{-0.1cm}
\begin{eqnarray}
\label{deltasolvalues}
\Delta m^2_{21} = 7.59^{+0.23}_{-0.18} \times 10^{-5}\ \eV^2, 
&\Delta m^2_{21} = (7.03 - 8.27) \times 10^{-5} \ \eV^2\,,\\
\label{sinsolvalues}
\sin^2 \theta_{12} = 0.318^{+0.019}_{-0.016},
&0.27 \leq \sin^2 \theta_{12} \leq 0.38\,,\\ 
\label{deltaatmvalues}
|\Delta m^2_{31}| = 2.40^{+0.12}_{-0.11} \times 10^{-3} \ \eV^2, 
&|\Delta m^2_{31}| = (2.07 - 2.75) \times 10^{-3} \ \eV^2\,,\\ 
\label{thetaatmvalues}
\sin^2\theta_{23} = 0.5^{+0.07}_{-0.06}, 
&0.36 \leq \sin^2\theta_{23} \leq 0.67\,.
\end{eqnarray}
Thus, we have  $|\Delta m^2_{31(2)}| >> \Delta m^2_{21}$,
$\Delta m^2_{21}/|\Delta m^2_{31}| \cong 0.03$, and
$|\Delta m^2_{31}| = |\Delta m^2_{32} - \Delta m^2_{21}| 
\cong |\Delta m^2_{32}|$.
Maximal solar neutrino mixing, i.e. 
$\theta_{12} = \pi/4$, is ruled out at more than 
6$\sigma$ by the data. Correspondingly, one has  
$\cos2\theta_{12} \geq 0.26$ (at $99.73\%$ C.L.).
A combined 3-neutrino oscillation analysis of the global data
gives \cite{MMTSchw10}:  
\vspace{-0.1cm}
\beq
\sin^2\theta_{13} < 0.031~(0.047)~~~{\rm at}~90\%~(99.73\%)~{\rm C.L.}
\label{th13glob}
\eeq
The results of the global analyzes include 
also a weak indication of nonzero 
$\sin^2\theta_{13}\sim 0.01$ (for a review 
see \cite{MMTSchw10}). 
If $\theta_{13} \neq 0$, the Dirac 
phase $\delta$ can generate
CP violation effects in neutrino oscillations 
\cite{Cabibbo78,BHP80,VBarg80CP}.
The size of the indicated leptonic CP violation effects 
depends on the magnitude of the currently unknown 
values of 
$\theta_{13}$ and 
$\delta$ \cite{PKSP3nu88}.

\vskip.3cm

 The existing data do not allow us to determine the 
sign of $\Delta m^2_{31(32)}$. The two possibilities,
$\Delta m^2_{31(32)} > 0$ or $\Delta m^2_{31(32)} < 0$,
as is well known, correspond to two different
types of neutrino mass spectrum:
with normal ordering (hierarchy (NO,NH)),
$m_1 < m_2 < m_3$, and 
with inverted ordering (hierarchy (IO,IH)),
$m_3 < m_1 < m_2$.  

\vskip.3cm

 Determining the nature - Dirac or Majorana, 
of massive neutrinos, getting more precise
information about the value 
of the mixing angle $\theta_{13}$,
determining the sign of $\deltaatm$,
or the type of the neutrino mass spectrum 
(with normal or inverted ordering (hierarchy)
\footnote{We use here and in what follows 
the generic terms ``normal hierarchical'' and 
``inverted hierarchical'' for the neutrino mass spectra
with normal ordering and inverted ordering, i.e., 
the spectra need not necessarily be hierarchical. 
We will use also the widely accepted 
term ``neutrino mass hierarchy'' for 
${\rm sgn}(\Delta m^2_{\rm atm})$ 
(i.e., for the neutrino mass ordering). 
}) and getting information about the status of the 
CP symmetry in the lepton sector 
are among the major and remarkably 
challenging goals of future studies in 
neutrino physics (see, e.g., 
\cite{STPNu04,Future,MMTSchw10}).
Establishing whether the neutrino mass spectrum 
is with normal or inverted hierarchy, i.e., 
measuring the sign of $\deltaatm$ and 
determining the nature of massive neutrinos,
in particular, are of fundamental importance 
for understanding the origin of neutrino 
masses and mixing (see, e.g., \cite{ThRMoh05}).

\vskip.3cm

    In the present article we
continue the studies of the possibility
to obtain information about the 
type of spectrum the light neutrino masses 
obey (i.e., about ${\rm sgn}(\deltaatm)$) 
in experiments with reactor antineutrinos.
This possibility was discussed first in 
\cite{PiaiP0103} and later was further 
investigated in \cite{SCSPMP03,Hano1,Hano2,YWang08}.
It is based on the observation that for 
$\cos2\theta_{12} \neq 0$ and $\sin\theta_{13}\neq 0$, 
the probabilities of $\bar{\nu}_e$ survival 
in the cases of NO (NH) and IO (IH) spectra 
differ \cite{PiaiP0103,BiNiPe02}: 
$P^{NH}({\bar \nu_e}\to{\bar \nu_e})\neq
P^{IH}({\bar \nu_e}\to{\bar \nu_e})$.
For sufficiently large $|\cos2\theta_{12}|$ and 
$\sin^2\theta_{13}$ and a baseline of 
several tens of kilometers, this difference 
in the  $\bar{\nu}_e$ oscillations
leads, in principle, to an observable
difference in the deformations 
of the spectrum of $e^+$ \cite{PiaiP0103}, 
produced in the inverse beta-decay reaction 
$\bar{\nu}_e + p \rightarrow e^+ + n$ 
by which the reactor $\bar{\nu}_e$ 
are detected. In \cite{SCSPMP03} 
the physics potential
of a reactor neutrino experiment 
with a relatively large detector at
a distance of several tens of kilometers 
has been analyzed in detail. 
More specifically, the strategies 
and the experimental set-up,
which would permit to measure
$\Delta m^2_{21}$ and $\sin^2\theta_{12}$
with a high precision,
get information on (or even measure) 
$\sin^2\theta_{13}$, and if 
$\sin^2\theta_{13}$ is sufficiently large
($\sin^2\theta_{13} \gtap 0.02$)
provide a high precision measurement
of $\Delta{m}^2_{\mbox{atm}}$ and determine 
the type of the neutrino mass hierarchy, 
have been discussed. The impact that 
i) the choice of the baseline $L$, 
ii) the effect of using a relatively low
$e^+-$ energy cut-off of $E_{th} \sim 1.0$ MeV,
iii) the detector's energy resolution, 
as well as iv) the statistical and systematical
errors, can have on the measurement
of each of the indicated neutrino oscillation parameters 
and on the determination of the neutrino mass hierarchy
have also been investigated in \cite{SCSPMP03}.

\vskip.3cm

 In \cite{Hano1} a Fourier analysis of reactor
$\bar{\nu}_e$  simulated data using 
the exponential Fourier transform (FT) was performed. 
It was found that the NH and IH neutrino mass 
spectra are distinguished by a relatively 
small shoulder beside the  $\Delta{m}^2_{\mbox{atm}}$  
modulation peak, which for the NH (IH) spectrum 
is to the left (to the right) of the peak. 
In the same study results of a statistical
analysis of the possibility to determine the 
neutrino mass hierarchy 
for different baselines, different values of
$\theta_{13}$ and different detector exposures (statistics)
were also presented. In that analysis the effects 
of the detector energy resolution were accounted for, 
but the systematic uncertainties 
and the uncertainties in the energy scale
and the neutrino oscillation parameters 
were not taken into account. 
The latter were included in an
unbinned maximum likelihood analysis 
performed in \cite{Hano2}.

\vskip.3cm

 It was noticed in \cite{YWang08} 
that the sine and cosine Fourier transforms 
of simulated reactor $\bar{\nu}_e$ data 
in the case of  NH and IH spectrum
show a difference 
in certain specific features
which can be used to distinguish between
the two types of  spectrum. 
The authors of \cite{YWang08}
include in their numerical simulations
the effects of the detector's energy resolution
and an uncertainty in the energy scale
(shift and shrink/expansion), which is independent of energy.
They do a statistical hierarchy analysis similar 
to that performed in \cite{Hano1}
and give results for different values of $\theta_{13}$, 
of the energy resolution and exposures.  
No systematic uncertainties or 
parameter marginalization were taken into account 
in this investigation. 
The possibility of an energy-dependent energy scale 
uncertainty was not considered either.

\vskip.3cm

   The present article is a natural continuation of the studies 
performed in \cite{PiaiP0103,SCSPMP03,Hano1,Hano2,YWang08}.
More specifically, we investigate further the behaviour 
of the sine and cosine Fourier transformed 
$e^+$ spectra taking into account, in particular,
the possibility of an energy-dependent 
energy scale uncertainty (assuming the shrink/expansion 
factor to have a linear dependence on the neutrino energy). 
In general, the mass hierarchy-dependent
features of the Fourier spectra 
of interest are changed in the case 
of an energy-dependent energy scale shift.
This might affect a statistical 
analysis using the FT method.
We perform also a $\chi^2$ analysis of 
the sensitivity to the neutrino mass hierarchy  
using simulated reactor $\bar{\nu}_e$ data. 
In this analysis we take into account a 
marginalization over the
relevant neutrino oscillation parameters, 
the detector resolution,
the energy scale uncertainty 
(both energy-dependent and independent) and
the systematic errors.
A $\chi^2$ analysis offers the advantage of a 
binned study in which the binning 
(the division of the L/E range into bins) is optimized 
on the basis of the energy resolution and 
the improvement in sensitivity 
so as to give the best
possible sensitivity to the neutrino mass hierarchy 
while being consistent with the detector's energy resolution.
The systematic uncertainties are included using the method of pulls.
We present results, in particular, 
for different values of the 
detector's energy resolution, 
exposure and $\theta_{13}$.

\vskip.3cm
 
 Let us note that the type of neutrino mass hierarchy, i.e.
${\rm sgn}(\deltaatm)$, can be determined 
by studying oscillations of neutrinos and
antineutrinos, say, $\nu_{\mu} \leftrightarrow \nu_e$
and $\bar{\nu}_{\mu} \leftrightarrow \bar{\nu}_e$,
in which matter effects are sufficiently large.
This can be done in long base-line 
$\nu$-oscillation experiments 
(see, e.g. \cite{Future,Blondel:2006su}). 
If $\sin^22\theta_{13}\gtap 0.05$
and $\sin^2\theta_{23}\gtap 0.50$,
information on ${\rm sgn}(\Delta m^2_{31})$
might be obtained in atmospheric 
neutrino experiments by investigating 
the matter effects in the subdominant transitions
$\nu_{\mu(e)} \rightarrow \nu_{e(\mu)}$
and $\bar{\nu}_{\mu(e)} \rightarrow \bar{\nu}_{e(\mu)}$ 
of atmospheric neutrinos which traverse the Earth \cite{JBSP203,SPTSchw06},
or by studying the ``disappearance'' of the 
atmospheric $\nu_{\mu}$ and $\bar{\nu}_{\mu}$
crossing the Earth \cite{SPTSchw06,Gandhi:2004bj}.
For $\nu_{\mu(e)}$ ({\it or} $\bar{\nu}_{\mu(e)}$) 
crossing the Earth core, a new type of resonance-like
enhancement of the indicated transitions
takes place due to the {\it (Earth) mantle-core
constructive interference effect
(neutrino oscillation length resonance (NOLR))} 
\cite{SP3198}\footnote{As a consequence of this effect
the indicated $\nu_{\mu(e)}$ ({\it or} $\bar{\nu}_{\mu(e)}$)
transition probabilities can be maximal \cite{106107}
(for the precise conditions of the mantle-core 
(NOLR) enhancement see \cite{SP3198,106107}).
Let us note that the Earth mantle-core (NOLR) enhancement of
neutrino transitions differs \cite{SP3198} from the MSW 
one.}. 
For $\Delta m^2_{31}> 0$, the neutrino transitions
$\nu_{\mu(e)} \rightarrow \nu_{e(\mu)}$
are enhanced, while for $\Delta m^2_{31}< 0$
the enhancement of antineutrino transitions
$\bar{\nu}_{\mu(e)} \rightarrow \bar{\nu}_{e(\mu)}$
takes place, which might allow 
to determine ${\rm sgn}(\Delta m^2_{31})$.
If neutrinos with definite mass are
Majorana particles, information about 
the ${\rm sgn}(\deltaatm)$ could be obtained 
also by measuring the effective neutrino Majorana
mass in neutrinoless double $\beta-$decay 
experiments~\cite{PPSNO2bb,BPP1}.
Information on the type of neutrino mass spectrum
can also be obtained in $\beta$-decay experiments 
having a sensitivity to neutrino masses  
$\sim \sqrt{|\deltaatm|}\cong 5\times 10^{-2}$ eV 
\cite{BMP06}
(i.e. by a factor of $\sim 4$ better sensitivity than 
that of the KATRIN experiment \cite{MainzKATRIN}).

%
\section{Preliminary remarks}
%

 We consider an experimental set-up with a nuclear 
reactor producing electron
antineutrinos by the $\beta$-decay of fission products 
of the isotopes U-235, U-238, Pu-239 and Pu-241. 
The $\bar{\nu}_e$ are assumed to be detected in a
single KamLAND-like \cite{KL162} liquid 
scintillator detector, located
at a distance of 60 Km from the reactor, 
by the inverse $\beta$-decay reaction: 

\be
\bar{\nu}_e + p \rightarrow e^{+} + n\,.
\ee
%
\noindent The visible energy of the detected positron is given by
\bea
E_{vis} = E + m_e - (m_n - m_p) \\
\simeq  E - 0.8~{\rm{MeV}}
\eea
%
\noindent Here $m_e,m_n$ and $m_p$ are the masses 
of the positron, neutron and proton, respectively, 
and   $E$ is the $\bar{\nu}_e$ energy.
The no-oscillation event rate spectrum 
is the product of the 
initial $\bar{\nu}_e$ flux spectrum
and the inverse  $\beta$-decay cross-section 
and is bell-shaped, with its peak at about 
$E_{vis} = 2.8$ MeV. 
In the present analysis we use the 
analytic expression for the $\bar{\nu}_e$ 
flux spectrum given in \cite{Vogel:1989iv}. The latter 
has a fit error of about 1.2$\%$ on the total event rate.   
The expression for the 
$\bar{\nu}_e + p \rightarrow e^{+} + n$
cross-section is taken from \cite{Vogel:1999zy}.
The threshold of the visible energy used is 
$E_{visth} = 1.0$ MeV (see further).

\vskip.3cm

 The event rate spectrum is given by the product of the 
no-oscillation spectrum and the ${\bar{\nu}_e}$ 
survival probability $P_{{\bar{e}}{\bar{e}}}$. 
In the convention we are using the expression for the
$\bar{\nu}_e$ survival probability in the case
of 3 flavor neutrino mixing and NH(IH) neutrino 
mass spectrum is given by\footnote{The Earth matter effects are
negligible for the values of the neutrino oscillation 
parameters ($\Delta m^2_{21}$ and $\deltaatm$), 
$\bar{\nu}_e$ energies and the short 
baseline $L\cong 60$ km we are interested in.}
\cite{PiaiP0103,BiNiPe02}:
\bea
\lefteqn{P_{NH(IH)}({\bar \nu_e}\to{\bar \nu_e}) 
\equiv P^{NH(IH)}_{\bar{e}\bar{e}}}
\nonumber\\
&& =\,~~ 1 - 2 \, \sin^2\theta_{13} \cos^2\theta_{13}\,
\left( 1 - \cos \frac{ \Delta{m}^2_{\mbox{atm}} \, L }{ 2 \, E } \right)
\nonumber \\
&& - \,~~\frac{1}{2} \cos^4\theta_{13}\,\sin ^{2}2\theta_{12} \,
\left( 1 - \cos \frac{ \Delta{m}^2_{\odot} \, L }{ 2 \, E } \right)
\label{P21sol}  \\
& & +\,~~ 2\,a^2_{NH(IH)}\, \sin^2\theta_{13}\,\cos^2\theta_{13}\,   
\left(\cos \left( \frac
{\Delta{m}^2_{\mbox{atm}} \, L }{ 2 \, E} - 
\frac {\Delta{m}^2_{\odot} \, L }{ 2 \,
E}\right)
-\cos \frac {\Delta{m}^2_{\mbox{atm}} \, L }{ 2 \, E} \right)\,,
\nonumber
\eea

\noindent where $\Delta{m}^2_{\odot} = \Delta{m}^2_{21}$ and
$a^2_{NH(IH)} =  \sin^{2}\theta_{12}~(\cos^{2}\theta_{12})$.
For the atmospheric neutrino mass squared 
difference $\Delta{m}^2_{\mbox{atm}}$
in the case of NH (IH) spectrum we have
$\Delta{m}^2_{\mbox{atm}} = \Delta{m}^2_{31}~(\Delta{m}^2_{23})$.
The properties of the $\bar{\nu}_e$ survival probability 
$P^{NH(IH)}_{\bar{e}\bar{e}}$
have been discussed in detail in 
\cite{PiaiP0103,SCSPMP03}. We only note here that
$P^{NH(IH)}_{\bar{e}\bar{e}}$ depends neither 
on the angle $\theta_{23}$
associated with the  atmospheric neutrino 
oscillations, nor on the CP violating phase 
$\delta$ in the PMNS matrix.
The fact that $\cos2\theta_{12}\neq 0$, 
$\cos2\theta_{12}\geq 0.26$ (at 3$\sigma$),
opens up the possibility
to get information about the 
neutrino mass spectrum if 
$\sin^22\theta_{13} \neq 0$:
$P^{IH}_{\bar{e}\bar{e}} - P^{NH}_{\bar{e}\bar{e}}
\propto \cos2\theta_{12}\sin^22\theta_{13}$.
This can be done, in principle, 
by studying the deformations of the observed event 
spectrum due to the $\bar{\nu}_e$ survival 
probability \cite{PiaiP0103}.

\vskip.3cm

 The detector energy resolution is taken into account 
assuming it has the standard Gaussian form:

\be
R(E,E_m) = 
\frac{1}{{\sqrt{2\pi}}\sigma} exp(-\frac{(E_m - E)^2}{2\sigma^2})\,.
\label{Ereso}
\ee
%
\noindent Here $E_m$ is the observed neutrino energy.
We have $E_m - E = E_{vism} - E_{vis}$, 
where  $E_{vism}$ is the measured $e^{+}$ energy.
The error for a scintillator detector is dominated 
by the photoelectron statistics,
and hence $\sigma/E_{vis}$ is proportional to $1/\sqrt{E_{vis}}$. 
We consider resolutions (i.e., $\sigma/E_{vis}$)
in the range of 
$2\%/\sqrt{E_{vis}} - 4\%/\sqrt{E_{vis}}$. 

\vskip.3cm

Further, we take into account the energy 
scale uncertainty of 
the detector by considering an energy scale 
shrink/expansion both with and without
energy dependence. This is parametrized as
\be
E_m^{'} = (1+a)E_m + b
\,,
\label{Em}
\ee
%
\noindent where $E_m$ is the neutrino energy after smearing and $E_m^{'}$ is the measured 
neutrino energy after including both the smearing and energy scale uncertainty.
The parameters $a$ and $b$ define 
the shrink/expansion and the shift 
of the energy scale, respectively. 
The parameter $a$ is taken to be 1$\%$
(unless otherwise specified) 
for the energy independent case, 
and 1$\%$ of $E_m$ (i.e., $a=0.01E_m$)
for the energy dependent case{\footnote{Accounting for the energy scale uncertainty on 
$E_{vism}$ leads to an additional shift
in $E_m^{'}$ which, however, does not have an effect on the spectrum features distinguishing between the NH and IH
neutrino mass spectra (see subsections 3.1 and 3.2).}}.  
Rigorously, in the energy dependent case, $a$ could have the form
$ a = cE_m +d$, corresponding to a combination of a non-linear and a linear dependence of $E_m^{'}$ on $E_m$.
However, it will be shown later that considering an energy dependent and an energy independent scale uncertainty
simultaneously in this way ($c,d$ non-zero) has the same effect as considering only an energy dependent scale uncertainty 
($c$ non-zero, $d$ zero).

\vskip.3cm

 The measured event rate spectrum, 
as a function of $L/E_m$,
is thus given by
\be
N(L/E_m) = \int R(E,E_m) \phi(E)\sigma(\bar{\nu}_e p \rightarrow e^{+}n;E)P^{NH(IH)}_{{\bar{e}}{\bar{e}}} dE\,,
\label{event}
\ee
%
\noindent where $\phi(E)$ is the $\bar{\nu}_e$
flux spectrum,  $\sigma(\bar{\nu}_e p \rightarrow e^{+}n;E)$ is the inverse 
$\beta$-decay cross-section 
and $P^{NH(IH)}_{{\bar{e}}{\bar{e}}}$ is the ${\bar{\nu}_e}$ 
survival probability defined earlier. 

\vskip.3cm

  The final statistics (total number of events) 
is a product of the event rate, 
the reactor power, the detector 
active mass and exposure time. The
exposure is thus expressed in the unit kT GW yr. The KamLAND-like 
large underwater detector planned within the project 
Hanohano \cite{Dye:2006gx},
can have a mass of up to $\sim 10$ kT and use a 
reactor having a power of $\sim 5$ GW. 
Hence we consider exposures in the range of 
200-800 kT GW yr. A $100\%$ efficiency of the 
detector is assumed. This gives, for example, 
a statistics of about $10^4$ events 
(with oscillations) when an exposure of 200 kT GW yr
is considered. Because of the high statistics, 
the geo-neutrino flux background
at lower energies becomes insignificant \cite{SCSPMP03}
and it is possible to use the relatively 
low visible energy threshold of $E_{visth} = 1.0$ MeV 
mentioned earlier. 

\vskip.3cm

In the statistical analysis we take into account
the systematic uncertainties relevant to
a detector of the type assumed by us. 
We consider 5 sources of systematic errors 
(3 related to the detector and 2 due to 
the geo-neutrino flux) \cite{Hano1,Hano2}:\\

i) The efficiency error, or the uncertainty in the predicted 
event rate, which can be between 1 to 5 $\%$.\\

ii) The uncertainty in the detector energy resolution estimation, 
which can be up to 10 $\%$.\\
 
iii) The energy scale uncertainty, which is around 1 $\%$.\\
 
iv) The uncertainty in the total detectable geo-neutrino flux.\\
 
v) The uncertainty in the ratio of the geo $\bar{\nu}_{e}$ fluxes 
from the decays of U-238 and Th-232.

\vskip.2cm
 
  We find during the course of the study that the effects 
of the indicated systematic and geo-neutrino 
uncertainties on the neutrino 
mass hierarchy sensitivity are not significant.

\vskip.3cm
 
Finally, we comment 
on the prospects of high 
precision determination of the neutrino oscillation 
parameters which serve as input in our analysis. The 
oscillation parameters
$\deltasol$, $\sin^2\theta_{12}$ and
$|\deltaatm|$ 
are determined by the existing
data with a 3$\sigma$ error
of approximately 9\%, 17\% and
15\%,  
respectively.
These parameters can (and very
likely will) be measured with much
higher accuracy in the future. 
The highest precision in the 
determination of $|\Delta m^2_{31}|$
is expected to be achieved 
in the next several years
from the studies
of $\nu_{\mu}$-oscillations in
the T2K experiment with 
Super-Kamiokande detector
(T2K (SK)) \cite{T2K}:
if the true $|\Delta m^2_{31}| =
2.5\times 10^{-3}$~eV$^2$ 
(and true $\sin^2\theta_{23} = 0.5$), 
the uncertainty in 
$|\Delta m^2_{31}|$ is estimated to be 
reduced in this experiment to  
$10^{-4}~{\rm eV^2}$ or 4$\%$ at 90$\%$ C.L. 
\cite{T2K,TMU04}.
The Fermilab-Homestake beam experiment (LBNE) 
is expected to reduce this error to
less than 3$\%$ at 90$\%$ C.L. \cite{LBNE10}.
Further, reactor antineutrino experiments themselves 
may be able to provide a determination of $|\deltaatm|$
with an uncertainty of approximately 1$\%$ at 1$\sigma$, or 
$(3 - 4)\%$ at 3$\sigma$ \cite{SCSPMP03,Hano1,Hano2}.
In what concerns the CHOOZ angle
$\theta_{13}$, three
reactor $\bar{\nu}_e$ experiments 
with baselines $L\sim$~(1--2) km, 
which could improve 
the current limit
by a factor of (5--10), are 
under preparation: 
Double-CHOOZ \cite{DCHOOZ},
Daya-Bay \cite{DayaB} and RENO \cite{RENO} 
(see also \cite{MMTSchw10}).
The most precise measurement 
of  $\Delta m^2_{21}$ could be achieved 
\cite{SKGdCP04} using Super-Kamiokande 
doped with 0.1\% of gadolinium (SK-Gd) 
for detection of reactor $\bar{\nu}_e$ 
\cite{SKGdBV04}: getting the same flux of 
reactor $\bar{\nu}_e$ as KamLAND,
the SK-Gd detector will have 
approximately 43 times bigger 
$\bar{\nu}_e$-induced event rate 
than KamLAND. After 3 years of data-taking 
with SK-Gd,  $\Delta m^2_{21}$ could be determined with  
an error of 3.5\% at 3$\sigma$ 
\cite{SKGdCP04}. 
A dedicated reactor  
$\bar{\nu}_e$ experiment with a 
baseline $L\sim 60$ km, 
tuned to the minimum of the
$\bar{\nu}_e$ survival probability, 
could provide the most precise 
determination of $\sin^2\theta_{12}$ 
\cite{TH12}:
with statistics of $\sim 60$ kT GW yr 
and systematic error 
of 2\% (5\%), $\sin^2\theta_{12}$  
could be measured with 
an error of 6\% (9\%) at\footnote{The inclusion of the current uncertainty
in $\theta_{13}$ ($\sin^2\theta_{13}<$0.05)
in the analysis increases the 
quoted errors by (1--3)\% to approximately 9\% (12\%) 
\cite{TH12}.}
3$\sigma$ \cite{TH12}. 


\vskip.3cm

{\section{The effects of energy smearing 
and energy scale uncertainty 
on the reactor $\bar{\nu}_e$ 
event rate and Fourier spectra}}


 In this Section we investigate in detail 
how the inclusion of the detector energy 
resolution and/or the energy scale 
uncertainty affects 
the reactor $\bar{\nu}_e$ event spectra, 
the Fourier spectra and hence the 
hierarchy sensitivity. 
For the detector's energy resolution we use
the Gaussian form 
given in eq. (\ref{Ereso}).
We consider an energy scale shrink/expansion both with and without 
energy dependence, which is parametrized in the form
specified in eq. (\ref{Em}).
As we have already indicated, 
the parameter a in eq. (\ref{Em}) 
is taken to be 1$\%$
(unless otherwise specified) for the energy 
independent case, and 1$\%$ of E ({\it{i.e.}} 
a linear dependence on energy) 
for the energy dependent case.  

%
\subsection{Behaviour of the Event Rate Spectrum}
%

\begin{figure}[t]
{\centerline
{ 
\hspace*{1em} \epsfxsize=5.5cm\epsfysize=6.5cm
\epsfbox{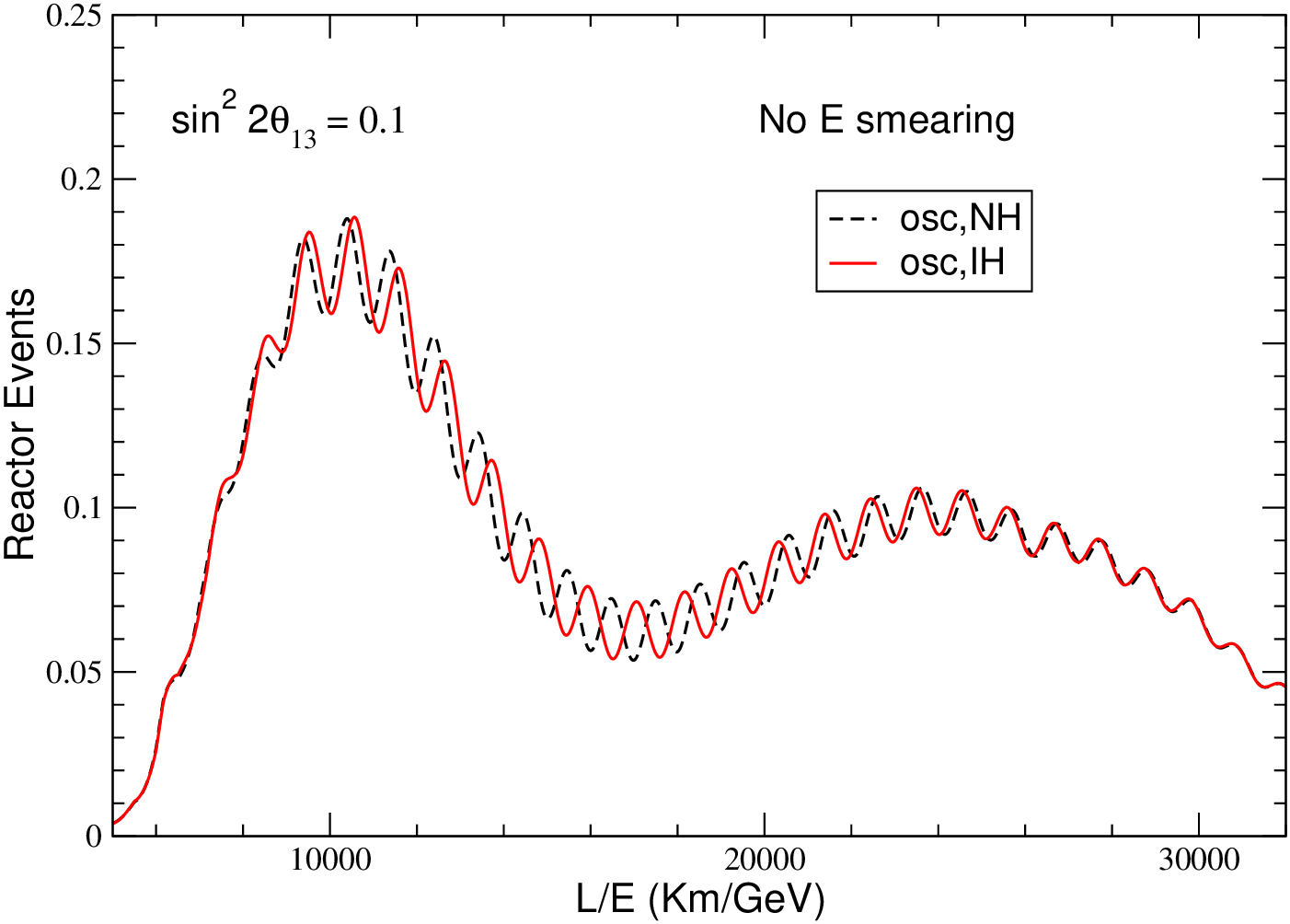}
        \hspace*{0.5ex}
\epsfxsize=5.5cm\epsfysize=6.5cm
\epsfbox{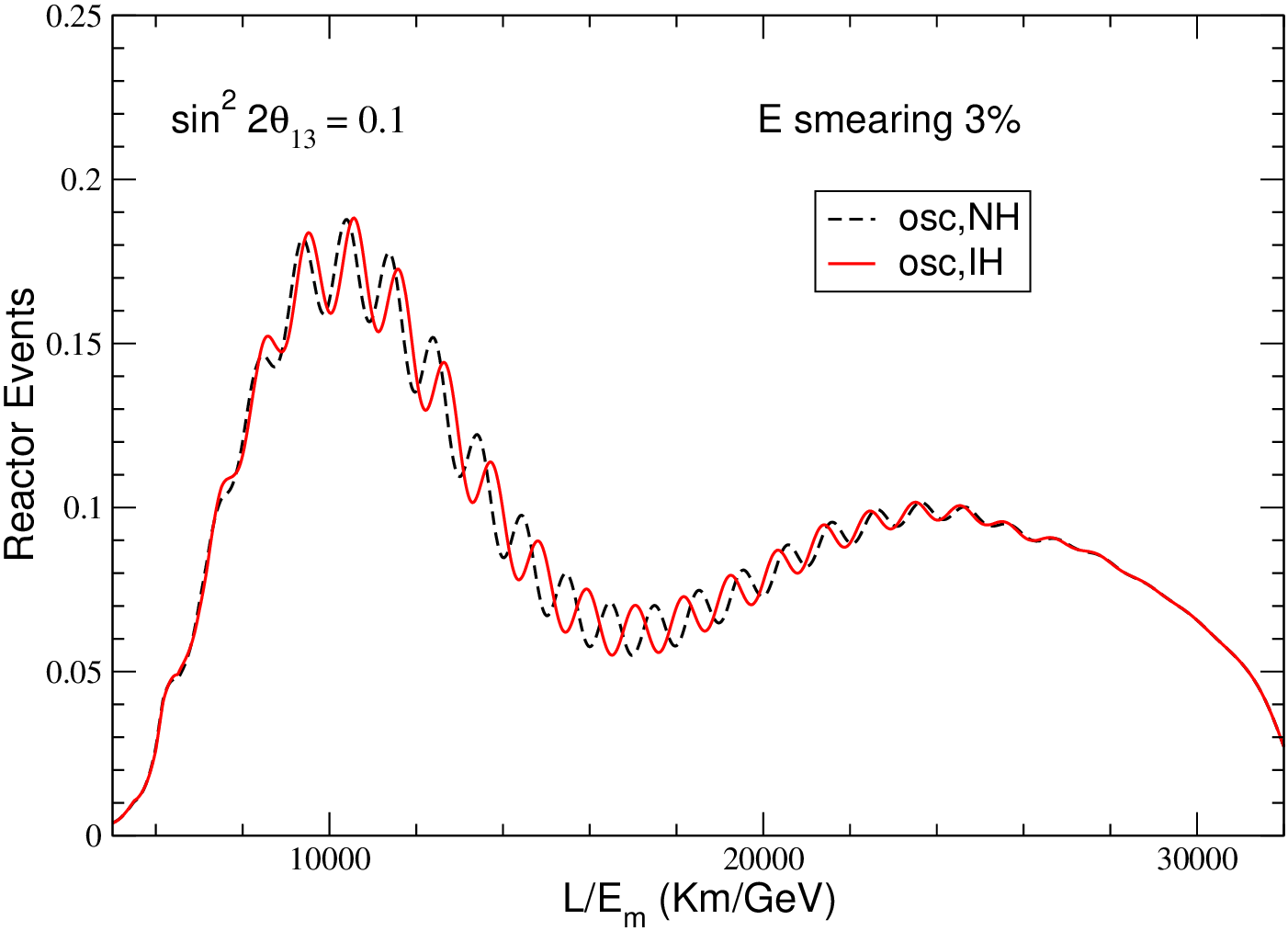}
        \hspace*{0.5ex}
\epsfxsize=5.5cm\epsfysize=6.5cm
\epsfbox{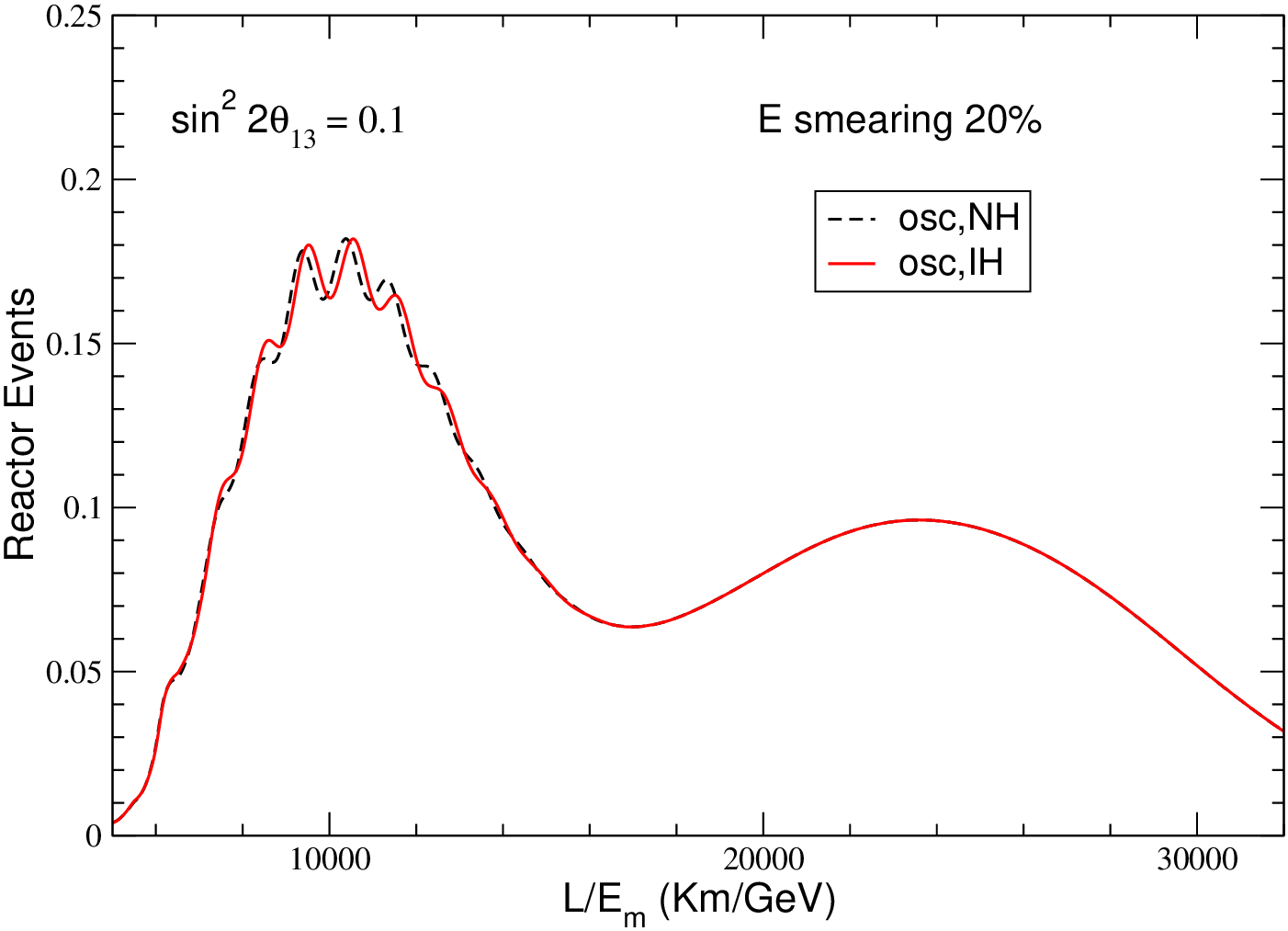}
}
\hskip 4cm
{\bf (a)}
\hskip 4cm
{\bf (b)}
\hskip 4 cm
{\bf (c)}
\caption[]{\footnotesize{Reactor event rate vs $\rm{L/E_m}$ 
for normal and inverted hierarchies, for {\bf{(a)}} ideal energy resolution, 
{\bf{(b)}} 3$\%$ energy resolution
and {\bf{(c)}} 20$\%$ energy resolution of the detector. 
} 
}
\label{fig1} 
}
\end{figure}

 Figure~\ref{fig1} illustrates the 
changes of the reactor event rate spectrum 
in the case of $\bar{\nu}_e$ oscillations 
when one varies the energy resolution
of the detector. This is done for
both the normal and inverted 
hierarchies, without including the effects of 
the energy scale shift. 
The spectrum plotted in Figure~\ref{fig1} is 
the normalized to 1 reactor $\bar{\nu}_e$ 
event rate spectrum:
\be
f(L/E_m) = \frac{N(L/E_m)}{\int_{(L/E_m)_{min}}^{(L/E_m)_{max}} N(x)~dx}\,,
\label{ff}
\ee
%
where $(L/E_m)_{min} = 5000$ Km/GeV, $(L/E_m)_{max} = 32000$ Km/GeV, 
and $N(L/E_m)$ is given by eq. (\ref{event}).
Figures (a), (b) and (c) show the 
spectrum $f(L/E_m)$ without energy 
smearing ({\it{i.e.}} assuming 
perfect detector energy resolution), with a 
realistic smearing 
of 3$\%$ and with a large smearing of 20$\%$, 
respectively. This and 
the subsequent event rate spectrum 
figures are obtained for $\sin^2 2\theta_{13} = 0.1$. 
The figures clearly show the effect of the 
energy resolution: the spectrum is 
slightly ``flattened'' towards the higher values 
of $\rm{L/E_m}$ in the case of resolution of 3$\%$ 
as compared to the unsmeared spectrum; it  
is smeared throughout and the hierarchy 
sensitivity is completely lost 
over almost the entire L/E range if the detector's
energy resolution is as poor as 20$\%$. 

\vskip.3cm

\begin{figure}[t]
\centerline
{
\epsfxsize=8.5cm\epsfysize=6.2cm
\epsfbox{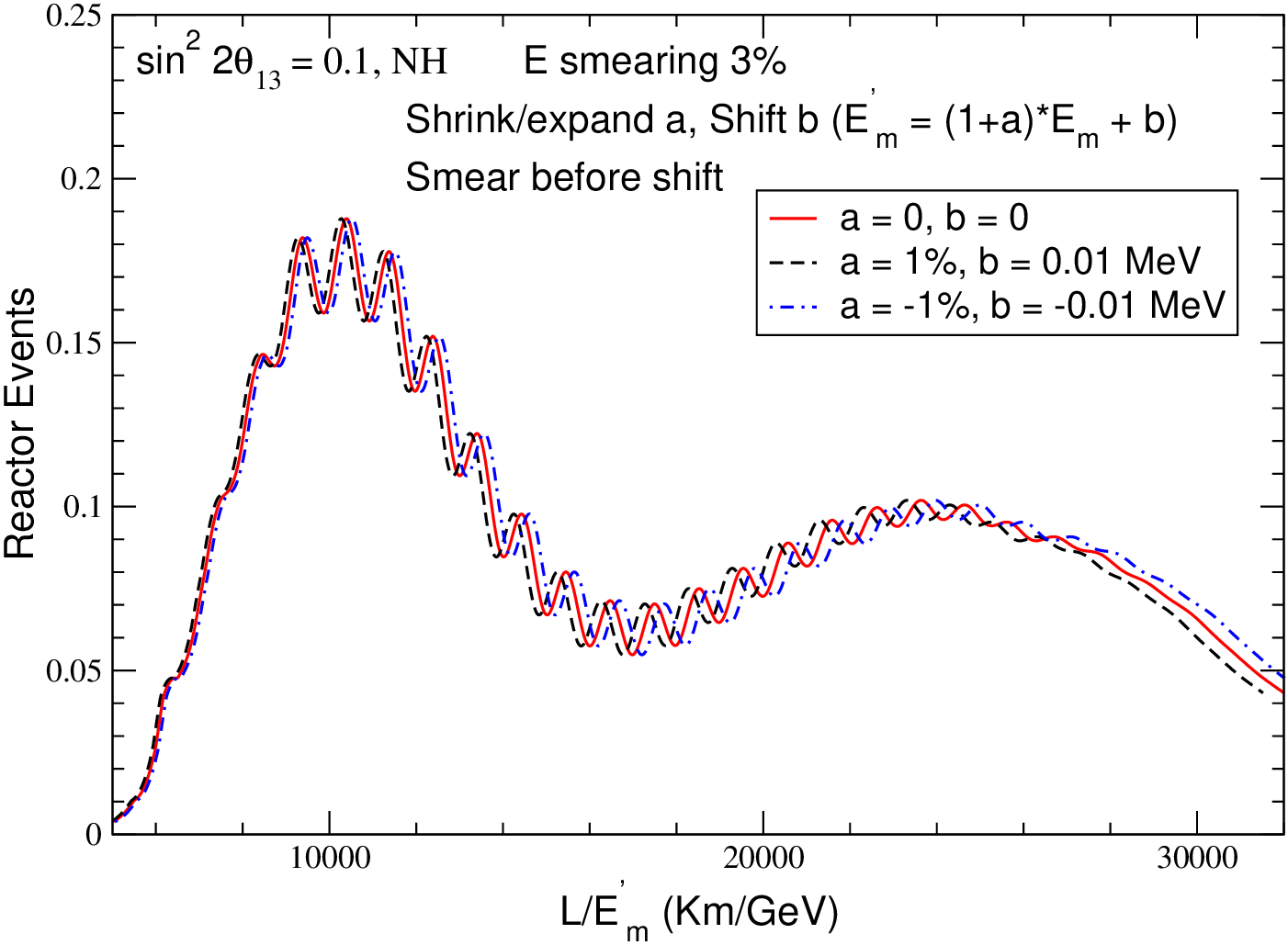}
        \hspace*{0.7ex}
\epsfxsize=8.5cm\epsfysize=6.2cm
\epsfbox{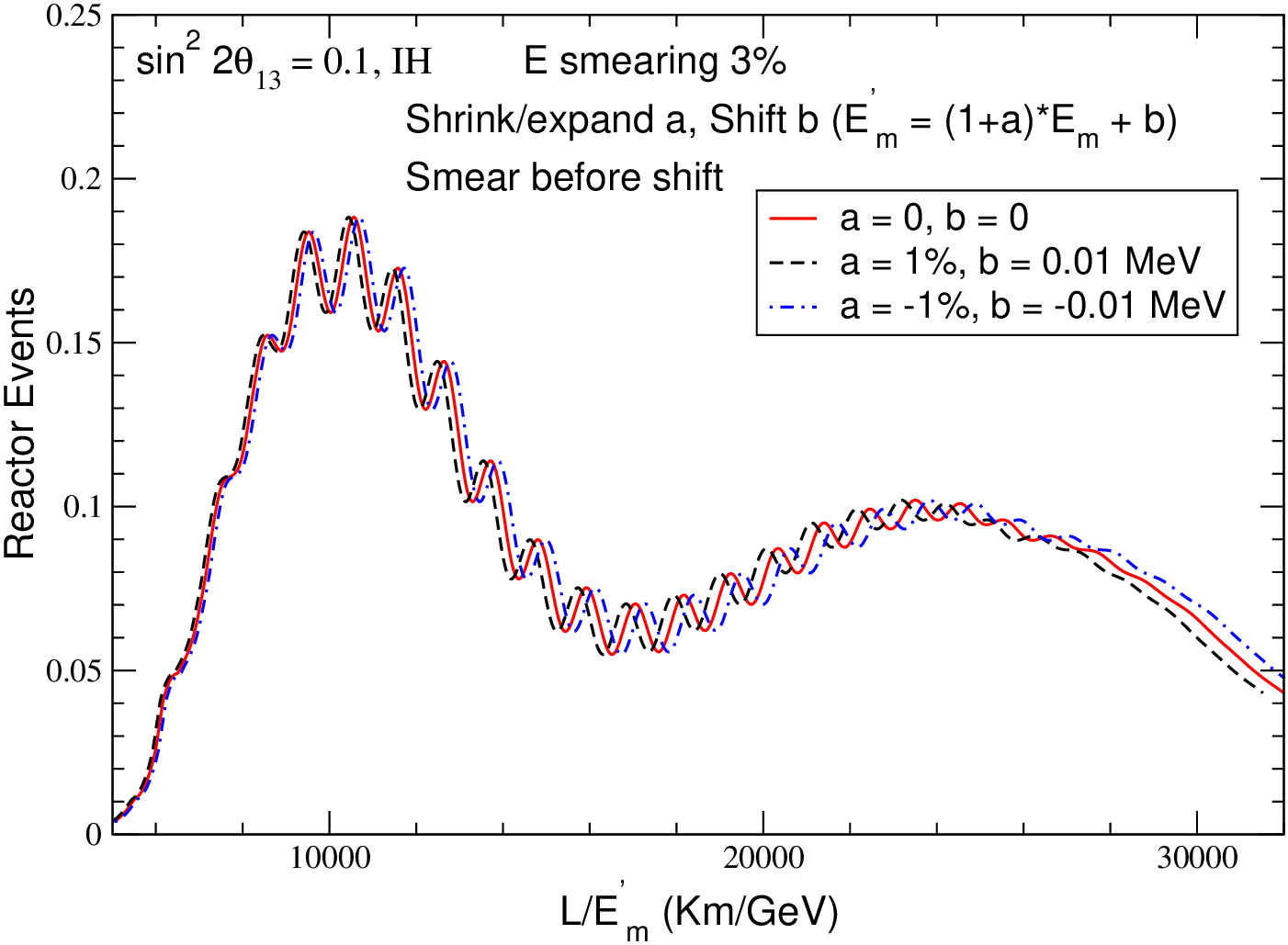}
}
\hskip 5cm
{\bf (a)}
\hskip 7cm
{\bf (b)}
\caption[]
{\footnotesize{{{\bf (a)}
Reactor event rate vs $\rm{L/E_m^{'}}$ for normal hierarchy, 3$\%$ energy resolution
of the detector and an energy-independent uncertainty (shrink/expansion and shift) in the energy scale.
The energy scale shift is performed on the neutrino energy $E_m$ after taking smearing into account.
{\bf (b)} 
The same as {\bf{(a)}} for inverted hierarchy. 
}}}
\label{fig2}
\end{figure}

Figs.~\ref{fig2} and \ref{fig3} show the behaviour 
of the reactor $\bar{\nu}_e$ event rate 
spectrum when the detector resolution and/or the energy 
scale uncertainty are taken into account.

%
\noindent In Figure~\ref{fig2}, the spectrum with 
an energy smearing of 3$\%$ and an energy-independent 
expand/shrink is plotted, for both
the cases of an expansion in energy scale, 
corresponding to $a= 1\%, b = 0.01$ MeV, 
and a shrink in energy scale, corresponding to $a = -1\%, b = -0.01$ MeV. 
The event spectra are seen to shift to the left and to 
the right, respectively. In Figure~\ref{fig3} we plot  
the same spectra for energy-dependent expansion
and shrink. The displacements in the spectra are seen 
to be larger in this case, and for this value of the expansion/shrink it leads 
to an effective flipping of the maxima/minima in the spectrum,
as compared to the spectrum without shrink/expansion.
Note that the effect is the same for the normal and 
inverted hierarchies. 
The changes in the event spectrum can be  
shown to be identical with 
those without energy smearing for both energy-independent 
and energy-dependent scale shifts. 



\subsection{Fourier analysis of the reactor 
$\bar{\nu}_e$ event rate spectrum}
%

The Sine and Cosine Fourier Transforms of 
the reactor 
$\bar{\nu}_e$ event rate spectrum are computed as  
a function of the ``frequency'' $\delta m^2$, 
varied in the range $2\times 10^{-3}~\rm{eV^2}$ to 
$2.8\times 10^{-3}~\rm{eV^2}$, using the best-fit 
values $|\Delta m^2_{31}| = 2.4\times 10^{-3}~{\rm eV^2}$ 
and $\Delta m^2_{21} = 7.6\times 10^{-5}~{\rm eV^2}$.
The expressions for the Fourier Transforms used 
by us read \cite{YWang08}:

\be
FCT(\omega) = \int_{(L/E_m)_{min}}^{(L/E_m)_{max}}f(L/E)~
\rm{\cos(\omega L/E)~d(L/E)}\,,
\label{FCT}
\ee
\be
FST(\omega) = \int_{(L/E_m)_{min}}^{(L/E_m)_{max}}f(L/E)~
\rm{\sin(\omega L/E)~d(L/E)}\,.
\label{FST}
\ee

Here $\omega = 2.54 \times \delta m^2~[eV^2]$, where 
$\delta m^2$ is in units of ${\rm eV^2}$,
and $L/E$ is in units of km/GeV.  

The values $|\Delta m^2_{31}| = 2.4\times 10^{-3}~{\rm eV^2}$ 
and $\Delta m^2_{21} = 7.6\times 10^{-5}~{\rm eV^2}$
appear in the normalised reactor $\bar{\nu}_e$
event rate spectrum  $f(L/E)$.
Hence, in the Fourier spectrum there is modulation 
due to both these frequencies. 
The modulation 
due to $\Delta m^2_{31}$ 
occurs near $\delta m^2 = 2.4\times 10^{-3}~{\rm eV^2}$, 
while that due to $\Delta m^2_{21}$ occurs near 
$\delta m^2 = 7.6\times 10^{-5}~{\rm eV^2}$. 
The values of the other neutrino oscillation
parameters used in the 
calculations are $\sin^2 2\theta_{12} = 0.87$ and 
$\sin^2 2\theta_{13} = 0.1$  (unless otherwise stated).


\vskip.3cm

  According to \cite{YWang08},
the main features in the FCT and FST spectra 
that allow to distinguish between the two types of 
neutrino mass spectrum are:

\vskip.2cm

(a) In the FCT spectrum, $(RV-LV)$ has opposite signs for
the NH and IH spectra, where $RV$ and $LV$ are 
the amplitudes of the right and 
left ``valleys'', i.e., of the minima located 
closest (i.e., immediately) to the right (RV) 
and to the left (LV) of the absolute 
modulation maximum,  in the Fourier spectra.
The right ``valley'' is deeper 
than the left ``valley'' for the NH spectrum, and
{\it vice versa} for the IH spectrum.

\vskip.2cm

(b) In the FST spectrum, $(P-V)$ has opposite signs for 
the NH and IH spectra, where $P$ and $V$
are the amplitudes of the absolute modulation maximum
(``peak'') and of the absolute modulation minimum (``valley'')
in the two event rate spectra.   
The amplitude of the ``peak'' is bigger than 
the amplitude of the ``valley'' for the NH spectrum, and
{\it vice versa} for the IH spectrum.

\vskip.3cm

The differences between the event rate spectra 
in the NH and IH cases 
can  thus be quantified by the following 
two asymmetries \cite{YWang08}:
\be
RL = \frac{RV-LV}{RV+LV}\,,
\label{RL}
\ee

\noindent for the FCT spectrum, and
\be
PV = \frac{P-V}{P+V}\,,
\label{PV}
\ee

\noindent for the FST spectrum. 
The $RL$ and $PV$ asymmetry features 
discussed above are illustrated in Figure \ref{FT4}.

\vskip.3cm


 We have analyzed the effects of the detector's 
energy resolution and energy scale uncertainty 
on the hierarchy-sensitive features of the FCT 
and FST spectra. Both the cases of energy-independent 
and energy-dependent energy scale 
uncertainty (``shrink/expansion'') have been 
considered. The magnitude of the 
shrink (expansion) was 
assumed to be (-1$\%$) ((+1$\%$)).

\vskip.3cm

  Our results are illustrated in 
Figs. \ref{FT5} - \ref{FT3}, in which we show 
the FCT and FST spectra, 
corresponding to the NH and IH 
neutrino mass spectra,
for different combinations 
of the detector's energy resolution and 
forms of the energy scale uncertainty.
The figures are obtained for
$\Delta m^2_{31}(IH) = -\Delta m^2_{32}(NH)$,
$\sin^2 2\theta_{13} = 0.02$, and the 
best-fit values of 
all other neutrino oscillation 
parameters.

\vskip.3cm

\begin{figure}[t]
\centerline
{
\epsfxsize=8.5cm\epsfysize=6.2cm
\epsfbox{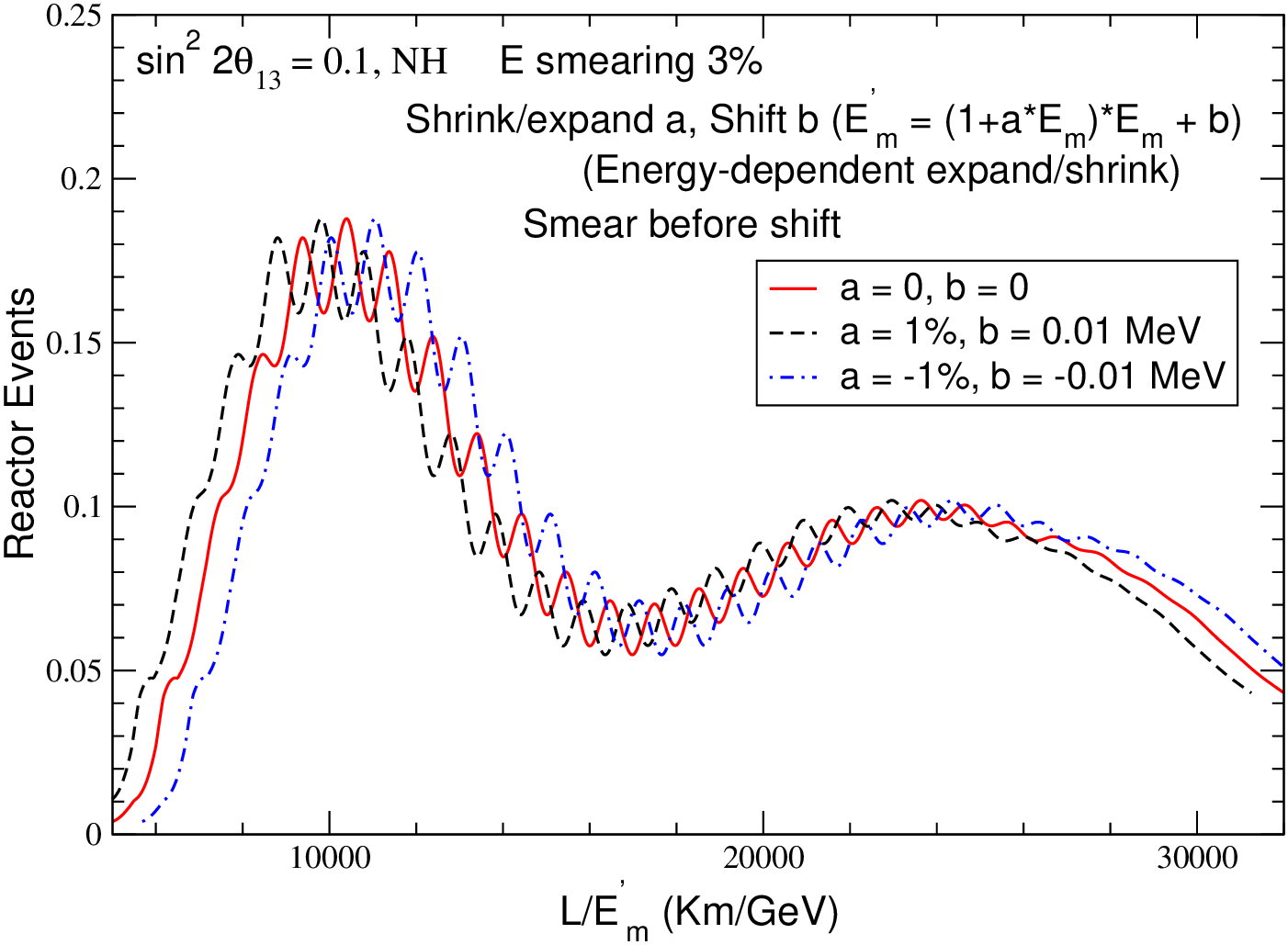}
        \hspace*{0.7ex}
\epsfxsize=8.5cm\epsfysize=6.2cm
\epsfbox{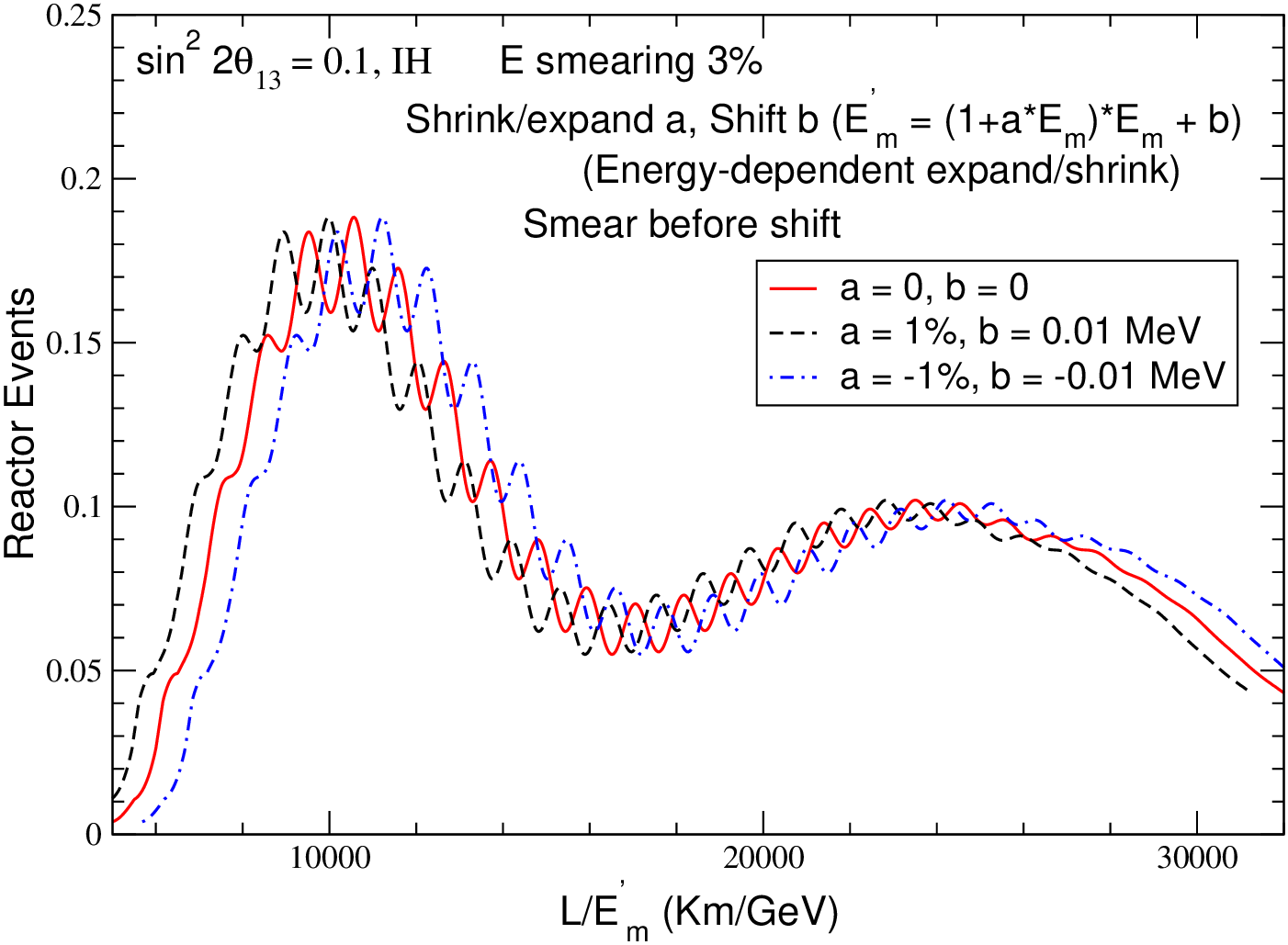}
}
\hskip 5cm
{\bf (a)}
\hskip 7cm
{\bf (b)}
\caption[]
{\footnotesize{{{\bf (a)}
Reactor event rate vs $\rm{L/E_m^{'}}$ for normal hierarchy, 
3$\%$ energy resolution
of the detector and an energy-dependent 
uncertainty (shrink/expansion and shift) in the energy scale.
The energy scale shift is performed on 
the neutrino energy $E_m$ after taking smearing into account.
{\bf (b)} 
The same as {\bf{(a)}} for inverted hierarchy.}} 
}
\label{fig3}
\end{figure}

Comparing the respective curves in Figs.~\ref{FT5}, 
\ref{FT6}, \ref{FT7} and \ref{FT8} demonstrates  
that with an energy-independent ``shrink'' performed 
on the measured energy (after smearing), 
the Fourier spectra get simply displaced to 
the left in all cases, with no change in the overall 
shape of the spectra. 
The shrink in the event versus energy 
spectrum leads to an ``expansion'' in 
the event versus $L/E_m$ spectrum, and as a consequence 
one obtains a given feature (maximum, minimum) 
at a smaller value of the oscillation frequency 
$\delta m^2$. This behaviour is accentuated 
if higher values of "shrink" are considered.
This leads  
to an overall left shift in the Fourier 
(frequency) spectra. 
Since the sensitivity to the hierarchy in 
the Fourier spectra is 
related to the relative 
positions and the amplitudes of the 
maxima and minima of the spectra,
the indicated changes do not affect results 
on the hierarchy sensitivity. 
In the case of an energy-dependent shrink, however,
the change in the Fourier spectra is more complicated 
and the shape gets distorted, as the figures clearly show.
This behaviour in both cases 
(energy-dependent or energy-independent shrink) 
is identical to the change in the corresponding 
spectra due only to an energy scale shift and 
no energy smearing, as observed by comparing Figs.~\ref{FT4}, 
\ref{FT5} and \ref{FT7}. 
The above comments hold true for both the FCT and 
FST spectra and for both the normal and 
inverted hierarchies.

\vskip.3cm

An energy-independent energy scale ``expansion'' of 1$\%$ 
gives a uniform right displacement to the Fourier spectra, 
as expected. 
This is because the event versus 
$L/E_m$ spectrum shrinks with an expansion
in the energy spectrum, leading to a shift of the 
Fourier spectral features to a higher frequency. 
With energy-dependent expansion of the 
measured energy scale, the spectrum shape is again changed 
(see Figure~\ref{FT3}).   

\vskip.3cm

 As discussed above,
the distinguishing feature 
of the NH and IH neutrino mass spectra
in the FCT spectrum, according to \cite{YWang08},
is the sign of the asymmetry $RL$ defined in 
eq. (\ref{RL}): we have $RL > 0$ ($RV > LV$) 
in the case of normal hierarchy and $RL < 0$ ($RV < LV$) for 
the inverted hierarchy. 
In the case of the FST spectrum, 
it was proposed in \cite{YWang08} 
to distinguish between the normal and 
inverted hierarchy neutrino mass spectra
by the sign of the asymmetry $PV$ 
defined in eq. (\ref{PV}): one has $PV > 0$
($P > V$) in the NH case and $PV < 0$ 
($P < V$) if the IH spectrum is realized. 
On the basis of our analysis
we can make the following 
observations.


\begin{figure}[t]
\centerline
{
\epsfxsize=8.5cm\epsfysize=6.2cm
\epsfbox{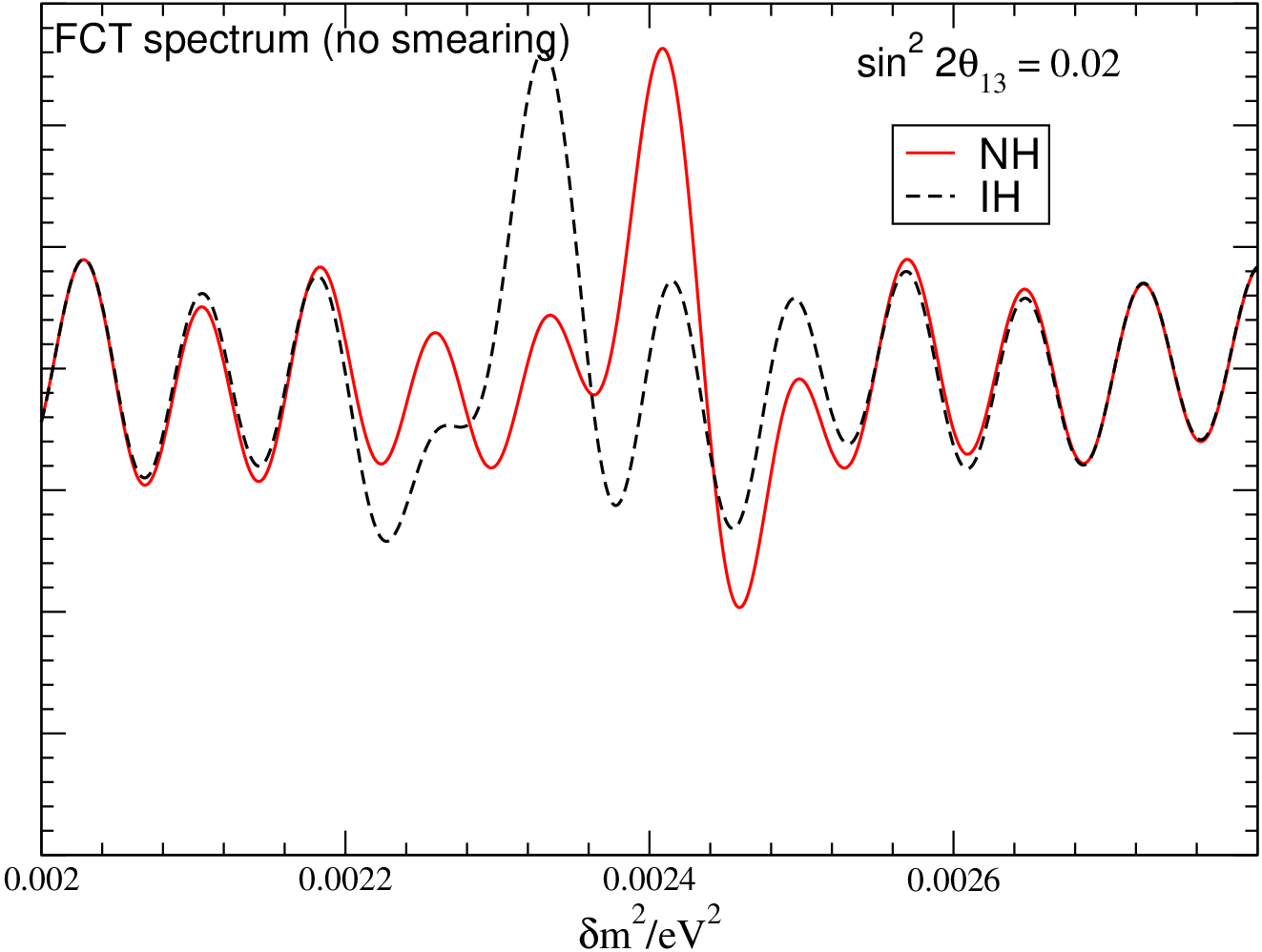}
        \hspace*{0.7ex}
\epsfxsize=8.5cm\epsfysize=6.2cm
\epsfbox{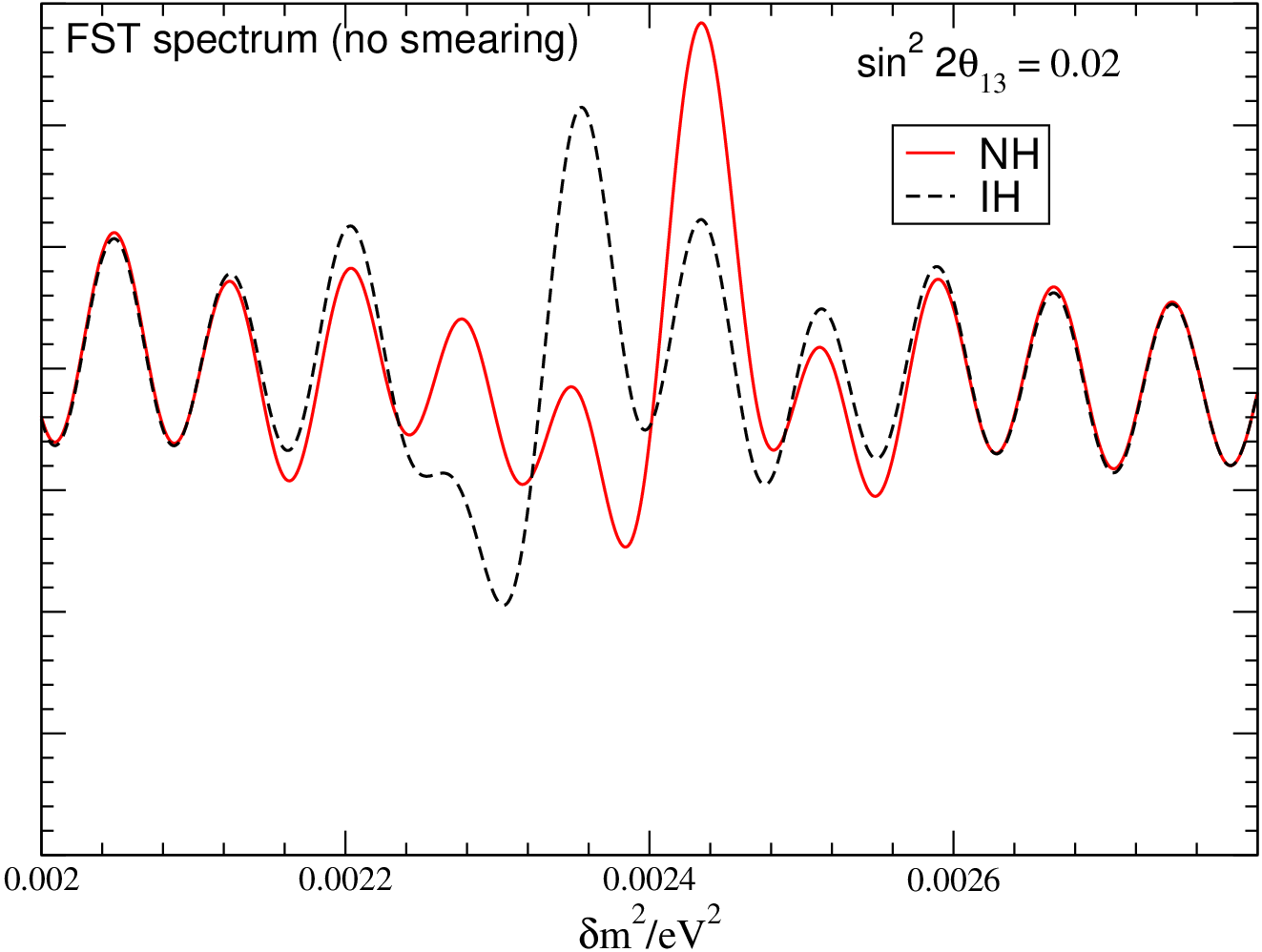}
}
\hskip 5cm
{\bf (a)}
\hskip 7cm
{\bf (b)}
\caption[]
{\footnotesize{{{\bf (a)}
Fourier cosine transformed (FCT)  
reactor $\bar{\nu}_e$ event rate spectrum vs 
$\delta m^2$ with power (y-axis) in arbitrary units, 
for $\sin^2 2\theta_{13} = 0.02$, normal hierarchy and 
inverted hierarchy,  ideal energy resolution
of the detector and no energy scale 
uncertainty.
{\bf (b)} 
Fourier sine transformed (FST) 
reactor $\bar{\nu}_e$ event rate spectrum vs 
$\delta m^2$ with power (y-axis) in arbitrary units, 
for $\sin^2 2\theta_{13} = 0.02$, normal hierarchy and 
inverted hierarchy,  ideal energy resolution
of the detector and no energy scale 
uncertainty.
}}}
\label{FT4}
\end{figure}


\vspace{0.4cm}
{\bf{FCT spectrum.}}
\begin{itemize}
\item Comparing Figure~\ref{FT4}(a) (no smearing, no scale shift) 
with Figure~\ref{FT5}(a) (3$\%$ smearing, no scale shift), 
the sign feature of the asymmetry $RL = (RV-LV)/(RV+LV)$, 
distinguishing between the NH and IH cases, is seen 
to be retained with a smearing of 3$\%$, 
with a somewhat reduced (increased) 
absolute magnitude of the asymmetry 
$RL$ in the NH (IH) case (we find $RL(NH) = 0.39$, 
$RL(IH) = -0.11$ in Figure~\ref{FT4}(a)
and $RL(NH) = 0.20$, $RL(IH) = -0.35$ in Figure~\ref{FT5}(a)).

\item Comparing Figure~\ref{FT4}(a) with Figure~\ref{FT5}(b) 
(no smearing, energy scale shift with energy-dependent shrink 
of 1$\%$), the $RL$ asymmetry feature distinguishing between 
the two hierarchies is no longer present with 
the inclusion of an energy-dependent energy scale 
shrink. Instead, the absolute modulation maxima 
in the NH and IH spectra appear to be replaced by absolute 
modulation minima, while the adjacent valleys (minima)
are replaced by adjacent peaks (maxima). We can define a quantity 
\be
RLP = \frac{RP-LP}{RP+LP}\,,
\label{RLP}
\ee
%
where $RP$ and $LP$ are the amplitudes of the right and left peaks 
adjacent to the absolute modulation minima.  
This is seen to have a significant positive value for the NH spectrum,
and a much smaller value close to zero for the IH spectrum.    

\vskip.2cm

This behaviour can be explained on the basis of Figure~\ref{fig3} (the unsmeared reactor event
spectrum as a function of $\rm{L/E}$ and with an energy-dependent scale shrink),
and Figure~\ref{FT5}(c), in which the FCT spectrum for NH (without smearing) is plotted for different
values of the energy-dependent shrink factor, varying from 0.1$\%$ ($a = 0.001 \times E_m$ in eq. (\ref{Em})) to 1$\%$ ($a = 0.01 \times E_m$).
It may be observed that there is a gradual left-shift in the FCT spectrum with an increase in the shrink factor,
as well as a change in its shape, with a progressive drop in the amplitudes of the maxima and an increase in the amplitudes
of the minima in the modulation region. This shift leads to what looks like a flipping of the maxima and minima
when the shrink reaches a value of 1$\%$. Note that this is not an actual inversion, but a feature caused by the left-shift
and shape change of the spectrum. We have observed that in the corresponding event spectrum (Figure~\ref{fig3}), there is a large right-shift
due to the energy-dependent shrink, which leads to an effective inversion of the maxima and minima (as compared to the
spectrum without shrink) for this value of the shrink factor. This is reflected in the left-shift and change in shape of the Fourier spectrum,
which leads to an effective flipping of the modulation maxima and minima for a shrink of 1$\%$ or more
(note that this is a continuous conversion as the value of the shrink increases).

\vskip.2cm

It may be pointed out that since this is a continuous change in the shape of the Fourier spectrum, the RL asymmetry feature
(eq. (\ref{RL})) is retained till a value of the energy-dependent shrink upto about 0.3$\%$, while the RLP asymmetry feature
in the changed spectrum (eq. (\ref{RLP})) becomes effective at values of about 0.7$\%$ or larger. For intermediate values of the shrink factor, it is
difficult to pinpoint a specific asymmetry feature.

\item Comparing Figure~\ref{FT5}(a) with Figure~\ref{FT6}(a) 
(3$\%$ smearing, scale shift with energy-independent 
shrink of 1$\%$ performed after smearing), 
the energy-independent shrink is seen to leave 
the hierarchy-sensitive feature almost unchanged, 
as expected from the preceding discussion. 

\item Comparing Figure~\ref{FT5}(a) with Figure~\ref{FT6}(b) 
(3$\%$ smearing, scale shift with energy-dependent 
shrink of 1$\%$ performed after smearing), 
the energy-dependent shrink is again observed to 
flip the modulation maxima to minima (as in Figure~\ref{FT5}(b)), 
and the hierarchy-sensitive feature can again be defined as $RLP$,
which in this case is still large and positive for the NH spectrum
and has a small negative value for the IH spectrum 
(we have $RLP(NH) = 0.40$, $RLP(IH) = -0.14$ in Figure~\ref{FT6}(b)).  

\item Comparing Figure~\ref{FT6}(a) and \ref{FT6}(b) with Figure~\ref{FT6}(c) 
(3$\%$ smearing, scale shift with energy-dependent shrink of 1$\%$ and energy-independent
shrink of 1$\%$ performed after smearing),  it is seen that the resulting FCT spectrum is almost identical to the 
spectrum in Figure~\ref{FT6}(b) obtained with only an energy-dependent shrink
factor of 1$\%$. 
This is because, as noted earlier, the energy-independent shrink factor leaves the spectrum
almost unchanged (Figure~\ref{FT6}(a)). Hence a combination of a linear
and a non-linear scale uncertainty leads to the same effects as a non-linear scale uncertainty.  
A similar behaviour is observed in the FST spectrum.

\end{itemize}

\begin{figure}[t]
\centerline
{
\hspace*{1em} \epsfxsize=5.3cm\epsfysize=6.5cm
\epsfbox{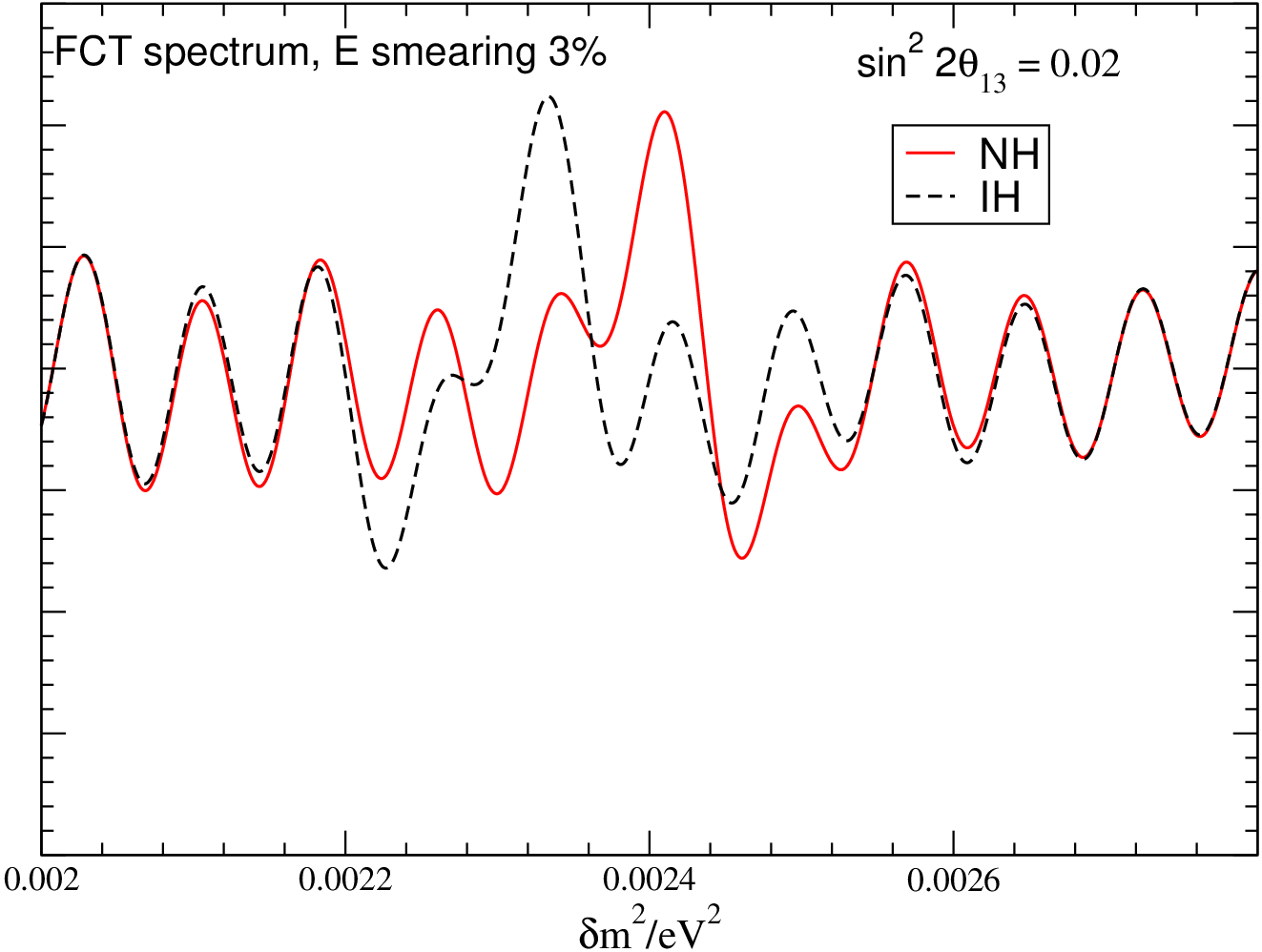}
        \hspace*{0.52ex}
\epsfxsize=5.3cm\epsfysize=6.5cm
\epsfbox{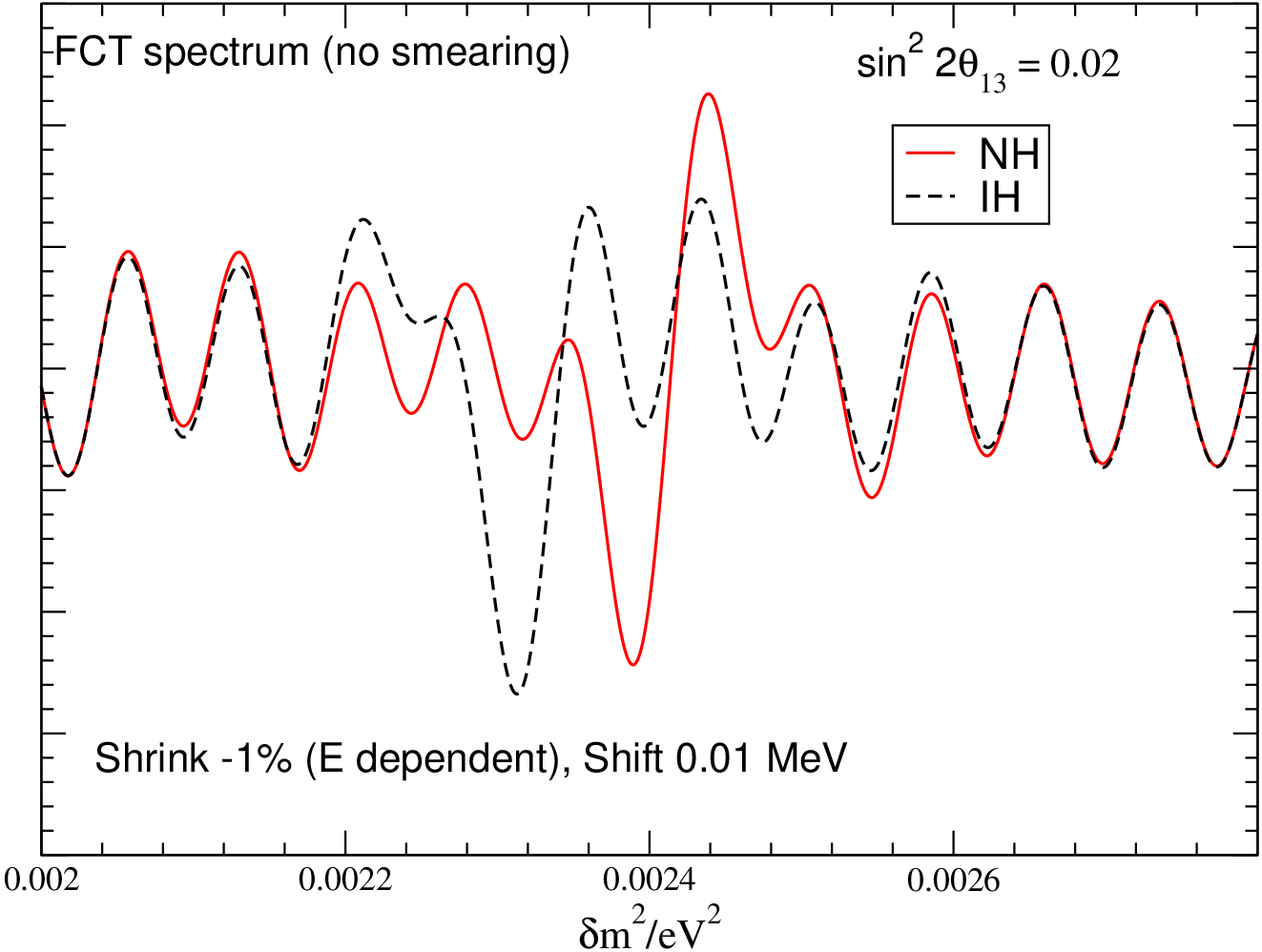}
        \hspace*{0.52ex}
\epsfxsize=5.3cm\epsfysize=6.5cm
\epsfbox{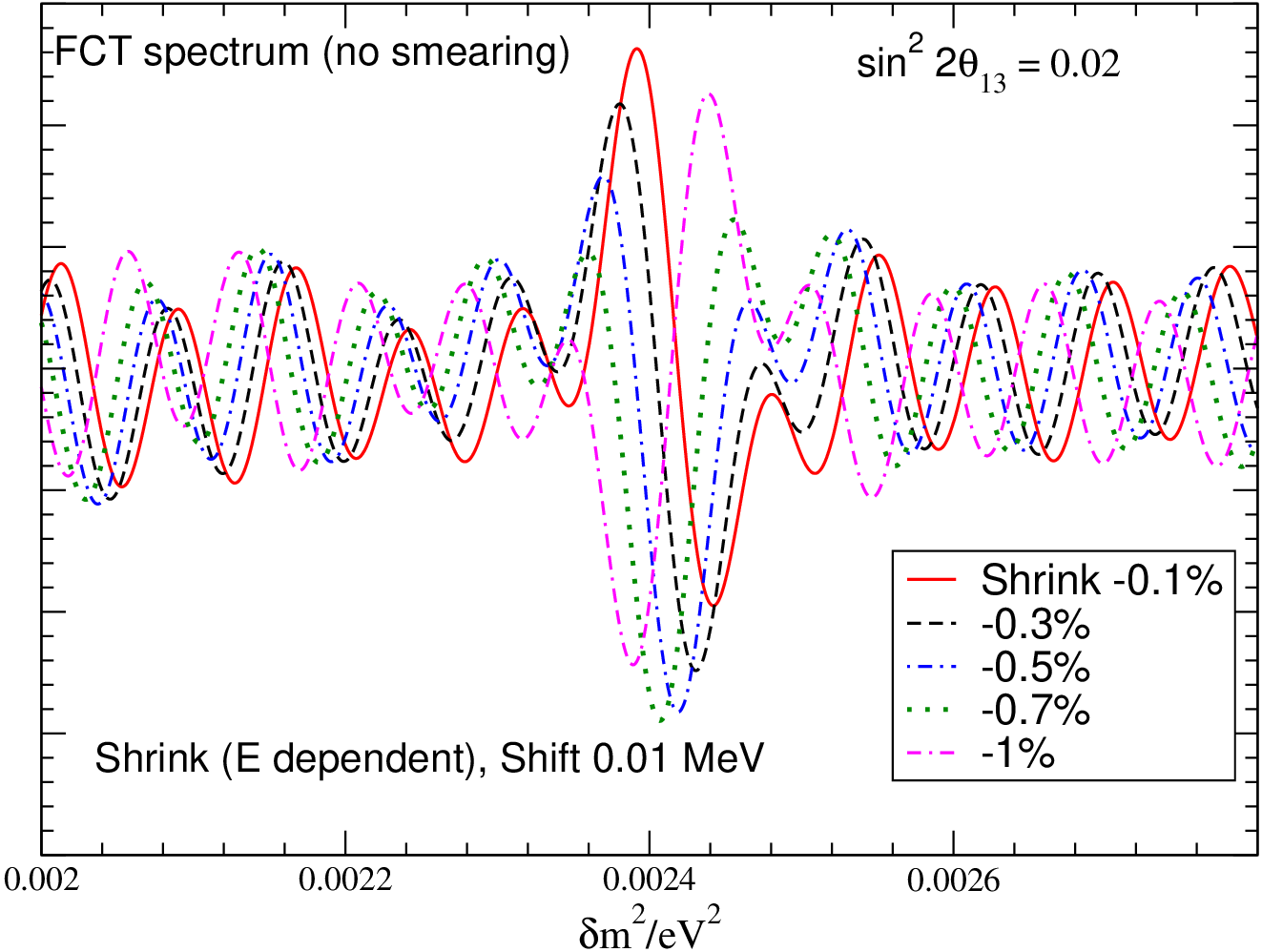}
}
\hskip 4cm
{\bf (a)}
\hskip 4cm
{\bf (b)}
\hskip 4 cm
{\bf (c)}
\caption[]
{\footnotesize{{{\bf (a)}
Fourier cosine transformed (FCT)  
reactor $\bar{\nu}_e$ event rate spectrum vs 
$\delta m^2$ with power (y-axis) in arbitrary units, 
for $\sin^2 2\theta_{13} = 0.02$, normal hierarchy and 
inverted hierarchy, 3$\%$ energy resolution
of the detector and no energy scale 
uncertainty.
{\bf (b)} 
Fourier cosine transformed (FCT)  
reactor $\bar{\nu}_e$ event rate spectrum vs 
$\delta m^2$ with power (y-axis) in arbitrary units, 
for $\sin^2 2\theta_{13} = 0.02$, normal hierarchy and 
inverted hierarchy,  ideal energy resolution
of the detector and energy-dependent uncertainty 
(shrink and shift) in the energy scale.
{\bf (c)}
FCT spectrum for $\sin^2 2\theta_{13} = 0.02$, normal hierarchy,  ideal energy resolution
of the detector and different values of energy-dependent uncertainty
in the energy scale.
}}}
\label{FT5}
\end{figure}

{\bf{FST spectrum.}}
\begin{itemize} 
\item Comparing Figure~\ref{FT4}(b) (no smearing, no scale shift) with 
Figure~\ref{FT7}(a) (3$\%$ smearing, no scale shift), 
the hierarchy-sensitive feature of the asymmetry 
$PV = (P-V)/(P+V)$  is seen to be retained. 
This feature is reflected in a large positive value of
the asymmetry $PV$ for the NH spectrum
and a value close to zero for the IH spectrum
(we have $PV(NH) = 0.32$, $PV(IH) = 0.04$ in Figure~\ref{FT4}(b),
and $PV(NH) = 0.41$, $PV(IH) = 0.07$ in Figure~\ref{FT7}(a)). 

\item Comparing Figure~\ref{FT4}(b) with Figure~\ref{FT7}(b) 
(no smearing, scale shift with energy-dependent shrink 
of 1$\%$), 
we see that the $PV$ asymmetry features 
corresponding to the NH and IH spectra 
are not present when 
the energy-dependent shrink is taken into account. 
Now the FST spectra 
show a behaviour similar to that of the FCT spectra with an energy-dependent shrink 
(Figure~\ref{FT5}(b) and Figure~\ref{FT6}(b)),
{\it{i.e.}} the absolute modulation maxima in the NH and IH cases 
are effectively replaced by absolute modulation minima and 
the adjacent right and left (minima) valleys are replaced by right and
left (maxima) peaks, $RP$ and $LP$. 
Here the quantity $RLP$ 
is close to zero for the NH spectrum and is 
significantly different from zero and negative
for the IH spectrum. This behaviour is explained in the same way
as for the FCT spectrum: it also appears for values of the shrink 
of about 0.7$\%$ or larger,
while the $PV$ asymmetry feature is retained for very small values of the shrink of upto
about 0.3$\%$. 

\item Comparing Figure~\ref{FT7}(a) with Figure~\ref{FT8}(a) 
(3$\%$ smearing, scale shift with energy-independent 
shrink of 1$\%$ performed after smearing), 
the $PV$ asymmetry feature, distinguishing between 
the NH an IH spectra,
remains largely unchanged, as expected. 

\item Comparing Figure~\ref{FT7}(a) with Figure~\ref{FT8}(b) 
(3$\%$ smearing, scale shift with energy-dependent 
shrink of 1$\%$ performed after smearing), 
we see that features distinguishing between 
the NH and IH spectra
in the case of a scale shift with an energy-dependent 
shrink of 1$\%$ are the same 
as in the unsmeared case discussed above
(when comparing Figure~\ref{FT4}(b) with Figure~\ref{FT7}(b)).  
 
\end{itemize}
 
  In the cases of energy-independent and energy-dependent 
energy scale expansion, the FCT and FST spectra 
exhibit the same features as those discussed 
above assuming energy-independent and energy-dependent 
energy scale shrink, respectively.
For the FCT spectrum this is illustrated 
in Figure~\ref{FT3}.  Comparing the curves with 
energy-dependent 
expansion in Figure~\ref{FT3}(a) and \ref{FT3}(b) 
shows that the RLP asymmetry feature is present. 
With an increase in the magnitude of the shrink/expansion uncertainty, 
the asymmetry features discussed above survive with a reduced amplitude, 
getting washed out if the uncertainty exceeds $\sim 5\%$.

\vskip.3cm

  The above properties of the Fourier spectra indicate that 
it should be possible, in principle, to extract 
information about the type of the spectrum 
the neutrino masses obey from the features 
present in the spectra, 
although the nature of the hierarchy-dependent features 
is changed in the case of an energy-dependent 
energy scale shrink/expansion. 


{\subsection
{The effect of the uncertainty of $\Delta m^2_{31}$ on the Fourier spectra}}


The effect of varying the atmospheric neutrino
mass-squared difference over its error range, 
in general, causes a change in the  
magnitude of the 
hierarchy-sensitive asymmetry 
features of the Fourier spectra. 
More specifically we note the following.

\vskip.2cm

\begin{itemize}

\item {\bf{FCT spectra.}} Comparing the NH and IH 
FCT spectra for different values of $\Delta m^2_{31}$ 
over its uncertainty range, 
it can be seen that the $RL$ asymmetry
feature of the NH and IH spectra is completely changed 
in the case of an energy-dependent 
scale shrink/expansion. 
Instead, the $RLP$ asymmetry feature discussed 
in the context of Figure~\ref{FT5}(b) and Figure~\ref{FT6}(b) appears.  
This feature is present throughout the considered 
range of $\Delta m^2_{31}$, though the magnitude of the effect 
varies over the range. 

\item {\bf{FST spectra.}} Comparing the NH and IH FST spectra, 
it can be seen that the $PV$ asymmetry feature in the
spectra is changed in the case of the 
energy-dependent energy scale shrink/expansion, 
and the $RLP$ asymmetry feature comes into play, 
as earlier noted in connection with
Figure~\ref{FT7}(b) and Figure~\ref{FT8}(b). 
It is present with different 
amplitudes throughout the considered 
range of $\Delta m^2_{31}$.  

\end{itemize}

\begin{figure}[t]
\centerline
{
\hspace*{1em} \epsfxsize=5.3cm\epsfysize=6.5cm
\epsfbox{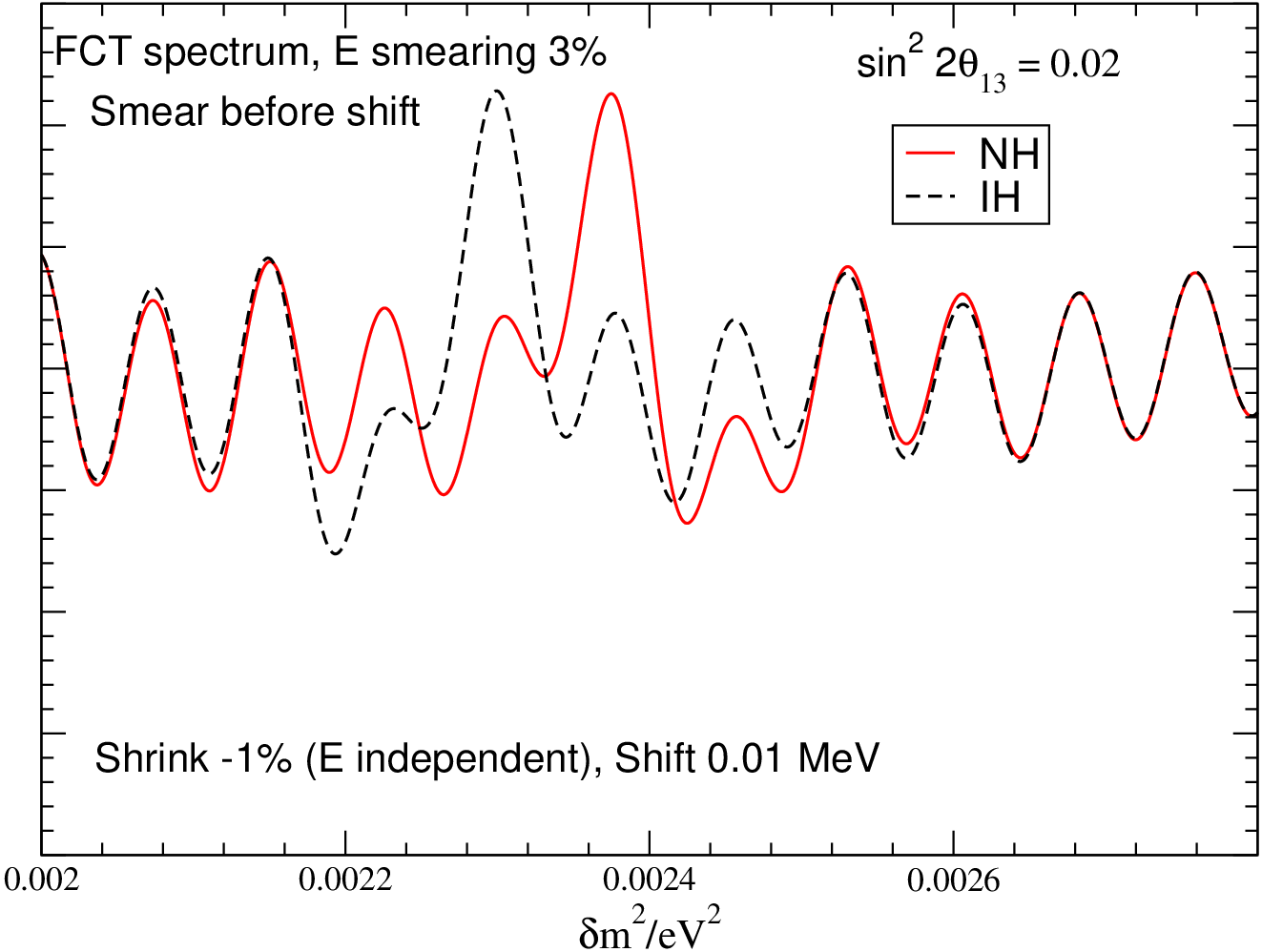}
        \hspace*{0.52ex}
\epsfxsize=5.3cm\epsfysize=6.5cm
\epsfbox{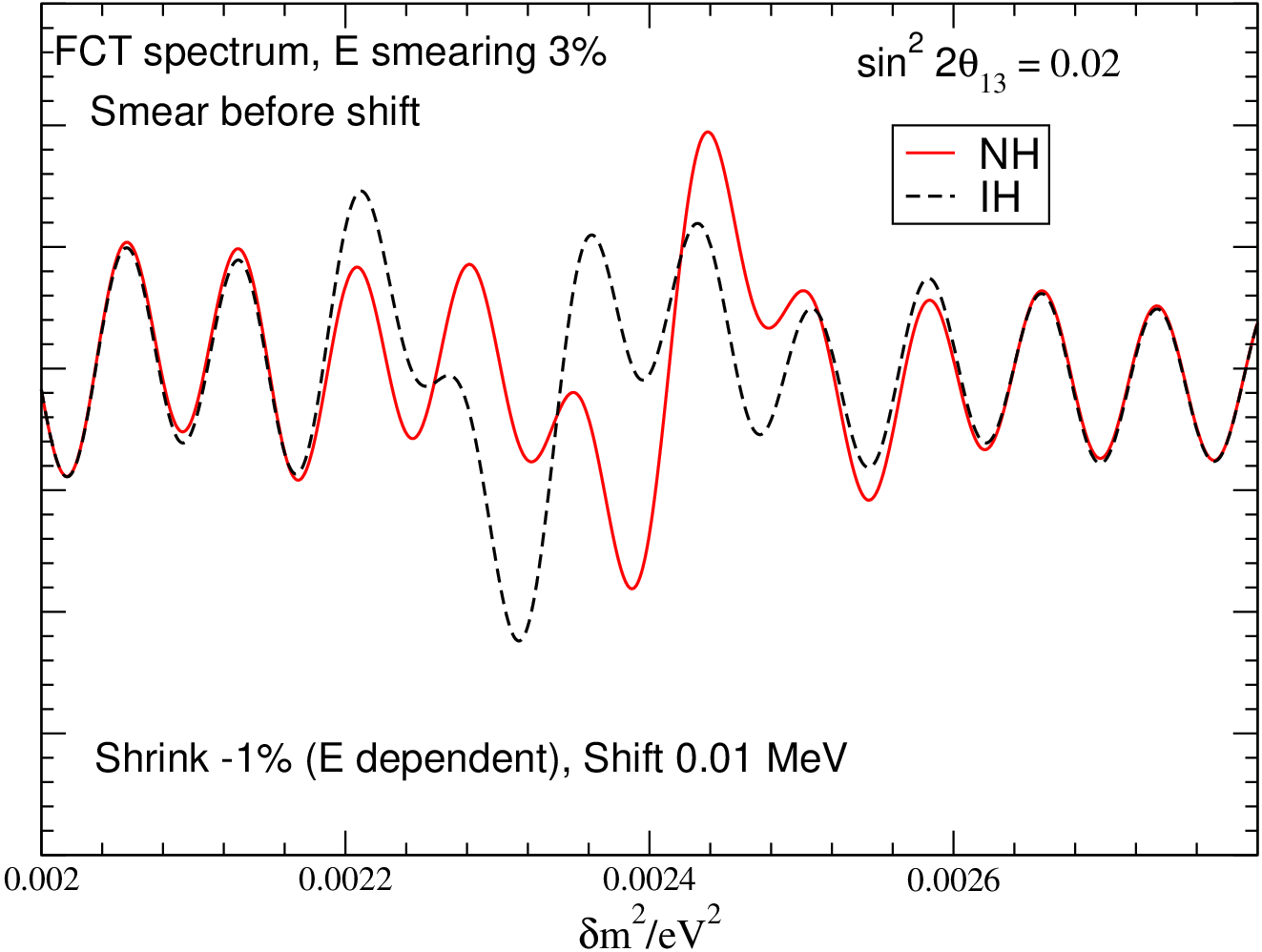}
        \hspace*{0.52ex}
\epsfxsize=5.3cm\epsfysize=6.5cm
\epsfbox{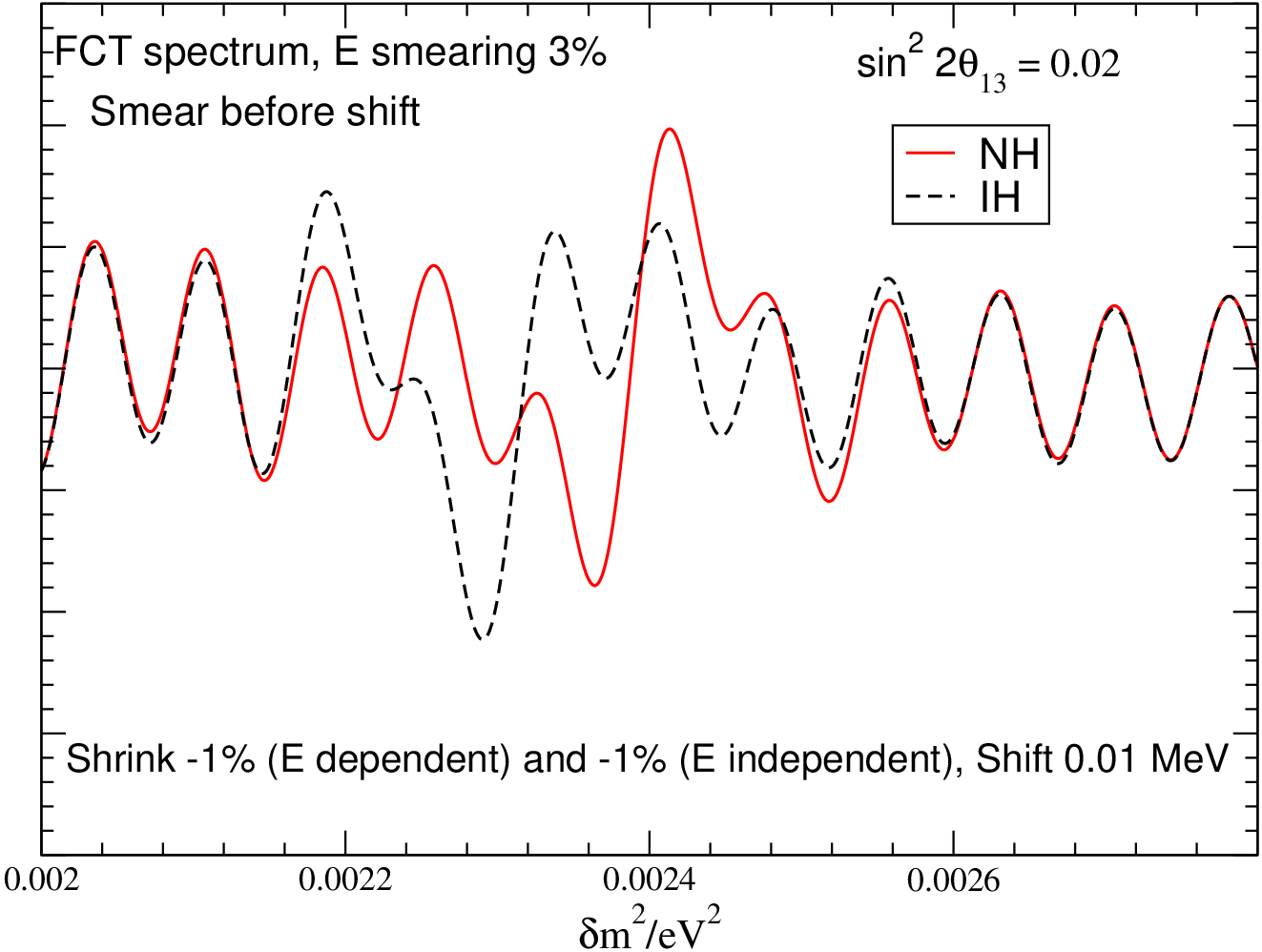}
}
\hskip 4cm
{\bf (a)}
\hskip 4cm
{\bf (b)}
\hskip 4 cm
{\bf (c)}
\caption[]
{\footnotesize{{{\bf (a)}
Fourier cosine transformed (FCT)  
reactor $\bar{\nu}_e$ event rate spectrum vs 
$\delta m^2$ with power (y-axis) in arbitrary units, 
for $\sin^2 2\theta_{13} = 0.02$, normal hierarchy and 
inverted hierarchy, 3$\%$ energy resolution
of the detector and energy-independent uncertainty 
(shrink and shift) in the energy scale. 
The energy scale shift is performed on the
neutrino energy $E_m$ after taking smearing into account.
{\bf (b)} 
The same as in {\bf (a)}, but for  
energy-dependent uncertainty (shrink and shift) in 
the energy scale.
{\bf (c)}
The same as in {\bf (a)} and {\bf (b)}, but for
a combination of an energy-dependent and energy-independent uncertainty (shrink and shift) in
the energy scale.
}}}
\label{FT6}
\end{figure}

{\section{$\chi^2$-Analysis of the sensitivity to the 
type of the neutrino mass spectrum}}

 In the present Section we perform a full $\chi^2$-analysis 
of the sensitivity to the type of the neutrino mass spectrum 
of a ``measured'' reactor $\bar{\nu}_e$ 
spectrum. This allows us to take into account in a 
systematic way the uncertainties in the 
knowledge of $|\Delta m^2_{\rm atm}|$, $\theta_{13}$, 
$\Delta m_{21}^2$, $\theta_{12}$, 
the uncertainty in the energy scale,
the systematic and geo-neutrino uncertainties, 
as well as the effects of the detector energy resolution. 
As is well known, the uncertainties 
in the values of $|\Delta m^2_{\rm atm}|$ and $\theta_{13}$, 
in particular, play a crucial role in 
the sensitivity to the neutrino mass hierarchy.

\vskip.3cm

  We perform a binned $\chi^2$ analysis which involves an 
optimization in binning, a marginalization over the 
relevant neutrino oscillation parameters, 
and incorporation of systematic errors by the 
method of pulls. 
We find that an energy scale 
shrink/expansion and/or shift at the level of $\sim 1\%$, 
even when energy-dependent, does not affect 
the sensitivity to the hierarchy, 
and that the inclusion of the systematic and 
geo-neutrino flux uncertainties has only a 
minimal effect on the sensitivity of interest. 
We present results for different values of 
$\sin^2\theta_{13}$, the detector 
exposure and the energy resolution.
A prior term is 
added to the sensitivity to take into account information 
from other experiments on parameter uncertainties, and it 
is shown that if the present error ranges are considered, 
this external information leads to only a 
slight improvement in the results.      

\vskip.3cm

  In order to compute the hierarchy 
sensitivity by the $\chi^2$-method, 
it is necessary to have binned event data.
For a set of ''experimental'' (observed) events $N_{ex}(i)$ 
and ''theoretical'' (predicted) events $N_{th}(i)$,
the standard Gaussian definition of the least squares sum of  
binned data reads:


\be
\chi^2_{stat} = \sum_i \frac{[N_{ex}(i) - N_{th}(i)]^2}{N_{ex}(i)}\,,
\ee


\noindent where only the statistical error 
$\sigma_{stat} = \sqrt{N_{ex}(i)}$ is taken into account, 
and i denotes the bin label. 
We simulate the ''experimental''
spectrum $N_{ex}$ for a fixed ''test'' or ''true'' 
hierarchy (performed with a normal hierarchy 
unless otherwise specified; 
the difference in results is minimal). 
All other parameters are also kept fixed at a set of ''test'' 
values  in $N_{ex}$. The theoretical 
spectrum $N_{th}$ is then generated 
with the {\it{other}}
hierarchy, called the ''wrong'' hierarchy. 
The $\chi^2$ thus obtained determines
the confidence level at which the ''wrong'' 
hierarchy can be {\it{excluded}} (i.e., 
the ``$\chi^2$ sensitivity'')
given the ''true'' hierarchy, the set of 
values of all other parameters used and 
the given values of errors, uncertainties, 
detector resolution, exposure, etc.

\vskip.3cm
   
   Errors other than the $\sigma_{stat}$, 
like the flux and geo-neutrino uncertainties 
and systematic errors, can be included using 
the method of pulls. Also,  
a comprehensive $\chi^2$ analysis 
requires a marginalization over the uncertainties
in the neutrino oscillation parameters, which can be done 
by varying the parameters in the theoretical spectrum $N_{th}$ and
choosing the minimum value of $\chi^2$ after taking 
into account this variation.

\vskip.4cm

\begin{figure}[t]
\centerline
{
\epsfxsize=8.5cm\epsfysize=6.2cm
\epsfbox{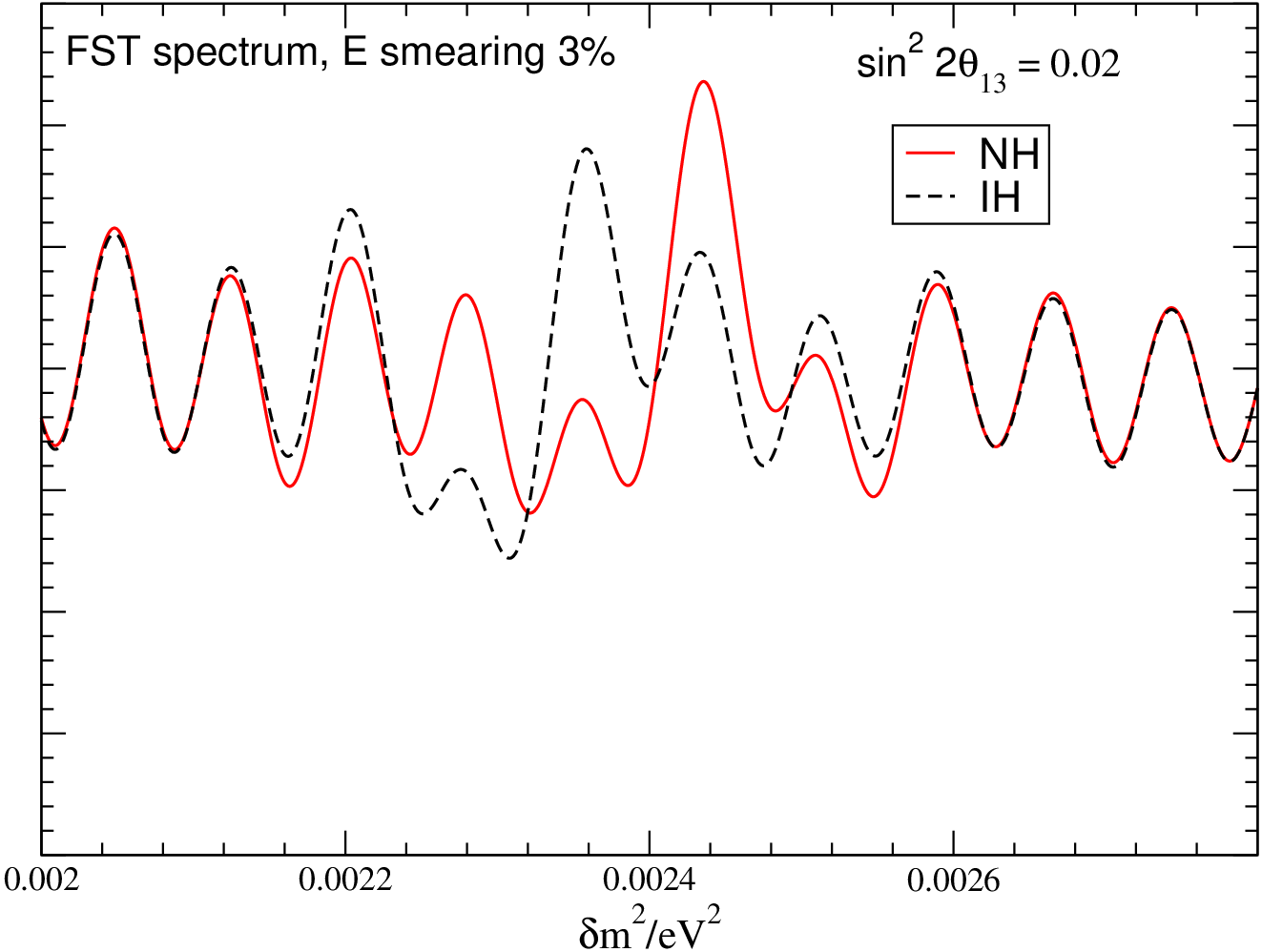}
        \hspace*{0.7ex}
\epsfxsize=8.5cm\epsfysize=6.2cm
\epsfbox{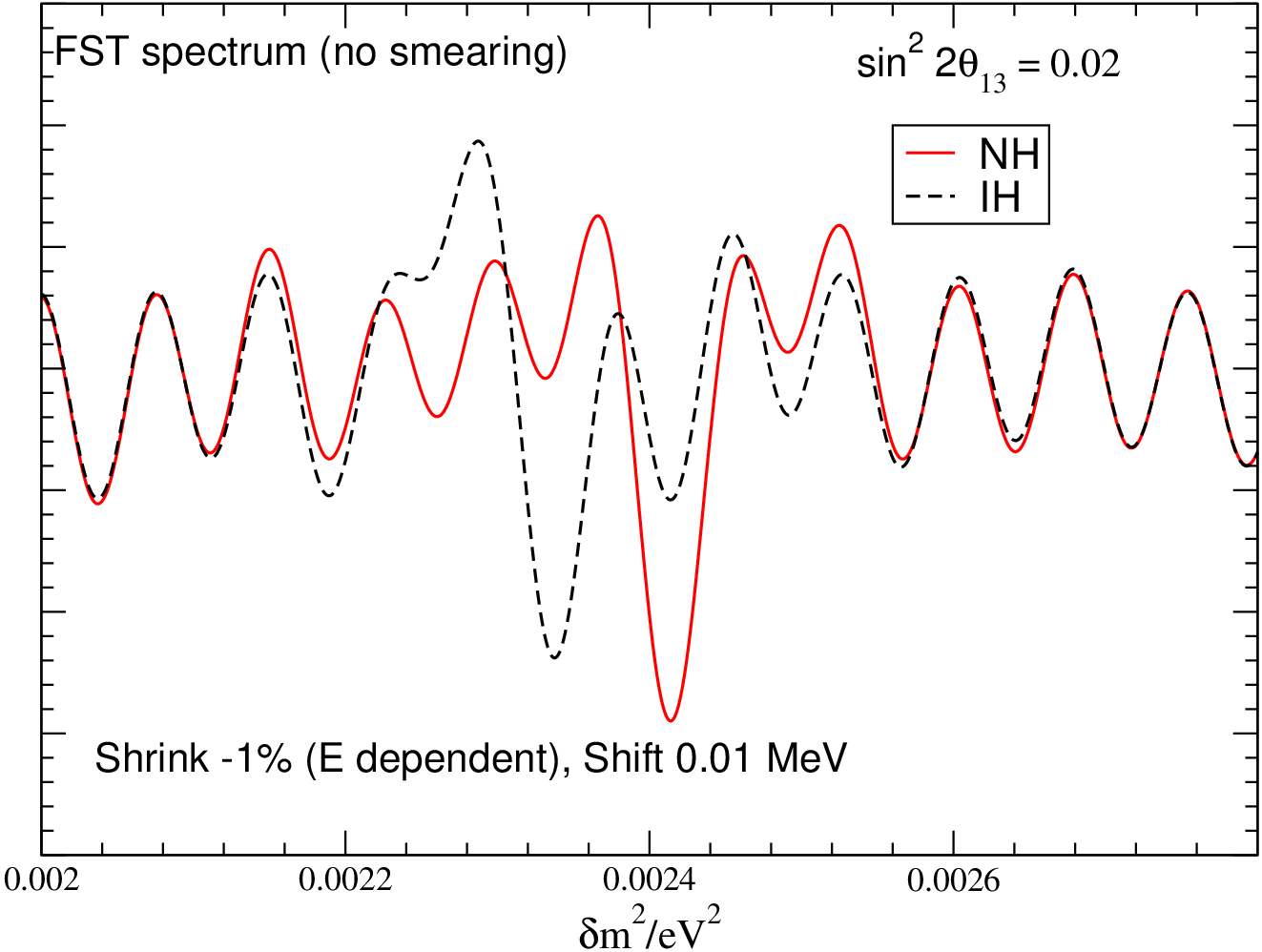}
}
\hskip 5cm
{\bf (a)}
\hskip 7cm
{\bf (b)}
\caption[]
{\footnotesize{{{\bf (a)}
Fourier sine transformed (FST)  
reactor $\bar{\nu}_e$ event rate spectrum vs 
$\delta m^2$ with power (y-axis) in arbitrary units, 
for $\sin^2 2\theta_{13} = 0.02$, normal hierarchy and 
inverted hierarchy, 3$\%$ energy resolution
of the detector and no energy scale 
uncertainty.
{\bf (b)} 
Fourier sine transformed (FST)  
reactor $\bar{\nu}_e$ event rate spectrum vs 
$\delta m^2$ with power (y-axis) in arbitrary units, 
for $\sin^2 2\theta_{13} = 0.02$, normal hierarchy and 
inverted hierarchy, for ideal energy resolution
of the detector and energy-dependent uncertainty 
(shrink and shift) in the energy scale.
}}}
\label{FT7}
\end{figure}

{\bf{Optimization of bin number.}} 
Figure~\ref{chisqvsbins} shows the behaviour of the 
$\chi^2$ sensitivity with an increase in the bin number. 
We plot in the figure the values of $\chi^2$ with fixed 
neutrino parameters for an exposure of 200 kT GW yr, 
$\sin^2 2\theta_{13} = 0.05$, $\Delta_{31}(NH) = 0.0024$, 
$\Delta_{31}(IH) = -\Delta_{31}(NH) + \Delta_{21}$ 
and a detector resolution of 3$\%$, 
for different numbers of L/E bins in the range  
${\rm{L/E}} = 5-32$ Km/MeV. The sensitivity is seen 
to improve dramatically with an improvement in 
the fineness of binning. However, the maximum bin number 
that can be used is restricted by the energy resolution 
of the detector. Hence, it becomes important to 
optimize the number of bins and choose a binning which 
is fine enough to give the best possible 
sensitivity while being consistent with the detector resolution. 

\vskip.3cm

In general, the bin width can be chosen to 
be of the same order as the resolution width, 
but not significantly smaller.  
Here, an energy resolution of 3$\%$ would mean 
a resolution width of $0.03 \times \sqrt{E_{vis}}$, 
or approximately 0.03 - 0.1 MeV, over the
given energy range of $E=1.8$ to 12 MeV. Hence we 
can choose to take approximately $10.2/0.07=145$ 
bins in this energy range. Therefore, we consider a 
150-bin analysis. 
The no-oscillation unbinned reactor event spectrum 
is used to generate a binned spectrum 
of events (the product of the 
no-oscillation event spectrum and 
the oscillation probability) in 
L/E bins of width 0.18 Km/MeV, {\it{i.e.}} 150 bins 
in the given L/E range of 5 - 32 Km/MeV. The simulated 
''predicted'' spectra
are then used to  
calculate the $\chi^2$ sensitivity.

\vskip.3cm


Figs.~\ref{binevent1} and \ref{binevent2} 
show the 150-bin event spectrum for both the 
normal and inverted hierarchies, with or without 
energy smearing and energy scale shift, using the 
no-oscillation spectrum as the unbinned data. 
The figures are obtained for $\sin^2 2\theta_{13}=0.05$ 
and a detector exposure of 200 kT GW yr. 
It can be seen that the NH and IH 
spectra show small differences through 
a greater part of the L/E range, 
which in a $\chi^2$ analysis 
can give a significant result since the 
procedure adds up the contributions from 
all the bins. A smearing of 3$\%$ washes 
out the sensitivity in part of the 
L/E range, as expected. The energy scale 
shift/shrink is seen to affect both 
the NH and the IH spectra identically.  

\vskip.3cm


\vskip.4cm

%
\subsection{Parameter marginalization} 
%

For a realistic analysis, one needs to take 
into account the ranges of uncertainty of the 
neutrino oscillation parameters, since they are 
not known to infinite precision. In order to do this, 
the values of the parameters (ideally all the 
neutrino parameters) are fixed at certain input 
(''true'') values in the ''observed'' event 
spectrum $N_{ex}(i)$ and varied over their present 
error ranges while computing the ''theoretical'' 
event spectrum $N_{th}(i)$, subsequently choosing 
the minimum value of $\chi^2$ after including a full variation. 

\vskip.3cm

\begin{figure}[t]
\centerline
{
\epsfxsize=8.5cm\epsfysize=6.2cm
\epsfbox{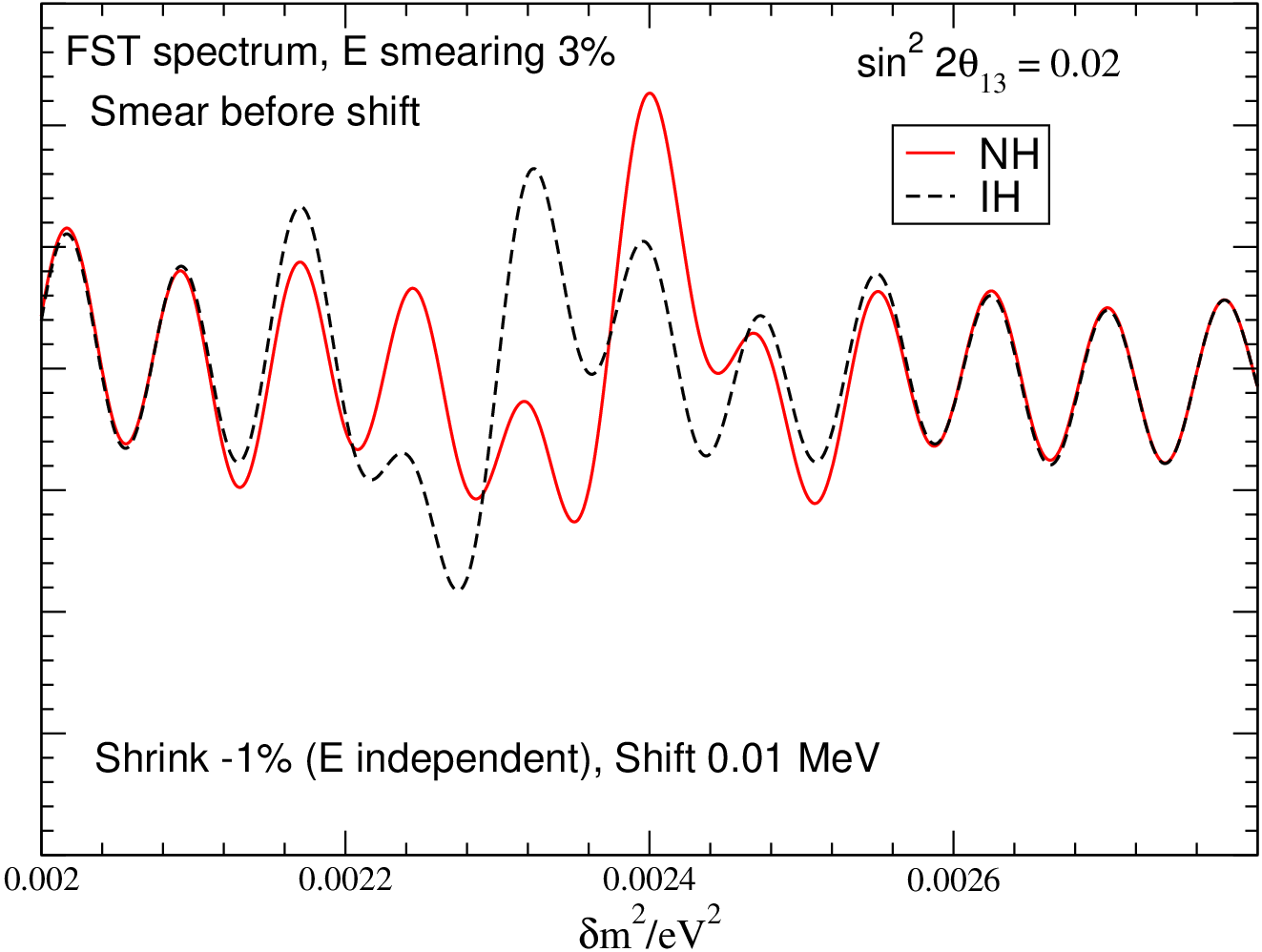}
        \hspace*{0.7ex}
\epsfxsize=8.5cm\epsfysize=6.2cm
\epsfbox{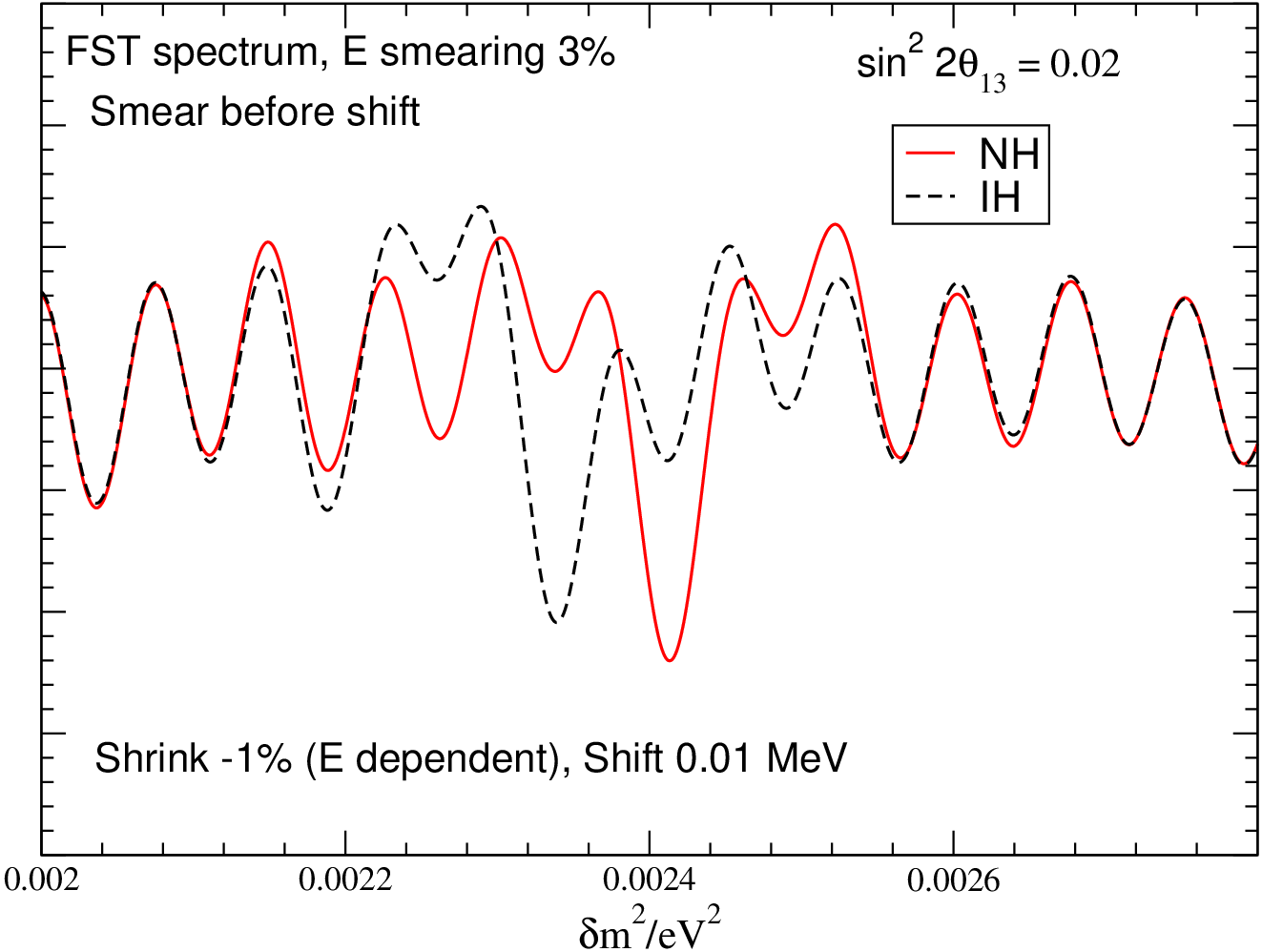}
}
\hskip 5cm
{\bf (a)}
\hskip 7cm
{\bf (b)}
\caption[]
{\footnotesize{{{\bf (a)}
Fourier sine transformed (FST)  
reactor $\bar{\nu}_e$ event rate spectrum vs 
$\delta m^2$ with power (y-axis) in arbitrary units, 
for $\sin^2 2\theta_{13} = 0.02$, normal hierarchy and 
inverted hierarchy, for 3$\%$ energy resolution
of the detector and energy-independent uncertainty 
(shrink and shift) in the energy scale.
The energy scale shift is performed on the
neutrino energy $E_m$ after taking smearing into account.
{\bf (b)} 
The same as in {\bf{a}}, but for energy-dependent uncertainty
(shrink and shift) in the energy scale.
}}}
\label{FT8}
\end{figure}

Practically, since the solar neutrino parameters 
$\theta_{12}$ and $\Delta m^2_{21}$ are already measured 
with a relatively high precision and the dependence of the 
oscillation probability on their variation is rather weak, 
it usually suffices to marginalize over 
the parameters $\theta_{13}$ and $|\Delta m^2_{31}|$. 
We have checked that a marginalization over 
$\theta_{12}$ and $\Delta m^2_{21}$ over their present 
3$\sigma$ ranges 
($\sin^2 \theta_{12} = 0.27 - 0.38$, 
$\Delta m^2_{21} = 7.0 \times 10^{-5} - 8.3 \times 10^{-5}$) 
does not affect the results. Also, the fineness of binning 
in the parameters being varied during the process 
of marginalization needs to be optimized,
since taking a coarse binning may give inaccurate 
results due to missing the actual point of 
minimal $\chi^2$, while making the binning 
more rigorous gives progressively improved 
results but also increases the computational time involved.    

\vskip.3cm

We consider the following error ranges for the two marginalized parameters:
i) $|\Delta_{31}|$ is allowed to vary
  in the range $2.3 \times 10^{-3} - 2.6
  \times 10^{-3}$ eV$^2$, and
ii) $\sin^2 2 \theta_{13}$ is varied from 0.0 to 0.15.

%
\subsection{The precision on
$\Delta m^2_{\rm atm}$ and its effect on the hierarchy sensitivity} 
%

When an experiment determines the 
atmospheric mass-squared difference, 
assuming that it does not also simultaneously 
determine the hierarchy, 
the question arises of what exactly it measures. 
We know that by definition, when the hierarchy 
is normal, the magnitude of $\Delta m^2_{31}$ 
is greater than that of $\Delta m^2_{32}$, 
since the third mass state lies above 
the states 1 and 2, while in the
case of an inverted hierarchy, $\Delta m^2_{32}$ 
is greater in magnitude than $\Delta m^2_{31}$, 
since the third state lies below the first two.
So, if the experiment measuring the mass-squared 
difference does not know the hierarchy, it is not 
possible for it to measure the quantity
$|\Delta m^2_{31}|$ or $|\Delta m^2_{32}|$. We can reasonably 
assume that it measures something in between, 
or an effective $\Delta m^2_{\rm{atm}}$
which is blind to the hierarchy, 
{\it{i.e.}} $|\Delta m^2_{\rm{atm}}(NH)| = |\Delta m^2_{\rm{atm}}(IH)|$. 
This is, in general, a linear combination of 
$\Delta m^2_{31}$ and $\Delta m^2_{32}$, i.e.
\be
\Delta m^2_{\rm{atm}} = c \Delta m^2_{31} + d \Delta m^2_{32}\,, 
\label{Delatm}
\ee
%
\noindent where c and d can vary from 0 to 1 and $c + d =1$.

\vskip.3cm

\begin{figure}[t]
\centerline
{
\epsfxsize=8.5cm\epsfysize=7.0cm
\epsfbox{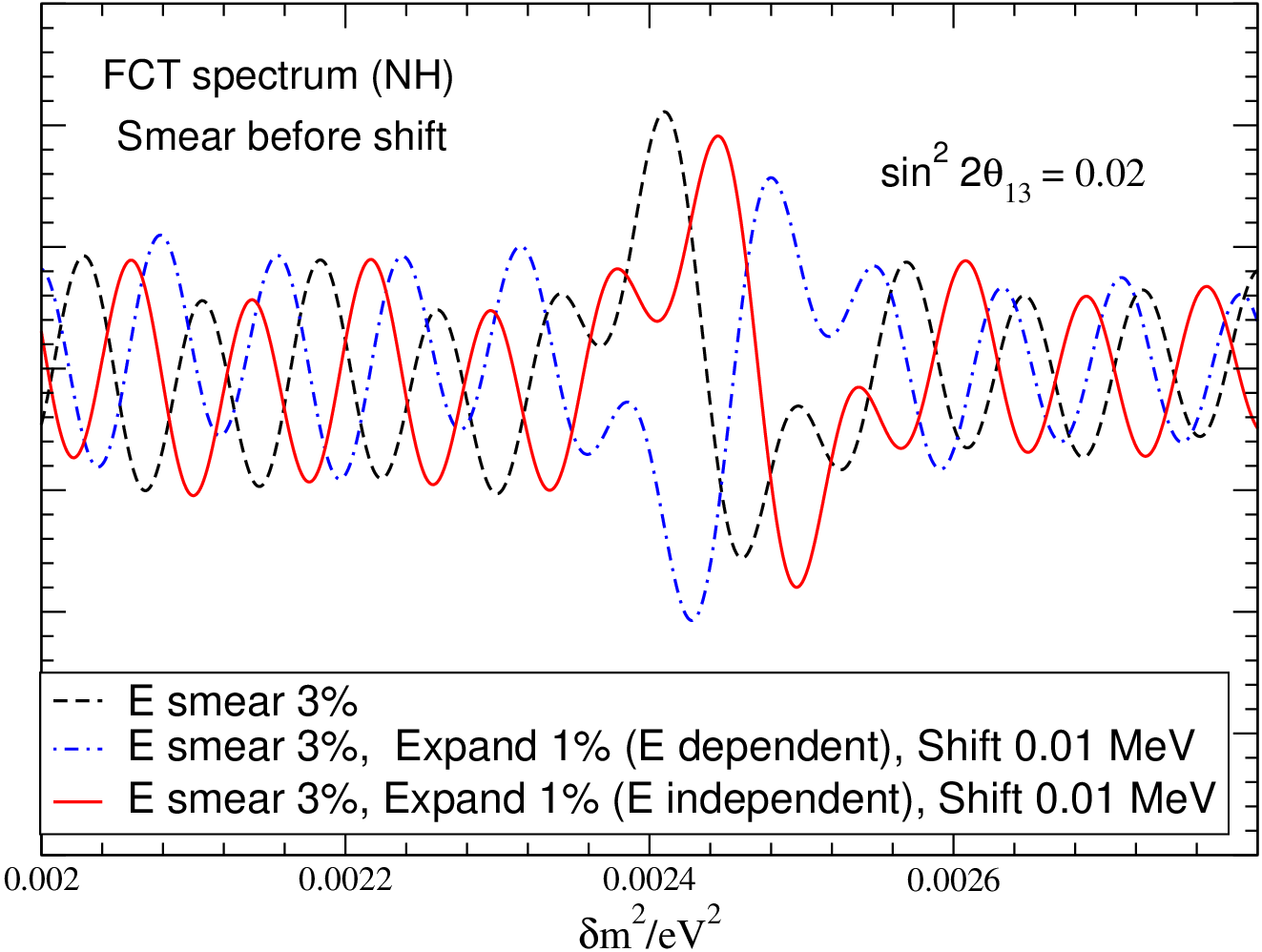}
        \hspace*{0.7ex}
\epsfxsize=8.5cm\epsfysize=7.0cm
\epsfbox{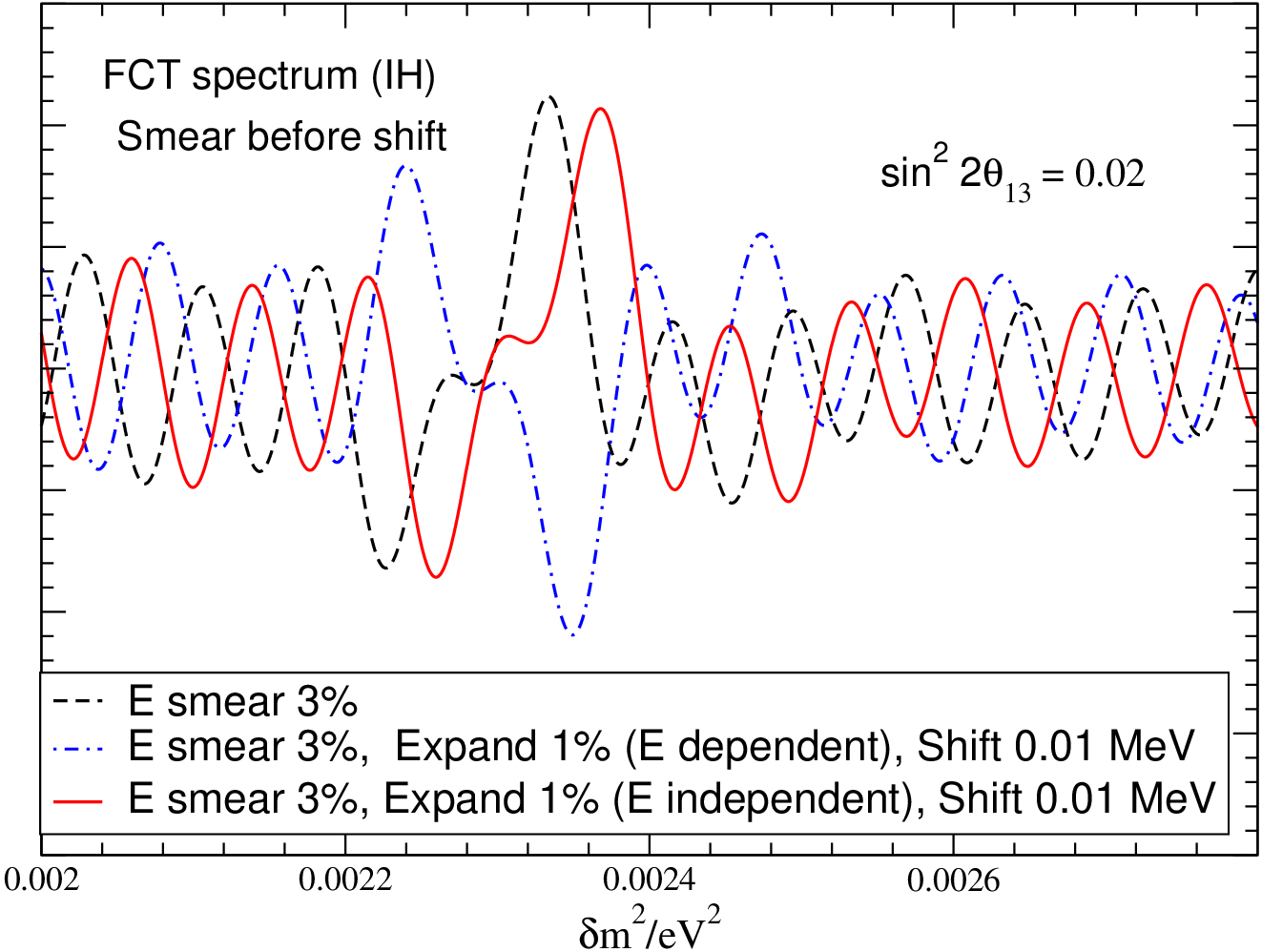}
}
\hskip 5cm
{\bf (a)}
\hskip 7cm
{\bf (b)}
\caption[]
{\footnotesize{{{\bf (a)}
Fourier cosine transformed (FCT)  
reactor $\bar{\nu}_e$ event rate spectrum vs 
$\delta m^2$ with power (y-axis) in arbitrary units, 
for $\sin^2 2\theta_{13} = 0.02$, normal hierarchy,
3$\%$ energy resolution of the detector and for 
both energy-dependent and energy-independent  
uncertainty (expansion and shift) in the energy scale.
The energy scale shift is performed on the 
neutrino energy $E_m$ after taking smearing into account.
{\bf (b)} The same as {\bf{(a)}} for inverted hierarchy. 
}}}
\label{FT3}
\end{figure}


Now when we perform the $\chi^2$ analysis for 
the hierarchy sensitivity,
we require, as inputs from some experimental 
measurement, the range of uncertainty in the 
atmospheric mass-squared
difference, as well as the values of $\Delta m^2_{31}$ 
and $\Delta m^2_{32}$ for both the normal and 
the inverted hierarchies, 
when computing the survival probability 
$P_{\bar{e}\bar{e}}$ in the two cases 
(the probability is the only neutrino 
parameter-dependent part in the event spectra
$N_{ex}$ and $N_{th}$). Hence we need to 
know how the magnitudes of $\Delta m^2_{31}$ and $\Delta m^2_{32}$ 
are related for the two hierarchies. From the definition 
of the measured $\Delta m^2_{\rm{atm}}$ and the fact
that it is equal in magnitude for NH and IH, 
it can be derived that the following relations hold:
\bea
|\Delta m^2_{31}(IH)| = \Delta m^2_{31}(NH) - 2d\Delta m^2_{21}\,, \nonumber \\ 
|\Delta m^2_{32}(IH)| = \Delta m^2_{32}(NH) + 2c\Delta m^2_{21}\,,
\label{DelNHIH}
\eea      
%
\noindent where c and d can vary from 0 to 1. 
In other words, the magnitude of $\Delta m^2_{31}(IH)$ 
(as derived from the measured mass-squared 
difference) can vary from anywhere between 
$\Delta m^2_{31}(NH)$ to $\Delta m^2_{31}(NH) - 2\Delta m^2_{21}$, 
while the magnitude of $\Delta m^2_{32}(IH)$ can be anywhere between
$\Delta m^2_{32}(NH)$ to $\Delta m^2_{32}(NH) + 2\Delta m^2_{21}$. 

\vskip.3cm

In some cases (for specific experiments 
and measurements localized in specific regions of L/E) 
it is possible to pinpoint the exact
linear combination being measured, since the 
relevant 3-flavour probability expressions may be reducible 
(with certain approximations) to effective 2-flavour 
forms which then define an effective 
mass-squared difference as the argument 
\cite{Minakata:2007tn}.{\footnote{In \cite{Minakata:2007tn},
an analysis of the possibility to determine the neutrino 
mass hierarchy is performed using specific values of the constants $c$ and $d$
in eq. (\ref{Delatm}) ($c = \cos^2 \theta_{12}$, $d = \sin^2 \theta_{12}$),
which are derived in the approximation of $\Delta m_{21}^2 L/4E << 1$.
For the range of $L/E$ considered by us we have $\Delta m_{21}^2 L/4E \sim 1$,
and thus the indicated approximation is not valid.}}  
In our analysis, we assume
the most general case of an input from 
an experiment where an unknown linear combination
is being measured. 

\vskip.3cm

The hierarchy sensitivity depends on the difference 
between the survival probability $P_{{\bar{e}}{\bar{e}}}$ 
for the two hierarchies, since the $\chi^2$ function 
is an artefact of this probability difference, averaged
over L/E bins. For different
values of the baseline L ({\it{i.e.}} different 
ranges of L/E), the $P_{{\bar{e}}{\bar{e}}}$ expression would give
a minimized value of $\Delta P_{{\bar{e}}{\bar{e}}} = P_{{\bar{e}}{\bar{e}}}(NH) - P_{{\bar{e}}{\bar{e}}}(IH)$ 
for different values of $\Delta m^2_{31}(IH)$ 
and $\Delta m^2_{32}(IH)$ in $P_{{\bar{e}}{\bar{e}}}(IH)$ 
(fixing a $\Delta m^2_{31}(NH)$ and $\Delta m^2_{32}(NH)$ 
in $P_{{\bar{e}}{\bar{e}}}(NH)$). 
In general, the minimum of $\Delta P_{{\bar{e}}{\bar{e}}}$ would occur 
for a point $|\Delta m^2_{31}(IH)| < \Delta m^2_{31}(NH)$ and 
$|\Delta m^2_{32}(IH)| > \Delta m^2_{32}(NH)$, for the same reason
as discussed above - this is how they are related in nature. 
So when performing the $\chi^2$ analysis, in addition
to marginalizing over the error range in $\Delta m^2_{\rm{atm}}$ 
(and hence in both $\Delta m^2_{31}(NH)$ 
and $\Delta m^2_{31}(IH)$), the possible variation 
in $|\Delta m^2_{31}(IH)|$ relative to $\Delta m^2_{31}(NH)$
as defined by  eq. (\ref{DelNHIH}) also has to be taken into account.    
        
\vskip.3cm

Thus, $|\Delta m^2_{31}(IH)|_{th}$ in the $N_{th}$ spectrum is 
varied from $\Delta m^2_{31}(NH)$ to
$\Delta m^2_{31}(NH) - 2\Delta m^2_{21}$, {\it{i.e.}} 0.0024 to 0.002248 
as well as over the error
range of $|\Delta m^2_{\rm{atm}}|$ 
(which is at present 0.0021 - 0.0028 at 3$\sigma$). 
Extending the range of marginalization does not 
change our  results.
What we need to check is
the value of the minimum $\chi^2$ obtained 
during this variation, at which point of $|\Delta m^2_{31}(IH)|$
it occurs, and whether it is zero or negligibly 
small at any point in this range. It is found that
(for the true value $\sin^2 2\theta_{13} = 0.05$ 
and a detector resolution of 4$\%$, 200 kT GW yr exposure,
and with a marginalization over $\theta_{13}$), 
the minimum $\chi^2$ is about 2, and occurs at
about $|\Delta m^2_{31}(IH)|_{th} = 0.002387$, as can 
be seen in Figure~\ref{chisqvsDel31}, which shows
the values of $\chi^2$ (with the above specifications) 
as a function of the magnitude of $\Delta m^2_{31}$ 
in the theoretical spectrum, choosing the hierarchy to be
normal in $N_{ex}$ and normal (dashed curve) 
or inverted (solid curve) in $N_{th}$. 

\vskip.3cm

\begin{figure}[t]
\centerline
{
\epsfxsize=8.5cm\epsfysize=7.0cm
\epsfbox{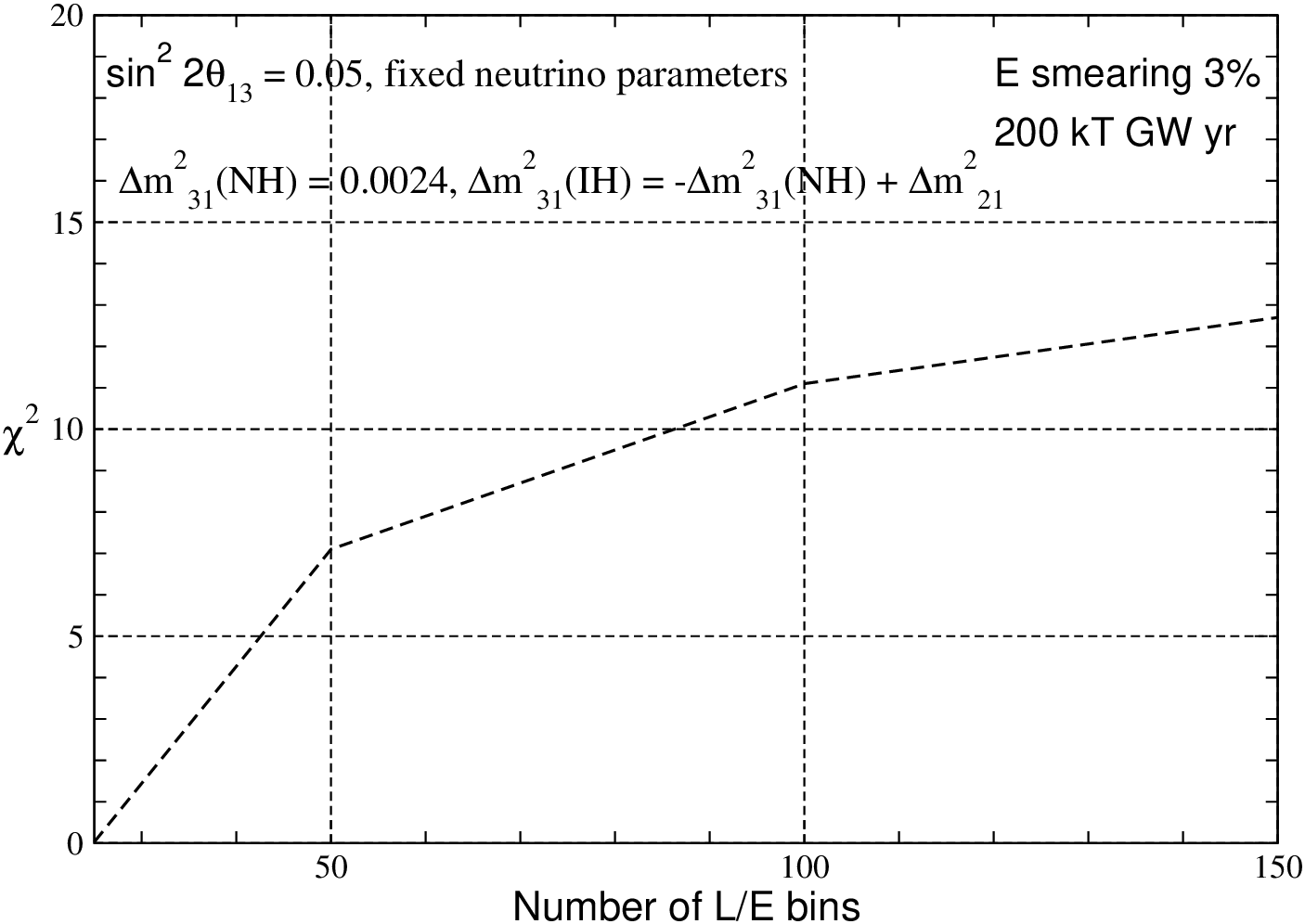}
}
\caption[]
{\footnotesize{
The hierarchy sensitivity $(\chi^2)_{stat}$ 
as a function of the number of L/E bins, 
for fixed neutrino oscillation parameters,
${\rm{\sin^2 2\theta_{13}}} = 0.05$ and detector's 
energy resolution of 3$\%$, statistics of 200 kT GW yr,
baseline of 60 Km and different L/E binnings in 
the range ${\rm{L/E}} = 5-32$ Km/MeV.
}}
\label{chisqvsbins}
\end{figure}

This verifies that for the hierarchy sensitivity
arising from the survival probability $P_{{\bar{e}}{\bar{e}}}$, 
in this L/E range, 
{\it the $\chi^2$ never vanishes at any point
of $|\Delta m^2_{31}(IH)|_{th}$, 
whichever experiment it may be derived from}. 
So, regardless of the precision
of $\Delta m^2_{\rm{atm}}$, there will be some non-zero 
hierarchy sensitivity given by this $\chi^2$, 
which would obviously be scaled up
with higher detector exposures and improved with 
better detector resolution. At the level
of the survival probability, this translates to 
the statement that the L/E spectra of $P_{\bar{e}\bar{e}}$  
for the normal and inverted hierarchies never become 
completely identical for any pair of possible values 
of $\Delta m^2_{31}(NH)$ and $\Delta m^2_{31}(IH)$ 
\footnote{This observation was also made in
ref. \cite{Hano2}.}.  
The point where they are most similar gives 
the minimum $\chi^2$.     
\vskip.3cm

%
\subsection{Results}
%

Table 1 lists the values of the hierarchy sensitivity 
$(\chi^2)_{stat}^{min}$ for different values of $\theta_{13}$ and 
the detector energy resolution, after a marginalization 
over the above parameter ranges, for an exposure of
200 kT GW yr, when a 150-bin analysis is performed. 
These results are with only statistical 
errors ({\it{i.e.}} no systematic uncertainties) taken into account. 
The hierarchy sensitivity in $\sigma$ is related to the 
1 d.o.f. $\chi^2$ here by the expression $\sigma = \sqrt{\chi^2}$. 

\vskip.3cm

\begin{figure}[t]
\centerline
{
\epsfxsize=8.5cm\epsfysize=7.0cm
\epsfbox{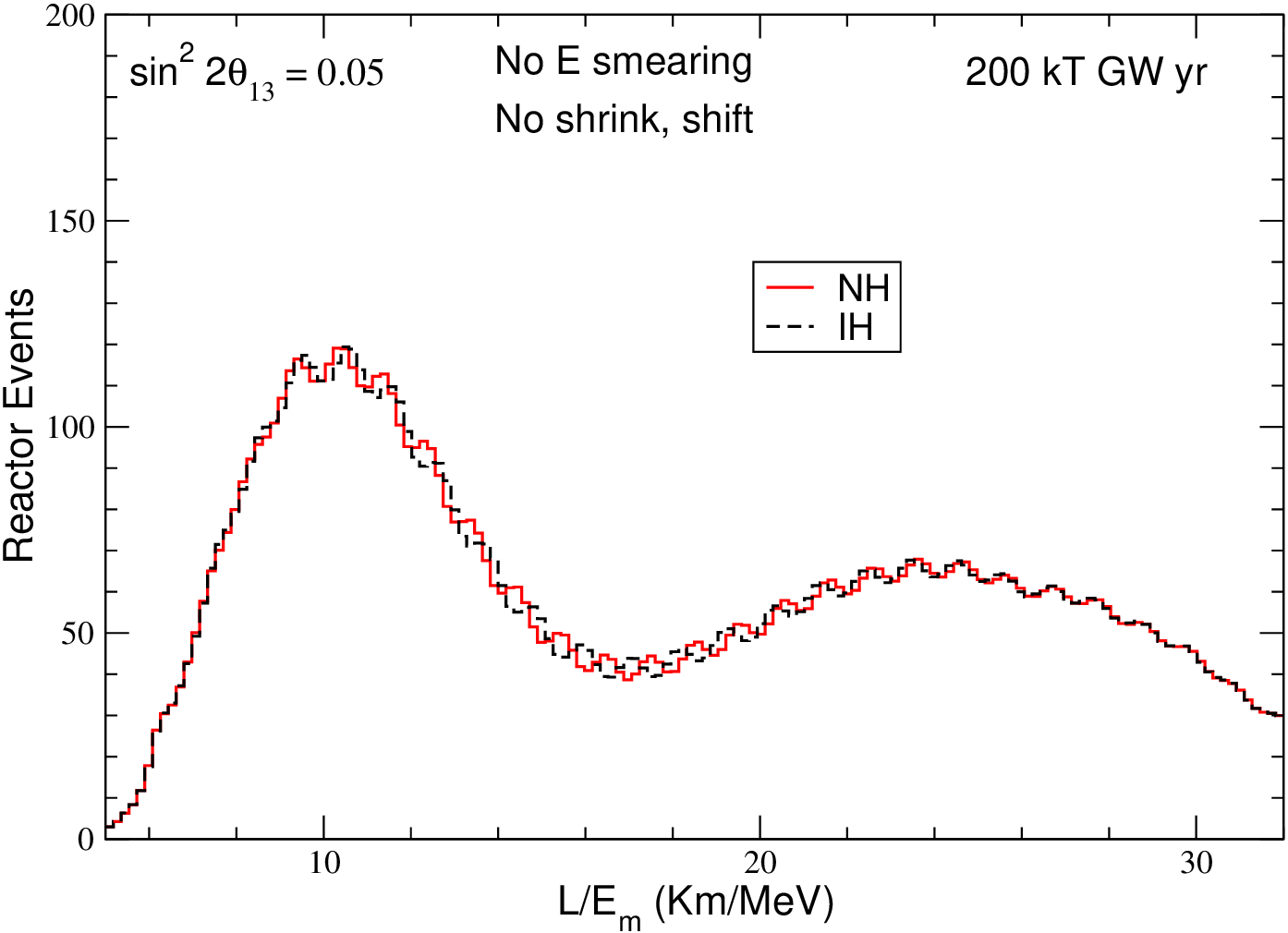}
        \hspace*{0.7ex}
\epsfxsize=8.5cm\epsfysize=7.0cm
\epsfbox{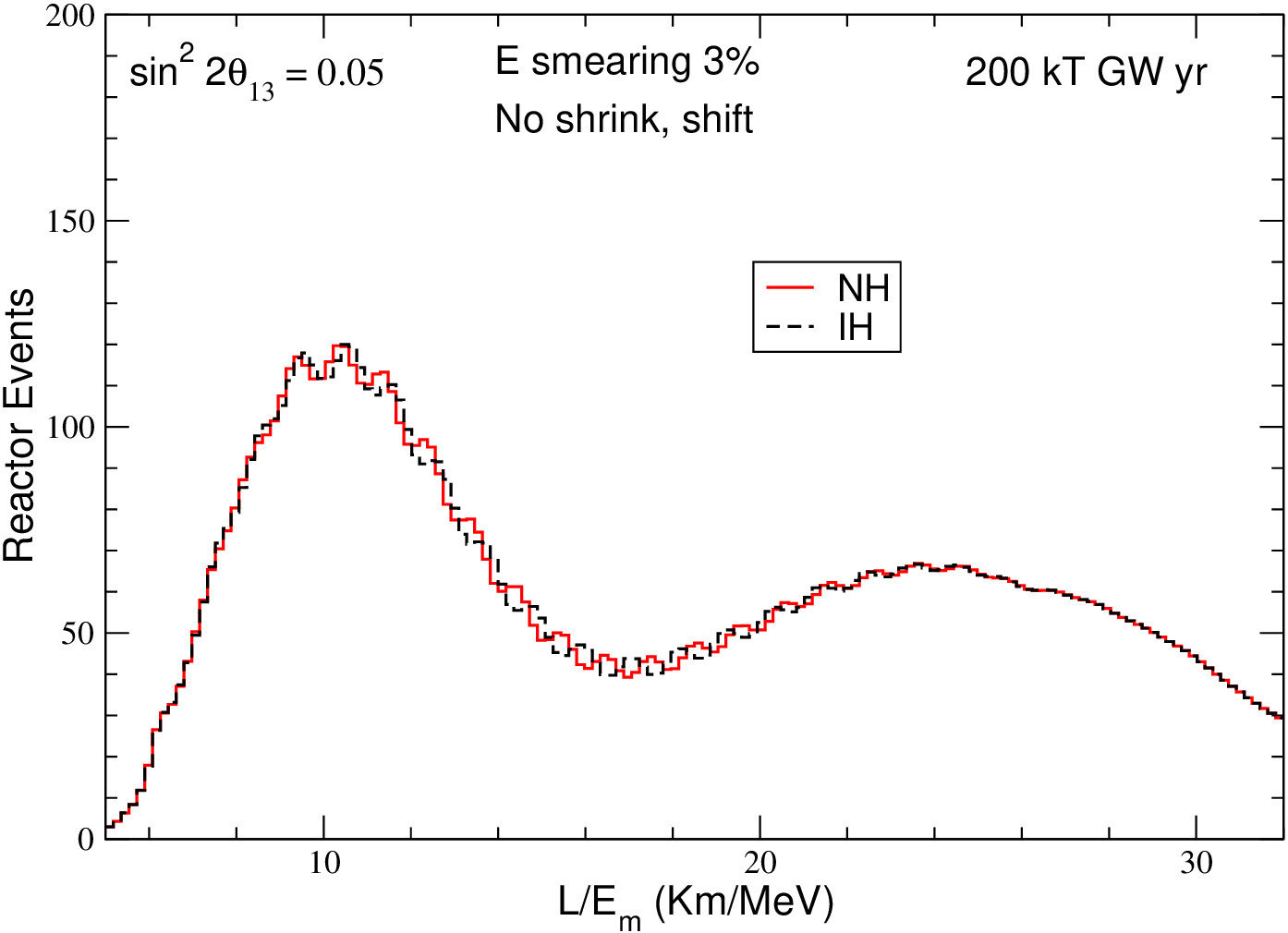}
}
\hskip 5cm
{\bf (a)}
\hskip 7cm
{\bf (b)}
\caption[]
{\footnotesize{{{\bf (a)}
Reactor event spectrum binned in 150 $\rm{L/E_m}$ bins 
for both the normal and inverted hierarchies, 
for $\sin^2 2\theta_{13} = 0.05$, ideal energy resolution 
of the detector (no smearing) and no energy scale shift. 
{\bf (b)} The same as {\bf{(a)}} for 3 $\%$ energy resolution of the detector. 
}}}
\label{binevent1}
\end{figure}

\begin{figure}[t]
\centerline
{
\epsfxsize=8.5cm\epsfysize=7.0cm
\epsfbox{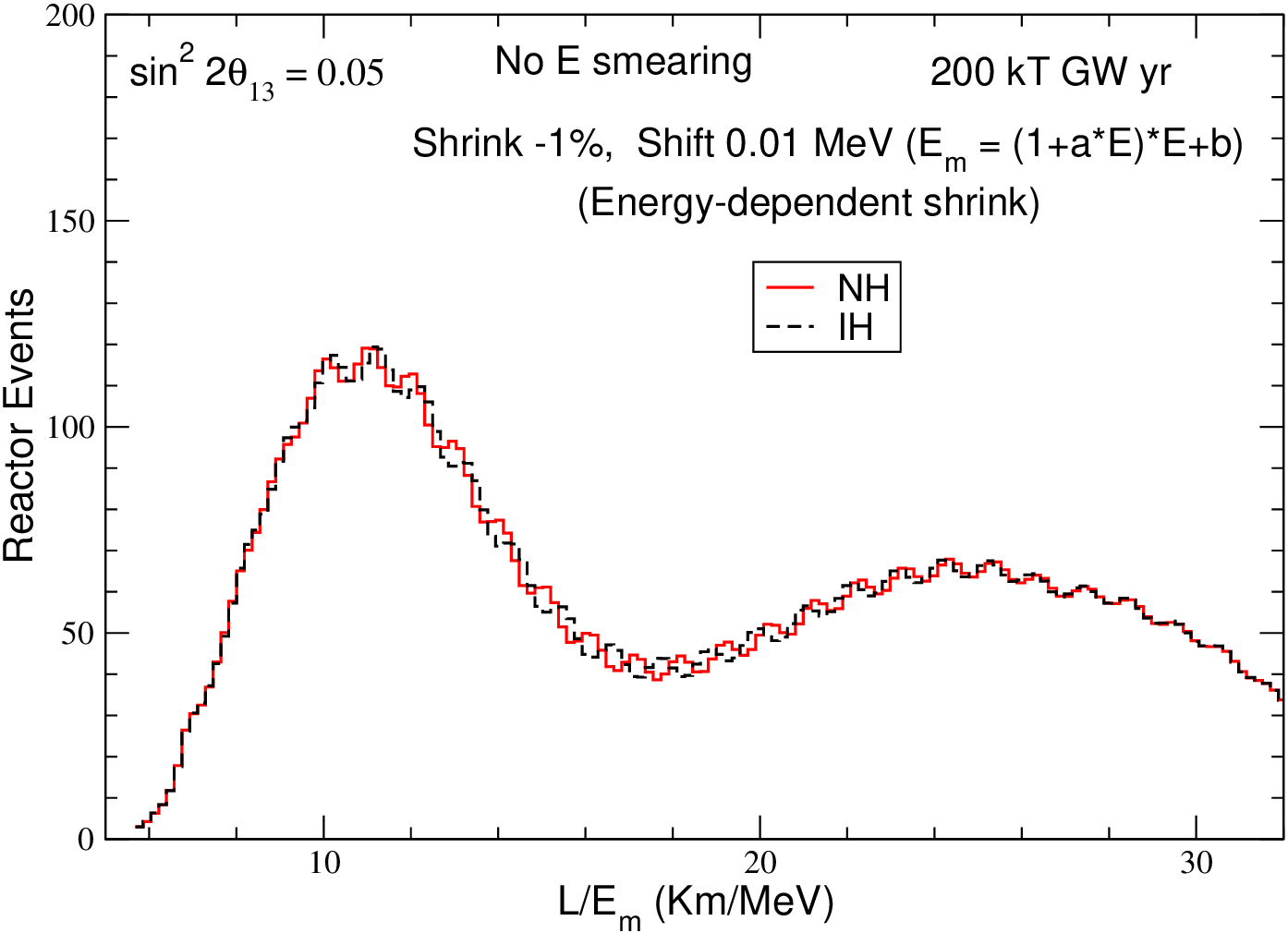}
        \hspace*{0.7ex}
\epsfxsize=8.5cm\epsfysize=7.0cm
\epsfbox{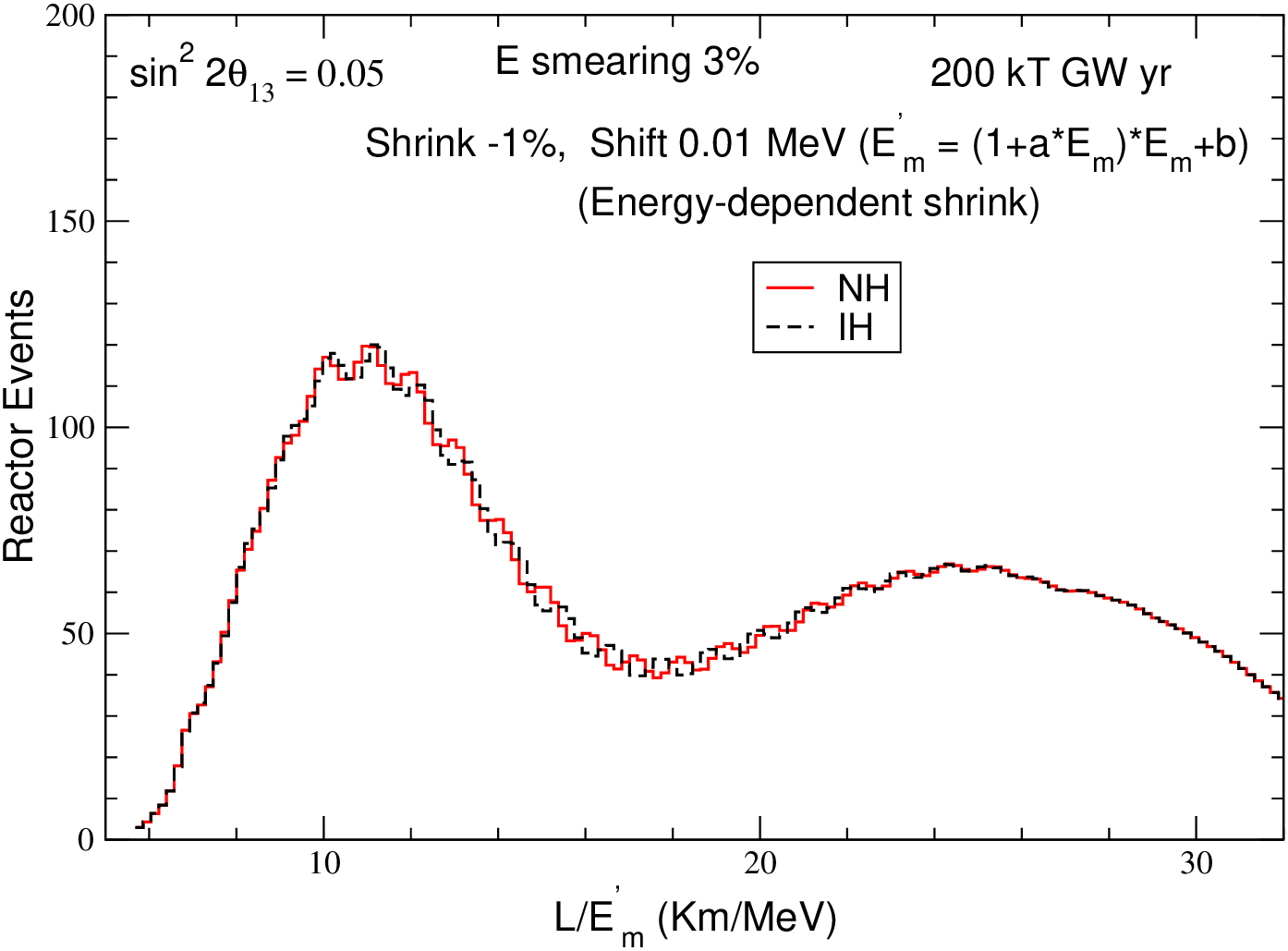}
}
\hskip 5cm
{\bf (a)}
\hskip 7cm
{\bf (b)}
\caption[]
{\footnotesize{{{\bf (a)}
Reactor event spectrum binned in 150 $\rm{L/E_m}$ bins  
for both the normal and inverted hierarchies, 
for $\sin^2 2\theta_{13} = 0.05$, for ideal 
energy resolution of the detector (no smearing) 
and an energy-dependent uncertainty 
(shrink/expansion and shift) in the energy scale. 
{\bf (b)} The same as {\bf{(a)}} for 3 $\%$ energy resolution of the detector. 
}}}
\label{binevent2}
\end{figure}

\begin{figure}[t]
\centerline
{
\epsfxsize=8.5cm\epsfysize=7.0cm
\epsfbox{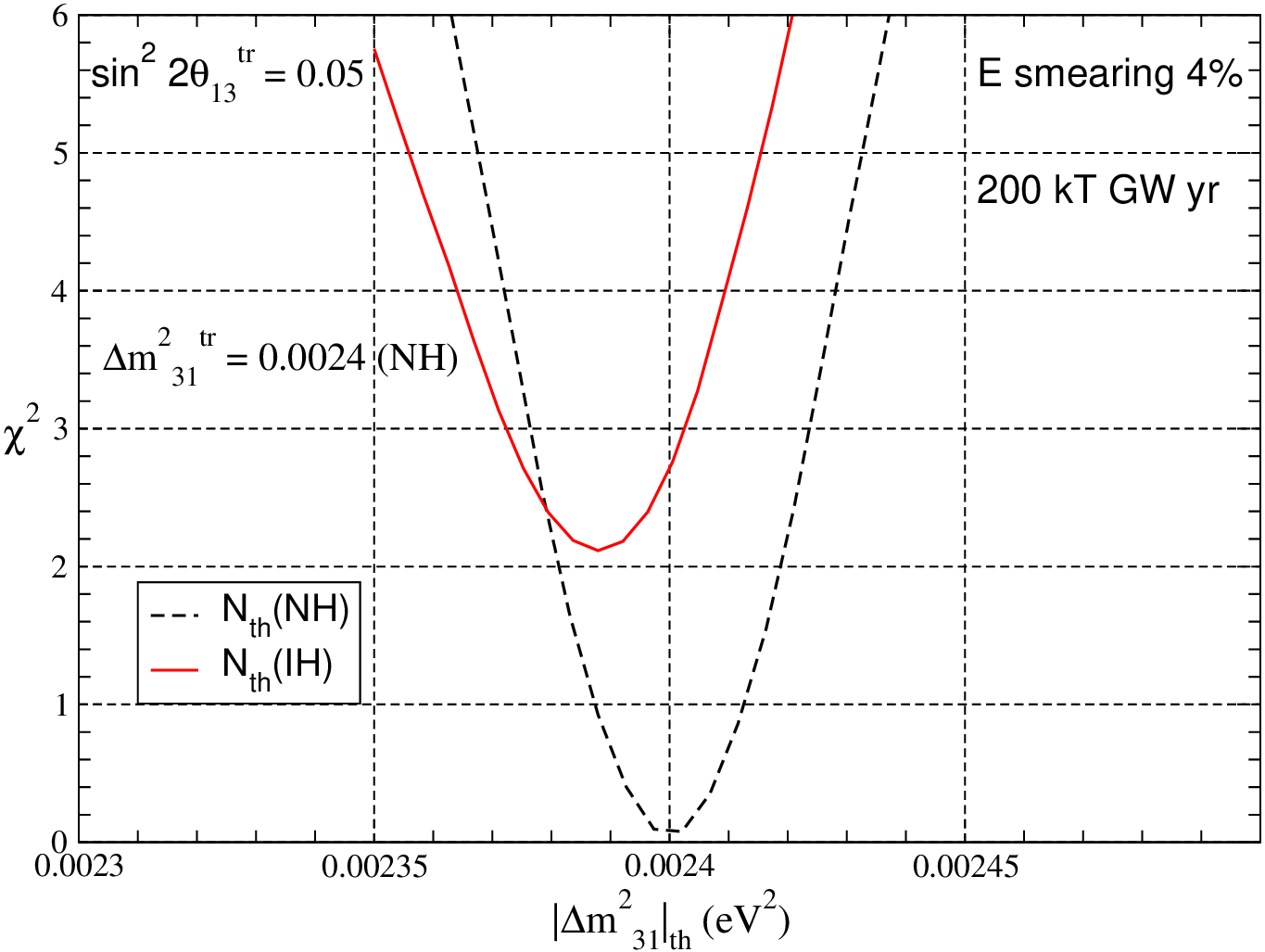}
}
\caption[]
{\footnotesize{
The function $(\chi^2)_{stat}$, marginalized over 
$\theta_{13}$, versus 
$|\Delta m^2_{31}|$ 
for ${\rm{\sin^2 2\theta_{13}^{true}}} = 0.05$ and 
a detector energy resolution of 4$\%$, 200 kT GW yr 
detector exposure and 60 Km baseline.  
The figure is obtained from a 150-bin analysis in 
the range ${\rm{L/E}} = 5-32$ Km/MeV.
The true hierarchy is chosen to be normal.
The dashed (solid) curve corresponds to the NH spectrum
(``wrong'' IH spectrum).
}}
\label{chisqvsDel31}
\end{figure}

{\bf{Energy scale uncertainty.}} We have checked that including 
the energy scale shift and shrink/expansion in the event 
spectrum has no effect on the hierarchy sensitivity, 
either with an energy-dependent or energy-independent 
shrink/expansion. This is because, as observed in 
Figs.~\ref{fig2}, \ref{fig3}, \ref{binevent1} and \ref{binevent2}, 
the effect of a scale shrink/expansion and shift  
is identical  on the event spectra for the 
normal and inverted hierarchies, 
irrespective of the different kinds of changes it 
produces in the spectrum for a specific hierarchy.
In other words, the shift or shape variations caused 
by an energy scaling do not lead to any change 
in the relative positions and behaviour of the 
NH and IH spectra. Hence the hierarchy sensitivity is unaffected.
\vskip.4cm
\begin{table}[t]
\begin{center} 
\begin{tabular}{| c || c | c | c | }
\hline 
 {\sf{$(\chi^2)_{stat}^{min}$}}  & {\sf {Energy resolution}} & & \\
        \hline
        \hline
{\sf {$\sin^2 2\theta_{13}^{\rm{true}}$}} & 2$\%$ & 3$\%$ & 4$\%$ \\
        \hline 
        \hline
         0.02 & 0.55  & 0.44  & 0.33   \\
         \hline
          0.05  & 3.50 & 2.79 & 2.11 \\
           \hline
            \end{tabular}
            \caption[]{\footnotesize{
Values of $(\chi^2)_{stat}^{min}$
marginalized over the parameters $\theta_{13}$ and 
$|\Delta m^2_{31}|$ for two values of $\sin^2 2\theta_{13}^{\rm{true}}$ 
and three values of the detector energy resolution, for
a detector exposure of 200 kT GW yr and a baseline of 60 Km.
The values are obtained in an analysis 
using 150 L/E bins in the range 5 - 32 Km/MeV.}} 
            \label{table2}
            \end{center}
            \end{table}

{\bf{Priors.}} Prior experimental information 
regarding the other neutrino parameters can be 
included in the analysis in the form of ''priors'',
defined as:
\begin{equation}
{{\chi^2_{prior}}}   =  
\left(\frac{|\Delta m^2_{\rm{atm}}|^{\mathrm{true}} - |\Delta m^2_{\rm{atm}}|}
{\sigma(|\Delta m^2_{\rm{atm}}|)}\right)^2  \nonumber \\
+ \left(\frac{{\sin^2 2\theta_{13}^{\mathrm{true}}} - \sin^2 2\theta_{13}}
{\sigma(\sin^2 2\theta_{13})}\right)^2 
\label{chisqfinal}
\end{equation}

\noindent Here $|\Delta m^2_{\rm{atm}}|$ and $\sin^2 2\theta_{13}$ 
are the values of the marginalized parameters 
in the $N_{th}$ spectrum, 
$|\Delta m^2_{\rm{atm}}|^{\mathrm{true}}$ and 
$\sin^2 2\theta_{13}^{\mathrm{true}}$ are the values 
fixed in the $N_{ex}$ spectrum, 
and $\sigma(|\Delta_{atm}|)$ and $\sigma(\sin^2 2\theta_{13})$ 
are the present 1$\sigma$ error ranges of the respective parameters,
here taken to be  
$\sigma(|\Delta m^2_{\rm{atm}}|)=
5$\%$ \times |\Delta m^2_{\rm{atm}}|^{\mathrm{true}}$ 
and $\sigma(\sin^2 2\theta_{13})=0.02$. 
This quantity serves as a penalty 
for moving away from the ''true'' value of a parameter, 
since this would obviously worsen the fit
with the (other) experiment(s) which measured the parameter. 
So adding the ''prior'' term to the $\chi^2$ and 
then performing the marginalization
effectively minimizes the $\chi^2$ over our data as well 
as that of the other experiment(s) which measured the parameters. 

\vskip.3cm

Table 2 lists the values of the hierarchy 
sensitivity $[(\chi^2)_{stat}^{min}]_{prior}$ 
for different values of $\theta_{13}$ and the 
detector energy resolution, after a marginalization 
over the above parameter ranges with priors 
taken into account, for the same values of 
detector exposure and event binning. 
There is a slight improvement in the results 
with the inclusion of priors. It may be noted 
here that if an improved 1$\sigma$ error of 
$\sigma(|\Delta m^2_{\rm{atm}}|)=1$\%$ \times |\Delta m^2_{\rm{atm}}|^{\mathrm{true}}$ 
in the atmospheric mass-squared difference is 
considered (which may be possible from future precision 
experiments), the improvement in the hierarchy 
sensitivity with the inclusion of the prior term 
is more pronounced. 
For example, the value of $[(\chi^2)_{stat}^{min}]_{prior}$ 
for ${\rm{\sin^2 2\theta_{13}^{true}}}=0.05$ and a 
detector resolution of 4$\%$ (second row, 
last column in Table 2) becomes 2.6 
in this case. Since $|\Delta m^2_{\rm{atm}}|$ is likely 
to be determined with increasingly better precision 
before the hierarchy ambiguity is resolved, 
it may be useful to include prior information 
in this way from measurements of $|\Delta m^2_{\rm{atm}}|$, 
when studying the hierarchy sensitivity of an experiment.  
 
\vskip.3cm

{\bf{Detector exposure.}} In Table 3, we give 
the values of the hierarchy 
sensitivity $[(\chi^2)_{stat}^{min}]$ for 
$\sin^2 2\theta_{13}^{\mathrm{true}}=0.02$, 
for 3 different values of the detector resolution 
and a scaling in the detector exposure. These 
results show the strong dependence of the 
sensitivity on the detector exposure, 
which is a function of the detector mass, 
power and time of running. In other words, 
the sensitivity is directly related to the 
statistics or total event number of the 
reactor experiment. Hence, a hierarchy 
sensitivity of $> 1.5\sigma$ may be possible 
even for $\sin^2 2\theta_{13}^{\mathrm{true}}=0.02$
with an exposure of 1000 kT GW yr and an 
energy resolution of 2$\%$, 
and this would improve further with a higher detector mass/power.
With a larger value, like $\sin^2 2\theta_{13}^{\mathrm{true}}=0.05$,
an exposure of 1000 kT GW yr may give a hierarchy sensitivity 
of $> 3 \sigma$ even for an energy resolution of 4$\%$.

\vskip.3cm
 
These results can be compared with the results 
for hierarchy sensitivity in \cite{Hano2}, 
where the detector exposure (in kT GW yr) required 
to obtain a sensitivity of 1$\sigma$ or $66.8\%$ C.L., 
is plotted as a function of the neutrino baseline 
or the energy resolution of the detector. 
The authors of \cite{Hano2} 
find that for an energy resolution of 2$\%$, 
for a baseline of 60 Km, $\sin^2 2\theta_{13}=0.05$
and best-fit values of other parameters, 
an exposure of about 220-230 kT GW yr will be required to 
obtain a sensitivity of 1$\sigma$.
Similar parameter values of the baseline, 
$\theta_{13}$, detector exposure and energy resolution give
$\chi^2_{stat} = 3.5$, or a 
sensitivity of 1.8$\sigma$, in our analysis.

\vskip.3cm

{\bf{Systematic errors.}} Apart from the uncertainties 
in the neutrino parameters, the systematic uncertainties related to 
the detector and geo-neutrinos also need to be 
included in a realistic analysis. In this case, 
we consider 5 sources of systematic uncertainties 
(3 from the detector and 2 from geo-neutrinos) 
\cite{Hano2}, as mentioned earlier, for which 
the following values are taken:

\vskip.3cm

\begin{itemize}
\item The efficiency error, 2$\%$.
\item The uncertainty in the estimation of the 
detector energy resolution, 8$\%$.
\item The linear energy scale uncertainty, 1$\%$.
\item The uncertainty in the total detectable terrestrial 
antineutrino flux, 10$\%$.
\item The uncertainty in the ratio of $\bar{\nu}_e$ 
from the decay of U-238 and Th-232, 10$\%$.
\end{itemize}

The last two errors may be quite large, 
but varying them to higher values has no significant 
effect on the results. 

\vskip,3cm

 We take into account the above uncertainties using 
the method of pulls (see, e.g., \cite{Fogli:2002pt}).
In this method, the inputs (quantities having 
systematic uncertainties) are 
allowed to deviate from their standard values 
in the computation of
$N_{th}(i)$. If the ${\rm{j}}$th input deviate 
from its standard value
by ${\rm{\sigma_j \xi_j}}$, where ${\rm{\sigma_j}}$ 
is the magnitude of the corresponding uncertainty, then the
value of $N_{th}(i)$ with the changed inputs is given by

\begin{equation}
N_{th}(i) = N_{th}(i)(std) + {\rm{\sum^{npull}_{j=1}
c_{i}^j \xi_j}}\,,
\label{ci}
\end{equation}

\noindent where $N_{th}(i)(std)$ is the theoretical rate for the
${\rm{{i}^{th}}}$ bin, calculated with the standard values of the inputs
and ${\rm{npull}}$ is the number of sources of uncertainty,
which in our case is 5.
The ${\rm{\xi_j}}$'s are called the "pull" 
variables and they determine
the number of ${\rm{\sigma's}}$ by which the ${\rm{j}}$th 
input deviates from its
standard value. In Eq.{\ref{ci}}, ${\rm{c_{i}^j}}$ is the change in
$N_{th}(i)$ when the ${\rm{j}}$ th input is changed 
by ${\rm{\sigma_j}}$ (i.e. by 1 standard deviation).
The shifted event rate defines a modified 
$\chi^2$ which is then minimized with respect to
the pull variables.

\vskip.3cm

Implementing this method with the error 
parameter values given above
is found to have only a minimal effect 
on the hierarchy sensitivity. 
For example, the value of $[(\chi^2)_{stat}^{min}]_{prior}$ 
for ${\rm{\sin^2 2\theta_{13}^{true}}}=0.05$ 
and a detector resolution of 4$\%$ 
(second row, last column in Table 2) 
changes only from 2.25 to 2.26 with the 
inclusion of systematic uncertainties. Hence 
we conclude that the hierarchy sensitivity
from a reactor antineutrino experiment is 
strongly dependent on the detector 
energy resolution, the exposure (statistics) 
and the value of the parameter $\theta_{13}$, 
but has a weak dependence on the values of 
the systematic errors of the detector, 
as long as they do not exceed $\sim 10\%$,
and on the flux uncertainty due to the 
geo-neutrinos. 

\begin{table}[t]
\begin{center} 
\begin{tabular}{| c || c | c | c | }
\hline 
 {\sf{$[(\chi^2)_{stat}^{min}]_{prior}$}}  & {\sf {Energy resolution}} & & \\
        \hline
        \hline
{\sf {$\sin^2 2\theta_{13}^{\rm{true}}$}} & 2$\%$ & 3$\%$ & 4$\%$ \\
        \hline 
        \hline
         0.02 & 0.57 & 0.46 & 0.37   \\
         \hline
          0.05  & 3.64 & 2.93 & 2.25   \\
           \hline
            \end{tabular}
            \caption[]{\footnotesize{Values of $[(\chi^2)_{stat}^{min}]_{prior}$
            marginalized over the parameters 
$\theta_{13}$ and $|\Delta m^2_{31}|$ with priors included,
for $\sin^2 2\theta_{13}^{\rm{true}} = 0.02;~0.05$ 
and three values of the detector's energy resolution.
The baseline, detector exposure and 
event binning are the same as those used to obtain Table 1.}} 
            \label{table3}
            \end{center}
            \end{table}
\vskip.4cm
\begin{table}[t]
\begin{center}
\begin{tabular}{| c || c | c | c || c | c | c | }
\hline
 {\sf{$(\chi^2)_{stat}^{min}$}} & $\sin^2 2\theta_{13}^{\rm{true}}=0.02$ & 
& & $\sin^2 2\theta_{13}^{\rm{true}}=0.05$ & & \\
 \hline
 \hline
{\sf{ Detector exposure, kT GW yr}} & {\sf {Energy resolution}} & & & & & \\
        \hline
        \hline
 & 2$\%$ & 3$\%$ & 4$\%$ & 2$\%$ & 3$\%$ & 4$\%$ \\
        \hline
        \hline
         200 & 0.55  & 0.44  & 0.33 & 3.50 & 2.79 & 2.11 \\
         \hline
          400  & 1.10 & 0.88 & 0.66 & 7.0 & 5.58 & 4.22 \\
           \hline
            600   & 1.65  & 1.32  & 0.98 & 10.50 & 8.37 & 6.33 \\
            \hline
             800  & 2.20 &  1.75 & 1.30 & 14.0 & 11.15 & 8.40 \\
            \hline
             1000  & 2.70 &  2.15 & 1.60 & 17.20 & 13.80 & 10.50 \\
             \hline
            \end{tabular}
            \caption[]{\footnotesize{
Values of $(\chi^2)_{stat}^{min}$
marginalized over the parameters $\theta_{13}$ and $|\Delta m^2_{31}|$
for several different detector exposures (in kT GW yr),
$\sin^2 2\theta_{13}^{\rm{true}}=0.02;~0.05$, three values of 
the detector's energy resolution
and a baseline of 60 Km, obtained
in an analysis using 150 L/E bins in the range 5 - 32 Km/MeV.
Including priors in the analysis increases
the sensitivity to the type of the neutrino mass spectrum. 
}}
            \label{table4}
            \end{center}
            \end{table}

%
\section{Conclusions}
%

 In the present article we have studied 
the possibility to determine the type of 
neutrino mass spectrum, i.e., ``the neutrino mass hierarchy'',
in a reactor $\bar{\nu}_e$ experiment with a relatively 
large KamLAND-like detector and an optimal baseline of 60 Km.
This possibility has been previously investigated in \cite{PiaiP0103},
and further in \cite{SCSPMP03} using the $\chi^2$-method, and in
\cite{Hano1,Hano2,YWang08} using the method of Fourier
transforms of simulated data 
and the method of maximum likelihood analysis. 
Here we first analyzed systematically the 
Fourier Sine and Cosine Transforms (FST and FCT) 
of simulated reactor antineutrino data with reference to their
specific neutrino mass hierarchy-dependent features 
discussed in \cite{YWang08}. 
In the second part of the study we performed a binned 
$\chi^2$ analysis of the sensitivity of the simulated data
to the mass hierarchy. We considered a detector 
with a mass of the order of 10 kT, similar 
to the one proposed for the Hanohano experiment \cite{Hano1},
using a $\bar{\nu}_e$ flux from a reactor having power 
of 5-10 GW and thus providing high statistical samples 
of $\sim 10^{4}$ or more events. The threshold of the 
measured $e^+$ (i.e., visible) 
energy was set to $E_{visth} = 1$ MeV.
We have considered values of detector's
energy resolution  $\sigma/E_{vis}$
in the interval $2\%/\sqrt{E_{vis}} - 4\%/\sqrt{E_{vis}}$;
in a few cases larger values have been 
utilized for clarifying and illustrative purposes.

\vskip.3cm

The investigation of the neutrino mass 
hierarchy sensitive features of the FST and FCT 
spectra was performed, in particular,
taking into account  the possibility 
of an energy scale uncertainty in the form of 
scale shrink/expansion and shift.
We have considered not only energy-independent,
but also energy-dependent scale factors, 
more specifically, scale factors which depend 
linearly on the energy. Our findings can be summarized as follows.\\

\noindent 1. The hierarchy-sensitive features in both the FCT and FST spectra discussed in 
\cite{YWang08} are progressively reduced in 
magnitude with the worsening 
of the detector's energy resolution 
(i.e., with the increasing of  $\sigma/E_{vis}$).\\

\vskip.02cm 

\noindent 2. An energy-independent energy scale 
uncertainty (shrink/expansion and shift), leaves 
these features substantially unchanged, 
since the shapes of the FST and FCT spectra 
suffer only sideways shifts.\\

\vskip.02cm 

\noindent 3. The asymmetry feature distinguishing between 
the two hierarchies discussed in \cite{YWang08}
is no longer present with the inclusion of 
an energy-dependent energy scale 
shrink/expansion (compare, e.g., 
Figure~\ref{FT4}(a) with Figure~\ref{FT5}(b)): 
the absolute modulation maxima 
in the cases of normal hierarchical (NH)
and inverted hierarchical (IH) spectra
are effectively replaced by absolute 
modulation minima for values of the shrink factor of about 0.7$\%$ or larger, 
while the adjacent valleys (minima) are replaced by adjacent peaks (maxima). 
We have defined the quantity 
\be
RLP = \frac{RP-LP}{RP+LP}\,,
\label{RLP1}
\ee
%
where $RP$ and $LP$ are the amplitudes of the right and left peaks 
adjacent to the absolute modulation minima.  
The asymmetry $RLP$ has a significant positive value for the
NH spectrum and a much smaller value close to zero for the
IH spectrum.  
For smaller values of the energy-dependent shrink factor, 
there is a continuous left-shift and change in shape of
the Fourier spectrum leading to a conversion from the $RL$
or $PV$ asymmetry feature to the $RLP$ asymmetry feature.
For values of the shrink factor between $\sim 0.4\% - 0.6\%$,
it is difficult to identify specific asymmetry feature.

\vskip.3cm

   These  properties of the Fourier spectra indicate that 
it should be possible, in principle, to extract 
information about the type of the spectrum 
the neutrino masses obey from the features 
present in the spectra, 
although the nature of the hierarchy-dependent features 
is changed in the case of an energy-dependent 
energy scale shrink/expansion. 

\vskip.3cm

 The effect of varying the atmospheric neutrino
mass-squared difference $\Delta m^2_{\rm atm}$ 
over its error range, causes, in general, a change in the  
magnitude of the hierarchy-sensitive asymmetry 
features of the FST and FCT spectra without 
eliminating them completely.

\vskip.3cm

 We have performed also a statistical study of the possible sensitivity
of such a reactor antineutrino experiment to the type of the 
neutrino mass spectrum. 
We adopted the method of a binned $\chi^2$ analysis, which
offers the advantages of a straightforward
incorporation of i) parameter
uncertainties, ii) detector characteristics like the energy resolution and
energy scale uncertainty, iii) systematic errors 
(for which we use the method of pulls),
iv) an optimized binning of data to reach the maximum possible
sensitivity while being consistent with the detector resolution,
and v) the inclusion of external information on the neutrino parameters
using priors.
The $\chi^2$ survey was performed using an exposure of 200 - 1000 kT
GW yr, and the results were presented for different values of the
detector resolution, detector exposure, and the true value of $\theta_{13}$,
with a marginalization over all neutrino parameters. The bin number was
optimized at 150. The true spectrum was assumed to be 
with normal ordering (NH). 
The results of this analysis can be summarized as follows.

\vskip.3cm
 
The hierarchy sensitivity depends strongly on the 
the true value of $\theta_{13}$, the energy resolution
of the detector, the detector exposure
and on the binning of the spectrum data.
It improves dramatically with an increase in
${\rm{\theta_{13}^{true}}}$, increases linearly with the exposure 
(due to the increase in statistics), 
and falls significantly with worsening 
resolution. For example, $(\chi^2)^{min}_{stat}$ 
for the ``wrong'' hierarchy improves from 0.55 for 
$\rm{sin^2 2\theta_{13}^{true}} = 0.02$, 
an energy resolution of 2$\%$ and a detector exposure of 200 kT GW yr 
(corresponding to a hierarchy sensitivity of less than 1$\sigma$), 
to 3.5 (a sensitivity of 1.8$\sigma$)
for $\rm{sin^2 2\theta_{13}^{true}} = 0.05$ for the same 
values of the resolution and exposure.
With an exposure of 1000 kT GW yr and 
the same values of the resolution and ${\rm{\theta_{13}^{true}}}$,
it increases to  2.7 (1.6$\sigma$).
On the other hand, if the energy resolution has a 
value of 3$\%$, the $(\chi^2)^{min}_{stat}$ falls
to 0.44 for the same values of ${\rm{\theta_{13}^{true}}}$ and exposure, 
and a significant sensitivity ($>2\sigma$)
can be achieved only if the exposure is scaled up to 
higher than 1000 kT GW yr.  

\vskip.3cm
 
A marginalization over the error ranges of the parameters
$\Delta m^2_{\rm{atm}}$ and $\theta_{13}$ has a significant effect 
in the case of $\Delta m^2_{\rm{atm}}$,
and a mild effect in the case of $\theta_{13}$. 
Varying the solar parameters $\Delta m^2_{21}$ and $\theta_{12}$ 
within their 3$\sigma$ ranges leaves
the results essentially unchanged. Moreover, since the 
currently measured value of the atmospheric neutrino 
mass-squared difference $\Delta m^2_{\rm{atm}}$ is, in general, in
between $\Delta m^2_{31}$ and $\Delta m^2_{32}$, it is
important to take into account the possible range of $\Delta m^2_{31}(IH)$
and $\Delta m^2_{32}(IH)$ with respect to the assumed true values of
$\Delta m^2_{31}(NH)$ and $\Delta m^2_{32}(NH)$, when computing the $\chi^2$
sensitivity. It is found that since the $\bar{\nu}_e$ 
survival probability $P_{{\bar{e}}{\bar{e}}}(L/E)$  
never becomes identical for any pair 
of possible values of $\Delta m^2_{31}(NH)$ and 
$\Delta m^2_{31}(IH)$ within the L/E range relevant for our analysis, 
the marginalized $\chi^2$ remains non-zero over the entire allowed range of 
$\Delta m^2_{31}(IH)$, and, if $\theta_{13}$ is sufficiently large,
it can assume significant values for the 
exposures and energy resolutions considered.  

\vskip.3cm
 
The sensitivity does not depend significantly on the energy scale 
uncertainty (up to a value of about 5$\%$), even in the case 
of a  scale uncertainty factor
which depends linearly on the energy. 
This is due to the fact 
that the scale shift affects the event spectra in the cases of  
the NH and IH neutrino mass spectra in the same way.
We found also that the effect of systematic errors 
(assumed to be smaller than $\sim 10\%$)
and geo-neutrino flux uncertainties 
is insignificant (less than 1$\%$).

\vskip.3cm
 
The number of L/E bins in the analysis strongly influences the
$\chi^2$ value for the ``wrong'' hierarchy. 
The value of $\chi^2$ increases three-fold when the bin
number is increased from 40 to 150. However, the allowed bin number is
constrained by the detector's energy resolution and the requirement
that the bin width is not smaller than the 
resolution width. Hence, the optimization of binning is important. 
Also, increasing the threshold of the visible energy 
in the analysis from $E_{visth} = 1.0$ MeV to $E_{visth} = 1.8$ MeV
({\it{i.e.}} putting a higher cut-off of 2.6 MeV on the ${\bar{\nu_e}}$
energy spectrum) significantly worsens the sensitivity, because of
the corresponding loss of statistics. 
If, for instance, we choose $E_{visth} = 1.8$ MeV 
and perform an analysis with 25 L/E bins,
we obtain a poor $(\chi^2)^{min}_{stat} = 0.8$ 
even for as high a value of $\theta_{13}$ as   
$\rm{sin^2 2\theta_{13}^{true}} = 0.1$, 
an exposure of 200 kT GW yr 
and an energy resolution of 3$\%$.
With  $E_{visth} = 1.0$ MeV, 150 L/E bins,
$\rm{sin^2 2\theta_{13}^{true}}=0.05$ and the same values of 
exposure and energy resolution,
we get  $(\chi^2)^{min}_{stat} = 2.8$. 
The worsening of the hierarchy sensitivity
with the increase of $E_{visth}$ to 1.8 MeV
occurs in spite of the fact that the 
increased threshold
excludes the contribution to the signal 
due to geo-neutrinos.  
This is because the total statistics has a much more
dramatic effect on the hierarchy sensitivity:
in the case of a sufficiently large statistics the
geo-neutrino uncertainties play essentially a 
negligible role. 

\vskip.3cm
 
The addition of external information in the form of priors has
only a minor effect on the sensitivity ($\sim$ 5$\%$) with the present
1$\sigma$ error range of 5$\%$ in $|\Delta m^2_{\rm{atm}}|$. 
The contribution of priors becomes
important if a prospective precision of 1$\%$ on
$|\Delta m^2_{\rm{atm}}|$ is considered, leading to an 
improvement of $\sim$20$\%$.
For example, for a $|\Delta m^2_{\rm{atm}}|$ error range of 5$\%$, 
${\rm{\sin^2 2\theta_{13}^{true}}}=0.05$ and a
detector resolution of 4$\%$,
the value of $[(\chi^2)_{stat}^{min}]_{prior} = 2.25$ 
(as compared to $(\chi^2)^{min}_{stat} = 2.11$ without priors),
but with an improved $|\Delta m^2_{\rm{atm}}|$ 
error range of 1$\%$ (which may be possible from future precision
experiments), the value of $[(\chi^2)_{stat}^{min}]_{prior} = 2.6$.
Since the neutrino 
parameters are likely to be measured with improved 
precision before the neutrino mass hierarchy 
is determined, it is useful to include prior information
from other experiments in this way. 

\vskip.3cm

Our results show that if 
$\sin^2 2\theta_{13}$ is sufficiently large, 
$\sin^2 2\theta_{13} \gtap 0.02$,
it would be possible to get a significant information
on, or even determine, the type of neutrino mass spectrum 
(i.e., the neutrino mass hierarchy) 
in a high statistics experiment with reactor $\bar{\nu}_e$ with 
a baseline of 60 km, using a relatively large 
KamLAND-like detector of mass $\sim 10$ kT,
having an energy resolution of 
$\sigma/E_{vis} \sim (2\%/\sqrt{E_{vis}} - 4\%/\sqrt{E_{vis}})$ 
and an exposure of at least 200 kT GW yr.
These requirements on the set-up are very challenging, 
but not impossible to realize.

%
\acknowledgments{We thank M. Roos for reading the manuscript 
and useful comments.
P.G. thanks S. Uma Sankar, Danny Marfatia, Srubabati Goswami and 
Raj Gandhi for useful discussions and Liang Zhan for helpful communication. 
This work was supported in part by the INFN program 
on ``Astroparticle Physics'' and
by the World Premier International Research Center 
Initiative (WPI Initiative), MEXT, Japan  (S.T.P.).}

\end{document}